\documentclass[fleqn,usenatbib]{mnras}

\usepackage{newtxtext,newtxmath}

\usepackage[T1]{fontenc}
\usepackage{ae,aecompl}

\usepackage{graphicx}	
\usepackage{amsmath}	
\usepackage{amssymb}	
\usepackage{multirow}
\usepackage{subfigure}
\usepackage{subfloat}
\usepackage{float}
\usepackage{morefloats}
\usepackage{longtable}
\usepackage{supertabular}
\usepackage{lscape}
\usepackage{multicol}

\newcommand{\total}{72}
\newcommand{\rhonolims}{$\rho\,=\,-0.28\pm0.12$}
\newcommand{\lims}{18}
\newcommand{\nolims}{54}
\newcommand{\rhoneg}{98.7\%}
\newcommand{\iterations}{20,000}

\newcommand{\MIRdominated}{38}
\newcommand{\MIRdominatedFIT}{$\rm{m}\,=\,-0.04\pm0.06$}

\newcommand{\lored}{40}

\newcommand{\srcperbin}{9}
\newcommand{\binFIT}{-0.06^{+0.05}_{-0.08}}
\newcommand{\medred}{0.038}

\newcommand{\ratioLTtwo}{36}
\newcommand{\ratioLTtwoFIT}{$\rm{m}\,=\,-0.08\pm0.05$}

\newcommand{\REFLdom}{14}
\newcommand{\REFLdomFIT}{$\rm{m}=-0.10^{+0.07}_{-0.08}$}

\newcommand{\LTmedzFIT}{$\rm{m}=-0.19^{+0.06}_{-0.07}$}
\newcommand{\GTmedzFIT}{$\rm{m}=-0.04\pm0.08$}

\title[An Iwasawa-Taniguchi Effect for Compton-thick AGN]{An Iwasawa-Taniguchi Effect for Compton-thick Active Galactic Nuclei}
\author[P. G. Boorman et al.]{
Peter G. Boorman,$^{1}$\thanks{P.G.Boorman@soton.ac.uk}
Poshak Gandhi,$^{1}$
Mislav Balokovi{\'{c}},$^{2,\,3}$
Murray Brightman,$^{2}$\newauthor 
Fiona Harrison,$^{2}$
Claudio Ricci$^{4,\,5,\,6,\,7}$
and Daniel Stern$^{8}$
\\
\\
$^{1}$Department of Physics \& Astronomy, Faculty of Physical Sciences and Engineering, University of Southampton, Southampton,\\
SO17 1BJ, UK\\
$^{2}$Cahill Center for Astronomy and Astrophysics, California Institute of Technology, Pasadena, CA 91125, USA\\
$^{3}$Harvard-Smithsonian Center for Astrophysics, 60 Garden Street, Cambridge,
MA 02138, USA\\
$^{4}$Instituto de Astrof\'{\i}sica, Facultad de F\'{i}sica, Pontificia Universidad Cat\'{o}lica de Chile, Casilla 306,\\
Santiago 22, Chile\\
$^{5}$N\'ucleo de Astronom\'ia de la Facultad de Ingenier\'ia, Universidad Diego Portales, Av. Ej\'ercito Libertador 441, Santiago, Chile\\
$^{6}$Kavli Institute for Astronomy and Astrophysics, Peking University, Beijing 100871, China\\
$^{7}$Chinese Academy of Sciences South America Center for Astronomy and China-Chile Joint Center for Astronomy,\\
Camino El Observatorio 1515, Las Condes, Santiago, Chile\\
$^{8}$Jet Propulsion Laboratory, California Institute of Technology, Pasadena, CA 91109, USA
}

\date{Accepted 2018 March 29. Received 2018 March 29; in original form 2017 September 28}

\pubyear{2018}

\begin{document}
\label{firstpage}
\pagerange{\pageref{firstpage}--\pageref{lastpage}}
\maketitle

\begin{abstract}
We present the first study of an Iwasawa-Taniguchi/`X-ray Baldwin' effect for Compton-thick active galactic nuclei (AGN).  We report a statistically significant anti-correlation between the rest-frame equivalent width (EW) of the narrow core of the neutral Fe\,K$\alpha$ fluorescence emission line, ubiquitously observed in the reflection spectra of obscured AGN, and the mid-infrared $12\,\mu \rm m$ continuum luminosity (taken as a proxy for the bolometric AGN luminosity).  Our sample consists of \total{} Compton-thick AGN selected from pointed and deep-field observations covering a redshift range of $z \sim 0.0014-3.7$.  We employ a Monte Carlo-based fitting method, which returns a Spearman's Rank correlation coefficient of \rhonolims{}, significant to \rhoneg{} confidence.  The best fit found is $\mathrm{log}(\mathrm{EW}_{\mathrm{Fe\,K}\alpha})\,\propto\,-0.08\pm0.04\,\mathrm{log}(L_{12\,\rm{\mu} \rm m})$, which is consistent with multiple studies of the X-ray Baldwin effect for unobscured and mildly obscured AGN.  This is an unexpected result, as the Fe\,K$\alpha$ line is conventionally thought to originate from the same region as the underlying reflection continuum, which together constitute the reflection spectrum.  We discuss the implications this could have if confirmed on larger samples, including a systematic underestimation of the line of sight X-ray obscuring column density and hence the intrinsic luminosities and growth rates for the most luminous AGN.
\end{abstract}


\section{Introduction}
\label{sec:intro}

X-ray continuum emission from active galactic nuclei (AGN) typically takes the form of a broadband powerlaw with a high-energy cut-off around 300\,keV \citep{Ballantyne14,Malizia14}, and originates from Comptonization of ultraviolet accretion disc photons in a hot X-ray corona \citep{Haardt91,Haardt93}.  Line of sight opacity alters this emission via photoelectric absorption and Compton scattering.  If properly accounted for, this can be used to predict the intrinsic spectral energy distribution of an AGN and thus indirectly study the circumnuclear environment of AGN.  Many studies have revealed that the vast majority of AGN are intrinsically obscured with hydrogen column densities ($N_{\rm H}$) greater than the Galactic value \citep[$N_{\rm H}\gtrsim 10^{22}\,\rm{cm}^{-2}$]{Risaliti99b,Burlon11,Ricci15}.  For $N_{\rm H}\lesssim10^{24}\,\rm{cm}^{-2}$, the intrinsic power law typically dominates over any other spectral features in the X-ray band.  As the column increases to $N_{\rm H}>1.5\times10^{24}\,\rm{cm}^{-2}$, the obscuring material becomes optically thick in X-rays to Compton scattering, in the \textit{Compton-thick} regime.  Here, the soft X-ray (\textit{E}\,$\lesssim 10$\,keV) spectrum is depleted and flattened due to the interplay of photoelectric absorption and Compton downscattering.  Depending on the orientation, geometry and column of the Compton-thick obscurer, the hard X-ray spectrum (\textit{E}$\,\gtrsim 10\,\rm{keV}$) can either be dominated by the direct intrinsic powerlaw component, absorbed along the line of sight (transmission-dominated Compton-thick AGN); or by a Compton-scattered reflection component, from intrinsic flux reprocessed by the obscurer into the line of sight (reflection-dominated Compton-thick AGN).

The geometrical configuration of the X-ray obscuring and reprocessing medium is typically assumed to be roughly axissymmetric but anisotropic \citep{Murphy09,Ikeda09,Brightman11b,Balokovic18}.  This is analogous to the putative torus in the Unified Model of AGN \citep{Antonucci93,Urry95,Netzer15} invoked to explain the infrared and optical emission observed from different classes of AGN as intrinsically a single class observed at different orientation angles.  The X-ray obscurer in Compton-thick AGN is what defines the spectral shape of the reprocessed reflection spectrum, which typically features two key components \citep{Lightman88,Reynolds99}:

\begin{enumerate}
\item A narrow Fe\,K$\rm{\alpha}$ fluorescence emission line arising from neutral (and hence cold) iron, with a characteristic energy of 6.4\,keV in the rest frame of the source.  This emission line is typically the most prominent in the X-ray spectra of AGN, due to a combination of the fluorescence yield and relative abundances of the gas located within the torus.

\item An underlying (flat) Compton scattered continuum with a broad `Compton hump' peaking at $\sim 30\,\rm{keV}$ formed from the combination of photoelectric absorption at \textit{E}\,$\lesssim 10\,\rm{keV}$ and Compton downscattering from higher energies.
\end{enumerate}

Modelling the strength and shape of the neutral Fe\,K$\rm{\alpha}$ fluorescence line together with the Compton hump can yield the line of sight obscuring column to a source.  This requires an observed X-ray spectrum spanning the Compton hump at $\sim 30\,\rm{keV}$ and the soft X-ray emission $\lesssim 10\,\rm{keV}$, to provide constraints on the continuum and reflection components.  However, many previous X-ray observations of AGN have typically been restricted to the \textit{E}\,$\lesssim 10\,\rm{keV}$ energy region (\textit{Suzaku} XIS, \textit{Chandra}, XMM-\textit{Newton}), completely missing the Compton hump for local sources.  This typically means that any attempt to fit AGN X-ray spectra in this energy region with the objective of constraining the line of sight $N_{\rm H}$ depends heavily on the Fe\,K$\rm{\alpha}$ fluorescence line alone, and can be uncertain.

Despite being an indicator of high obscuring columns, the equivalent width (EW) of the narrow core of the neutral Fe\,K$\alpha$ fluorescence line has been observed to anti-correlate with the underlying intrinsic X-ray continuum luminosity in samples of transmission-dominated AGN.  This effect was first reported by \citet{Iwasawa93} for a sample of 37 largely unobscured AGN, observed by the \textit{Ginga} satellite.  The best fit linear relation derived was of the form $\rm{log(EW}_{\rm{Fe\,K}\alpha})\,\propto\,-0.20\pm0.03\,\mathrm{log}(L_{2-10\,\rm{keV}})$.  This is sometimes referred to as the `X-ray Baldwin' effect due to the similarity with the study by \citet{Baldwin77} on the anti-correlation between the EW of the $\rm{C}\,\textsc{iv}\,1549\,\AA$ ultraviolet emission line and AGN continuum.  Here we refer to the X-ray Baldwin effect as the `Iwasawa-Taniguchi' effect.

The Iwasawa-Taniguchi effect has been explored in further detail for different AGN classes.  For example, \citet{Page04} reported an Iwasawa-Taniguchi effect of $\mathrm{log}(\mathrm{EW}_{\mathrm{Fe\,K}\alpha})\,\propto\,-0.17\pm0.08\,\mathrm{log}(L_{2-10\,\rm{keV}})$ for a sample of 53 type 1 AGN observed by XMM-\textit{Newton}, with the slope being consistent with that of \citeauthor{Iwasawa93}.  However, \citet{Jiang06} later reported a much shallower anti-correlation of $\mathrm{log}(\mathrm{EW}_{\mathrm{Fe\,K}\alpha})\,\propto\,-0.10\pm0.05\,\mathrm{log}(L_{2-10\,\rm{keV}})$ for a sample of 75 radio-quiet AGN observed by XMM-\textit{Newton} and \textit{Chandra}.  The authors attribute the reduction in slope of the anti-correlation to radio-loud contamination of previous AGN samples, proposing that radio-loud AGN could have an enhanced continuum contribution from a relativistic jet.  The authors further postulated that short-term variability of the primary X-ray source could, in part, contribute to the anti-correlation.  Despite the shallower gradient found, two measurements of the same gradient would be expected to differ by the separation between \citeauthor{Iwasawa93} and \citeauthor{Jiang06} $\sim8$\% of the time\footnote{\url{https://ned.ipac.caltech.edu/level5/Sept01/Orear/frames.html}}, and are thus not strongly inconsistent with each other.  \citet{Bianchi07} later studied the Iwasawa-Taniguchi effect for a sample of 157 radio-quiet unobscured type 1 AGN, including narrow line Seyfert 1s (which share some spectral characteristics with obscured AGN).  In contrast to \citet{Jiang06}, the authors found a somewhat steeper anti-correlation of $\mathrm{log}(\mathrm{EW}_{\mathrm{Fe\,K}\alpha})\,\propto\,-0.17\pm0.03\,\mathrm{log}(L_{2-10\,\rm{keV}})$, fully consistent with the original Iwasawa-Taniguchi effect and \citet{Page04}.  \citeauthor{Bianchi07} further suggest an additional strong anti-correlation between the Fe\,K$\rm{\alpha}$ fluorescence line EW and Eddington ratio.  Indeed, \citet{Ricci13b} tested the positive relation between the photon index and Eddington ratio found for AGN \citep{Lu99,Shemmer06,Risaliti09,Brightman13,Trakhtenbrot17}, even into the Compton-thick regime \citep{Brightman16}, finding that this could contribute to the Iwasawa-Taniguchi effect.  This is because a lower Eddington ratio (and thus photon index, resulting in a flatter spectrum) would lead to more photons at the energy required to generate iron K$\alpha$ fluorescence, giving a larger EW.

Individual source variability has been shown to considerably affect the strength of the anti-correlation, with \citet{Shu12} finding a reduction in the observed slope from $\mathrm{log}(\mathrm{EW}_{\mathrm{Fe\,K}\alpha})\,\propto\,-0.22\,\mathrm{log}(L_{2-10\,\rm{keV}})$ to $\mathrm{log}(\mathrm{EW}_{\mathrm{Fe\,K}\alpha})\,\propto\,-0.13\,\mathrm{log}(L_{2-10\,\rm{keV}})$, after accounting for the time-averaged Fe\,K$\rm{\alpha}$ strength in a sample of 32 AGN with $N_{\rm H}\lesssim10^{23}\,\rm{cm}^{-2}$, observed multiple times by the \textit{Chandra} high-energy grating (HEG).

The conventional Iwasawa-Taniguchi effect describes the strength of the Fe\,K$\rm{\alpha}$ line relative to the intrinsic continuum (readily available for unobscured AGN), but a difficulty is introduced when trying to study the effect for obscured sources, which by definition start to lack a prominent transmitted intrinsic component in the iron line flux, to measure the EW against.  \citet{Ricci14} report a significant detection of the Iwasawa-Taniguchi effect for two separate samples of Seyfert 1s and 2s, consistently of $\mathrm{log}(L_{\mathrm{Fe\,K}\alpha}/L_{\rm{10-50\,keV}})\,\propto\,-0.11\pm0.01\,\mathrm{log}(L_{10-50\,\rm{keV}})$.  Type 2 Seyferts are typically observed to be obscured in the optical and often X-rays also (e.g., \citealt{Koss17}).  Thus the work of \citeauthor{Ricci14} was the first study into the effect for obscured sources, in which the higher $10-50\,\rm{keV}$ energy range was used to describe the intrinsic continuum and Fe\,K$\alpha$ EW since photoelectric absorption is minimised for photons at harder energies.  Interestingly, the authors postulate that the consistency of slopes between Seyfert 1s and 2s could indicate that the physical mechanism responsible for the Iwasawa-Taniguchi effect is unaffected by orientation under Unification schemes.  For a breakdown of the results into the Iwasawa-Taniguchi effect from the different works mentioned above, see Table 1 of \citet{Ricci13a}.

Numerous physical scenarios have been considered to explain the observed Iwasawa-Taniguchi effect, with one of the most favoured being an intrinsic luminosity-dependent covering factor of neutral obscuring gas surrounding the AGN.  This effect was first suggested in \citet{Lawrence82} \& \citet{Lawrence91}, dubbed the `receding torus', and has been observed in various large AGN samples.  This idea is strengthened by the results from multiple studies reporting an increased number density of obscured AGN at lower X-ray luminosities \citep{Ueda11,Lusso13,Merloni14,Georgakakis17}.  Simulations of torus reprocessing of X-ray emission have also shown that the Fe\,K$\alpha$ line EW can be dramatically enhanced when the observer is exposed to less intrinsic flux than the reprocessor \citep{Krolik94}, which is physically attained with higher covering factors of the central engine.

\begin{figure}
\centering{}
\includegraphics[angle=0,width=1.\columnwidth]{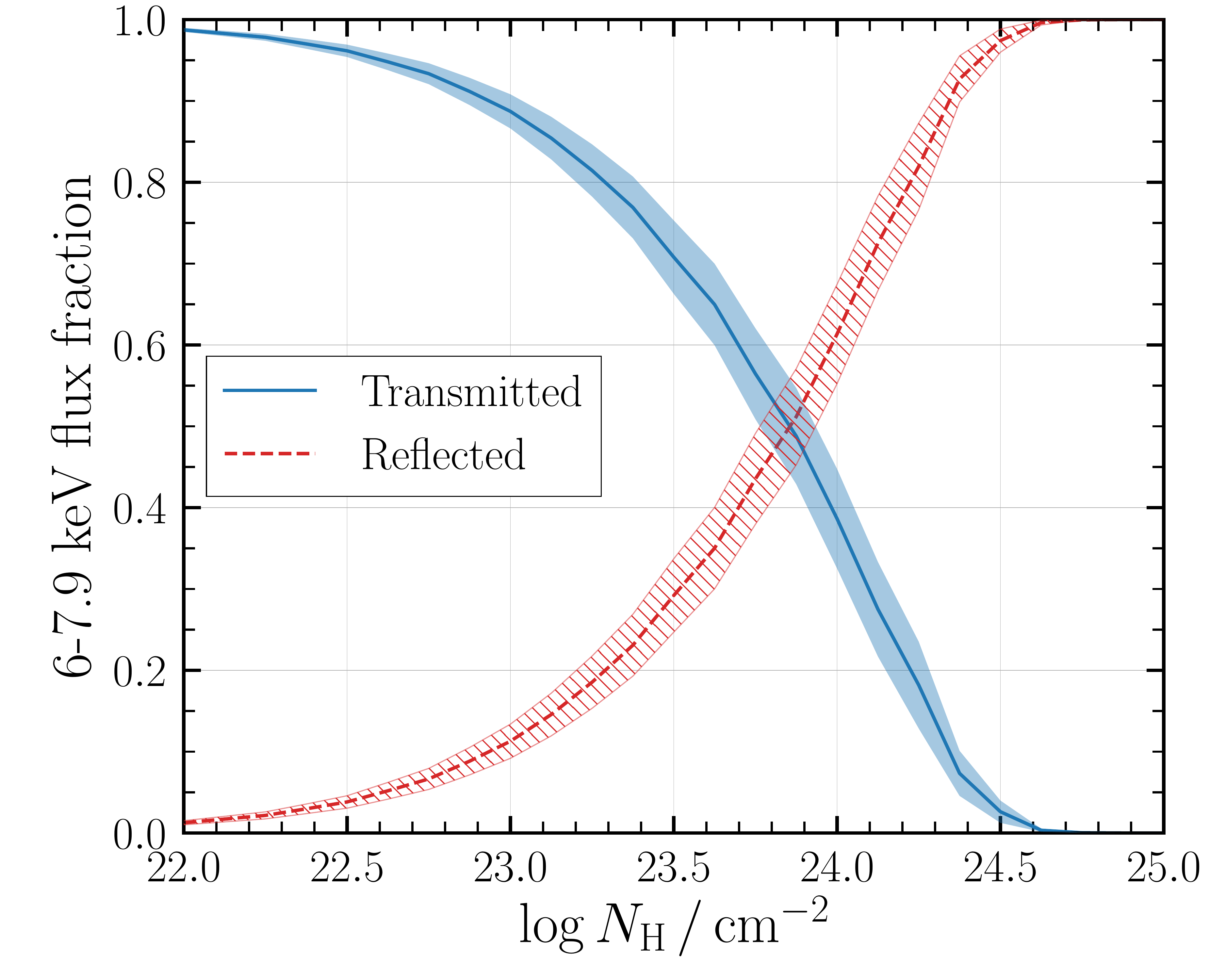}
\caption{\label{fig:transmitted_flux} The contribution to the total observed flux in the iron line region (approximated to $6-7.9\,$keV) from the direct transmitted component (blue) and Compton-scattered reflected component (red).  This was simulated using the \texttt{borus02} X-ray reprocessor model \citep{Balokovic18}, for a spherically distributed obscurer with polar cutouts.  For each column density, the average flux ratio for a series of covering factor/inclination angle combinations is plotted with the confidence region showing the range between the minimum and maximum found around the average.}
\end{figure}

A receding torus model provides a possible explanation for the Iwasawa-Taniguchi effect in which the observed spectrum contains a dominant unscattered component, as is the case for transmission-dominated obscured systems.  The prominence of the direct transmitted component would scale with intrinsic luminosity, resulting in the narrow Fe\,K$\alpha$ line (arising from the reflection component) being diminished by the brightened intrinsic power law.  To illustrate the contribution to the observed flux from the transmitted component vs. the reflected component from an anisotropic X-ray reprocessor, Figure \ref{fig:transmitted_flux} shows the relative contribution to the total line flux (approximated here to $6-7.9\,$keV) from the transmitted component (blue) and reflected component (red).  This was simulated with the \texttt{borus02\_v170709a} (\texttt{borus02}\footnote{available at \url{http://www.astro.caltech.edu/~mislavb/download/index.html}}) model \citep{Balokovic18}, in which the obscurer is spherically distributed with polar cutouts.  For each column density, we plot the average flux ratio for a series of covering factor/inclination angle combinations, and sources are predicted to become reflection-dominated in the Fe\,K$\alpha$ line for $\rm{log}\,\textit{N}_{\rm H}\,\gtrsim\,23.6\,\rm{cm}^{-2}$.

\begin{figure}
\centering{}
\includegraphics[angle=0,width=1.\columnwidth]{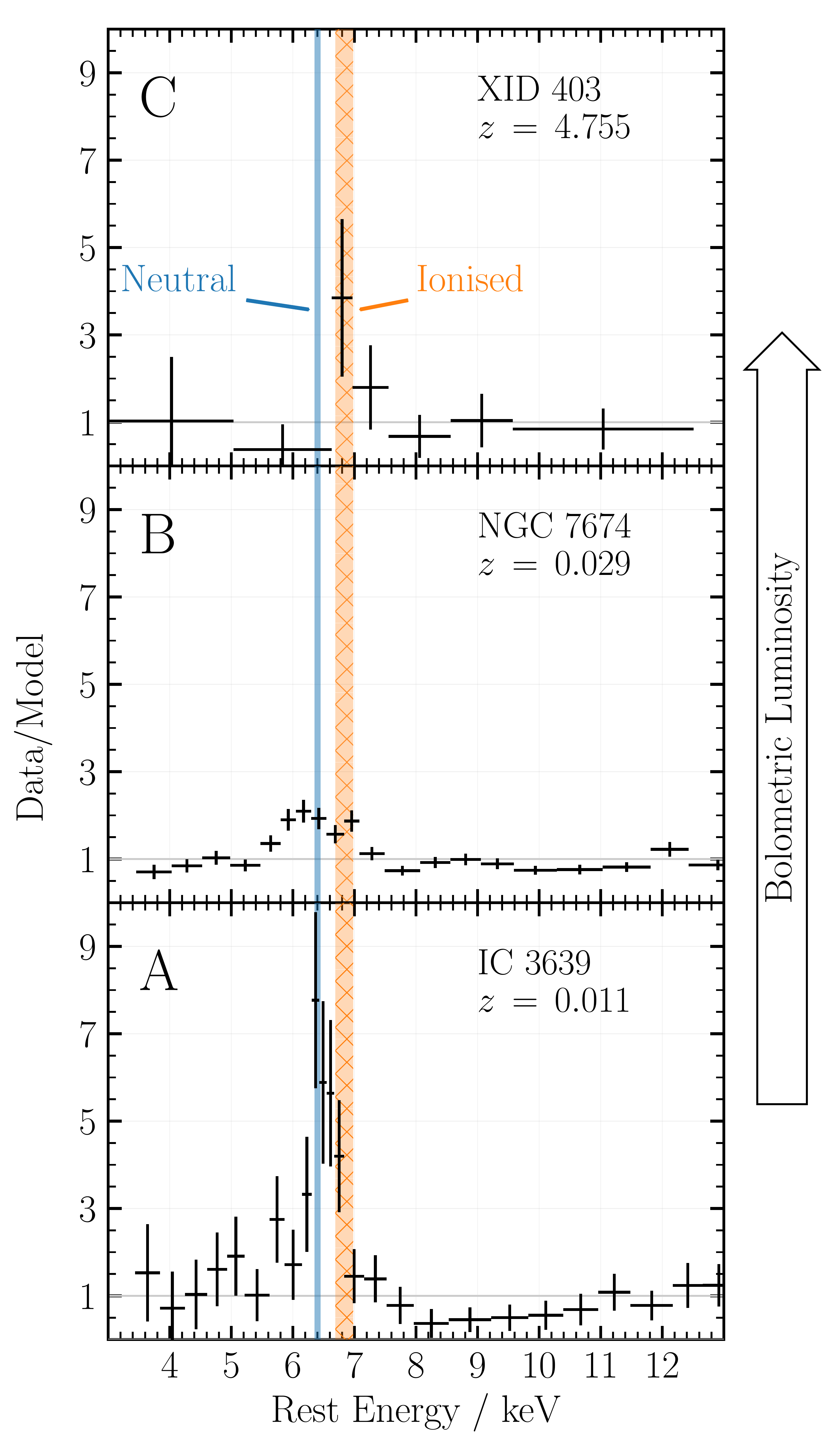}
\caption{\label{fig:ctagn} Three Compton-thick case studies illustrating the motivation for the studying a Compton-thick Iwasawa-Taniguchi effect.  Panel A: IC 3639 \citep{Boorman16}: a local Compton-thick AGN at $z\,=\,0.011$ with bolometric infrared luminosity, $\mathrm{log}(L_{8-1000\,\rm{\mu m}}\,[L_{\odot}])\sim10.9$, and one of the strongest neutral Fe K$\alpha$ lines reported in the literature.  Panel B: NGC 7674 \citep{Gandhi17}: a local Compton-thick AGN, and luminous infrared galaxy with bolometric infrared luminosity, $\mathrm{log}(L_{8-1000\,\rm{\mu m}}\,[L_{\odot}])\sim11.6$.  Contrastingly, this source has the lowest neutral Fe K$\alpha$ EW reported for local Compton-thick AGN.  Furthermore, the spectrum clearly shows a contribution to the residuals in the ionised energy region.  Panel C: LESS J033229.4-275619 \citep{Gilli14}: the highest redshift Compton-thick AGN currently known, with $\mathrm{log}(L_{8-1000\,\rm{\mu m}}\,[L_{\odot}])\sim12.8$.  The spectrum plotted is from a 7\,Ms observation of \textit{Chandra} Deep Field South (CDFS).  The neutral emission is not detected, yet a considerable flux contribution can be seen in the ionised emission line energy region.  Each source was fitted with a redshifted powerlaw in \texttt{xspec}, and the resulting data/model \texttt{ratio} is shown.  For all panels, the narrow core of the neutral line is shown with the orange region for $E=6.35-6.45$\,keV, and the ionised line region is shown in blue hatch for the $E=6.69-6.98$\,keV region (to encompass the 6.7 and 6.97 keV ionised lines).}
\end{figure}

Recent dedicated studies into specific X-ray-obscured AGN appear to show a trend of decreased neutral Fe\,K$\rm{\alpha}$ line EW, with increasing luminosity.  Here we highlight three Compton-thick case studies for comparison; also illustrated in Figure \ref{fig:ctagn}:
\begin{enumerate}
\item[1.] Local low luminosity Compton-thick Seyferts typically show prominent lines.  One of the strongest observed Fe\,K$\rm{\alpha}$ line EWs found to date was for IC 3639 \citep{Boorman16}; a reflection-dominated Compton-thick AGN with infrared bolometric luminosity (in the $8-1000\,\rm{\mu} \rm{m}$ wavelength range) of $\mathrm{log}(L_{8-1000\,\rm{\mu m}}\,[L_{\odot}])\sim10.9$ and $\mathrm{EW}_{\mathrm{Fe\,K}\alpha}\sim3\,\rm{keV}$, relative to the observed underlying reflection continuum.
\item[2.] On the other hand, NGC 7674 \citep{Gandhi17} is a heavily Compton-thick Seyfert 2, with a higher infrared bolometric luminosity of $\mathrm{log}(L_{8-1000\,\rm{\mu m}}\,[L_{\odot}])\sim11.6$.  Yet the source has an observed EW of the neutral line of $\mathrm{EW}_{\mathrm{Fe\,K}\alpha}\sim0.4\,\rm{keV}$: the lowest constrained EW of the Fe\,K$\alpha$ line detected for any bona-fide Compton-thick AGN to date.
\item[3.] At the highest luminosities, \citet[LESS J033229.4-275619]{Gilli11,Gilli14} is the most distant ($z\sim4.75$) Compton-thick AGN classified to date, with infrared bolometric luminosity $\mathrm{log}(L_{8-1000\,\rm{\mu m}}\,[L_{\odot}])\sim12.8$.  Interestingly, the neutral Fe\,K$\rm{\alpha}$ fluorescence line is not detected in the observed X-ray spectrum obtained from the 4\,Ms \textit{Chandra} Deep Field South observation, yet with a prominent ionised Hydrogen-like iron line at $\sim6.9\,$keV to $\sim2\,\sigma$ confidence, with rest-frame EW\,=\,$2.8^{+1.7}_{-1.4}\,$keV.  In fact, there is increasing observational evidence for prominent ionised iron lines in luminous infrared galaxies (LIRGs: $\mathrm{log}(L_{8-1000\,\rm{\mu m}}\,[L_{\odot}])>11$) \citep{Iwasawa09}.
\end{enumerate}

We note that although the contribution to the infrared flux from star formation will increase with bolometric flux, the AGN contribution also increases.  This means a higher infrared flux should indicate a more intrinsically luminous AGN.  These three case study sources are illustrated in Figure \ref{fig:ctagn}, in which we plot the data/model \texttt{ratio} for each source after fitting a powerlaw to the observed spectrum.  Although NGC 7674 appears to show a large component to the observed flux around $6-6.5\,$keV, the narrow core of the neutral Fe\,K$\rm{\alpha}$ line is considerably weaker.  The panels have been binned for clarity.

This paper presents the first study into an Iwasawa-Taniguchi effect for Compton-thick AGN, with Fe\,K$\rm{\alpha}$ EWs measured relative to the observed continuum vs. rest-frame mid-infrared $12\,\rm{\mu} \rm m$ luminosity ($L_{\rm{12\,\mu m}}$; taken as a proxy for the intrinsic AGN bolometric luminosity).  The cosmology adopted for computing luminosity distances is $H_{0}$\,=\,67.3\,km\,s$^{-1}$\,Mpc$^{-1}$, ${\rm \Omega}_{\Lambda}$\,=\,0.685 and ${\rm \Omega}_{\rm M}$\,=\,0.315 \citep{Planck14}\footnote{Redshift-dependent distances are used for consistency across the full sample.  Only a handful of the closest AGN have redshift-independent distances which scatter around our adopted luminosity distances.} The paper is organised as follows: Section \ref{sec:sample} describes our source selection and the sample used in our statistical analysis.  Section \ref{sec:method} then describes our method for clarifying candidate Compton-thick AGN, as well as for determining the $L_{\rm{12\,\mu m}}$ and Fe\,K$\rm{\alpha}$ EW values.  We then discuss our fitting procedure.  Section \ref{sec:results} comprises our main results, followed by the discussion and implications of the effect if confirmed on future larger Compton-thick AGN samples, in Section \ref{sec:discussion}.  We summarise our findings in Section \ref{sec:summary}.

\section{THE SAMPLE}
\label{sec:sample}

Our primary goal whilst collating Compton-thick candidates from the literature was to cover a broad redshift (and hence luminosity) range.  Furthermore,  X-ray spectra encompassing the observed frame neutral Fe\,K$\rm{\alpha}$ fluorescence line, seen at $6.4/(1+z)\,\rm{keV}$ in the rest-frame, were required.  In order to robustly quantify the EW required a detection of the underlying observed continuum, neighbouring the line centroid.  Below we include details of the high and low redshift subsamples we include in our work.

\subsection{High redshift}
\label{sec:highred_sample}
For higher redshift (or fainter) sources, \textit{Chandra} observations were ideal due to low background and optimal sensitivities in the $0.5-8.0\,\rm{keV}$ energy range.  At high redshift, the k-corrected Compton hump also shifts to the observed \textit{Chandra} energy range.  A considerable contribution to our sample thus includes the \citet{Brightman14} compilation of Compton-thick AGN candidates collated from archival deep \textit{Chandra} surveys.  The original sample includes $\sim$\,100 Compton-thick candidates.  A source was only retained for our study if it met the following criteria:

\begin{enumerate}
\item $>50$ total X-ray counts detected in the \textit{Chandra} energy band.
\item A spectroscopic redshift.
\item A line of sight column density of $N_{\rm H}\geq1.5\times10^{24}\,\rm{cm}^{-2}$ at 90\% confidence, determined by \citet{Brightman14}.
\item Infrared detection by the \textit{Wide-field Infrared Survey Explorer (WISE)}\footnote{A `reliable' \textit{WISE} detection corresponds to a detection with S/N\,$>$\,5.  See \url{http://wise2.ipac.caltech.edu/docs/release/allsky/expsup/sec5_3.html} for further details.} or \textit{Spitzer Space Telescope} to enable a reliable $L_{\rm{12\,\mu m}}$ estimate.
\end{enumerate}

Of the resulting candidates, a further two were excluded due to a disagreement with our Compton-thick classification (COSMOS 0661 \& COSMOS 1517: Section \ref{sec:method}), leaving a total of 27 sources from \citet{Brightman14}.  An additional five high redshift sources come from further Compton-thick studies by \citet[BzK 4892]{Feruglio11,Corral16}, \citet[XMMID 324]{Georgantopoulos13}, \citet[XMMID 2608, XMMID 60152]{Lanzuisi15} and \citet[IRAS F15307+3252]{Hlavacek-Larrondo17}.  In total, 32 sources make up our high redshift subsample of Compton-thick candidates.

\subsection{Low redshift}
\label{sec:lowred_sample}
A major contribution to our low redshift subsample comes from \citet{Ricci15}.  The sample consists of 55 Compton-thick AGN candidates selected from the \textit{Neil Gehrels Swift}/Burst Alert Telescope (BAT) 70-month catalogue, all within the local Universe (average $z=0.055$).  Of these 55, we rejected 19 sources without publicly available \textit{NuSTAR} observations.  \textit{NuSTAR} \citep{Harrison13} is the first true hard X-ray imaging instrument in the $3-79\,\rm{keV}$ energy range, encompassing the full underlying reflection continuum for low redshift AGN, and thus ideal for studying Compton-thick candidates.  By combining with soft X-ray observations, many works have constrained the $N_{\rm H}$ values for numerous obscured, Compton-thick and changing-look AGN to date (e.g. \citealt{Arevalo14}, Circinus Galaxy; \citealt{Balokovic14}, NGC 424, NGC 1320, IC 2560; \citealt{Gandhi14}, Mrk 34; \citealt{Teng14}, Mrk 231; \citealt{Annuar15}, NGC 5643; \citealt{Bauer15}, NGC 1068; \citealt{Ptak15}, Arp 299; \citealt{Boorman16}, IC 3639; Megamaser sample; \citealt{Masini16a}, Mrk 1210; \citealt{Ricci16}, IC 751; \citealt{Ricci17a}, WISE J1036 +0449; \citealt{Annuar17}, NGC 1448; \citealt{Gandhi17}, NGC 7674), hence our preference for \textit{NuSTAR} availability.

An additional three sources from the \citet{Ricci15} sample were excluded due to a disagreement with our mid-infrared diagnostic Compton-thick classification (2MASX J09235371-3141305; MCG -02-12-017; NGC 6232, Section \ref{sec:method}).

The last contribution to our low redshift subsample comes from the \citet{Gandhi14} compilation of bona-fide Compton-thick AGN, updated to include IC 3639 \citep{Boorman16}, NGC 1448 \citep{Annuar17} and NGC 7674 \citep{Gandhi17}, whilst excluding changing-look candidates: Mrk 3 \citep[find a Compton-thin column density to 90\% confidence]{Ricci15}, NGC 4102, NGC 4939\footnote{Our own analysis of the archival XMM-\textit{Newton} EPIC/PN spectrum as compared to the more recent \textit{NuSTAR} FPMA \& FPMB spectra strongly indicate a changing-look AGN for these sources.}, NGC 4785 \citep{Gandhi15a,Marchesi17} and NGC 7582 \citep{Rivers15}.  In total, \lored{} sources make up our low redshift subsample of Compton-thick candidates.  Full details of the \total{} (low + high redshift) Compton-thick candidates in our sample are included in Table \ref{tab:sample}.

\section{METHOD}
\label{sec:method}

\subsection{Infrared luminosities}
\label{subsec:infraredL}

In selecting a suitable proxy for the bolometric luminosity of each source, we adhered to the following criteria: (1) the bolometric luminosity could not be derived from the spectral energy region responsible for the neutral Fe\,K$\rm{\alpha}$ line nor from the continuum surrounding the line that would be used to derive an EW, and (2) the proxy should be prominent and well detected for Compton-thick AGN.

We used the infrared contribution to the broadband spectra of our AGN sample, which is considered to have sizeable contributions in this wavelength range due to reprocessing of the primary intrinsic AGN emission.  Since typical AGN contributions to composite galaxy spectra dominate at $\sim6-20\,\mu\rm m$ \citep{Mullaney11}, we used the rest-frame $12\,\rm{\mu} \rm m$ luminosity of each source.

To determine the rest-frame $12\,\mu\rm m$ luminosity, we utilised the infrared spectral template of \citet{Mullaney11} to interpolate the rest-frame $12\,\mu\rm m$ flux from \textit{observed-frame} flux measurements as close to $12\,\mu\rm m$ as possible.   For high-quality infrared observations, we use the \textit{WISE} and \textit{Spitzer} Multiband Imaging Photometer (MIPS).  \textit{WISE} had four imaging channels onboard (W1, W2, W3 \& W4) corresponding to $\lambda=3.35\,\mu\rm m,\,4.60\,\mu\rm m,\,11.56\,\mu\rm m,\,22.09\,\mu\rm m$, respectively \citep{Wright10}, whereas \textit{Spitzer}/MIPS was capable of imaging in spectral bands centered on $\lambda=\,24\,\mu\rm m,\,70\,\mu\rm m,\,160\,\mu\rm m$.  For $z\lesssim0.84$ a robust interpolation could be made from W3 and W4 observations.  However, for higher redshift sources in which the k-correction shifts the rest-frame $12\,\mu\rm m$ luminosity to wavelengths beyond W4 (or for poorly constrained/faint observations from \textit{WISE}), we use \textit{Spitzer}/MIPS.

For archival \textit{WISE} observations, we use the AllWISE Source Catalog\footnote{http://irsa.ipac.caltech.edu/cgi-bin/Gator/nph-dd} to get profile-fitted magnitudes and the NASA Extragalactic Database (NED)\footnote{http://ned.ipac.caltech.edu} to search for archival \textit{Spitzer}/MIPS observations.

To test how representative the \citet{Mullaney11} template was for predicting the 12\,$\mu$m luminosity for the AGN in our sample with $L_{2\,-\,10\,\rm{keV}}<10^{42}\,\rm{erg\,s}^{-1}$ or $L_{2\,-\,10\,\rm{keV}}>10^{44}\,\rm{erg\,s}^{-1}$, we compared the interpolated luminosities with those predicted from the type 2 AGN template from \citet{Polletta07}, which were derived over a wider range of luminosities.  On average, the offset between the interpolated luminosities from the two templates was only $\sim0.06$\,dex.

\subsection{Compton-thick confirmation of sample}
\label{subsec:selection}

Strong correlations between mid-infrared and intrinsic X-ray emission have been found with ground-based high angular resolution observations of AGN, around $12\,\mu\rm m$ \citep{Horst08,Levenson09,Gandhi09,Asmus15}.  A similar correlation has been found as a function of large aperture $6\,\mu\rm m$ luminosity, with akin results \citep{Lutz04,Mateos15,Stern15,Chen17}, and at 5.8\,$\mu$m \citep{Lanzuisi09}.  The $12\,\mu\rm m$ luminosity correlation has been used with considerable success for identifying candidate Compton-thick AGN.  Correcting X-ray absorption in Compton-thick sources acts to increase the observed X-ray luminosity to values consistent with the relation.  We refer the reader to \citet{Boorman16} for the effects of absorption correction on X-ray luminosities relative to their observed mid-infrared luminosities for the \citet{Gandhi14} compilation of bona fide Compton-thick AGN.  Here we use the study of the X-ray vs. $12\,\mu\rm m$ correlation reported in \citet{Asmus15} to classify our sample as candidate Compton-thick.

\begin{figure*}
\centering
\includegraphics[width=1.\textwidth]{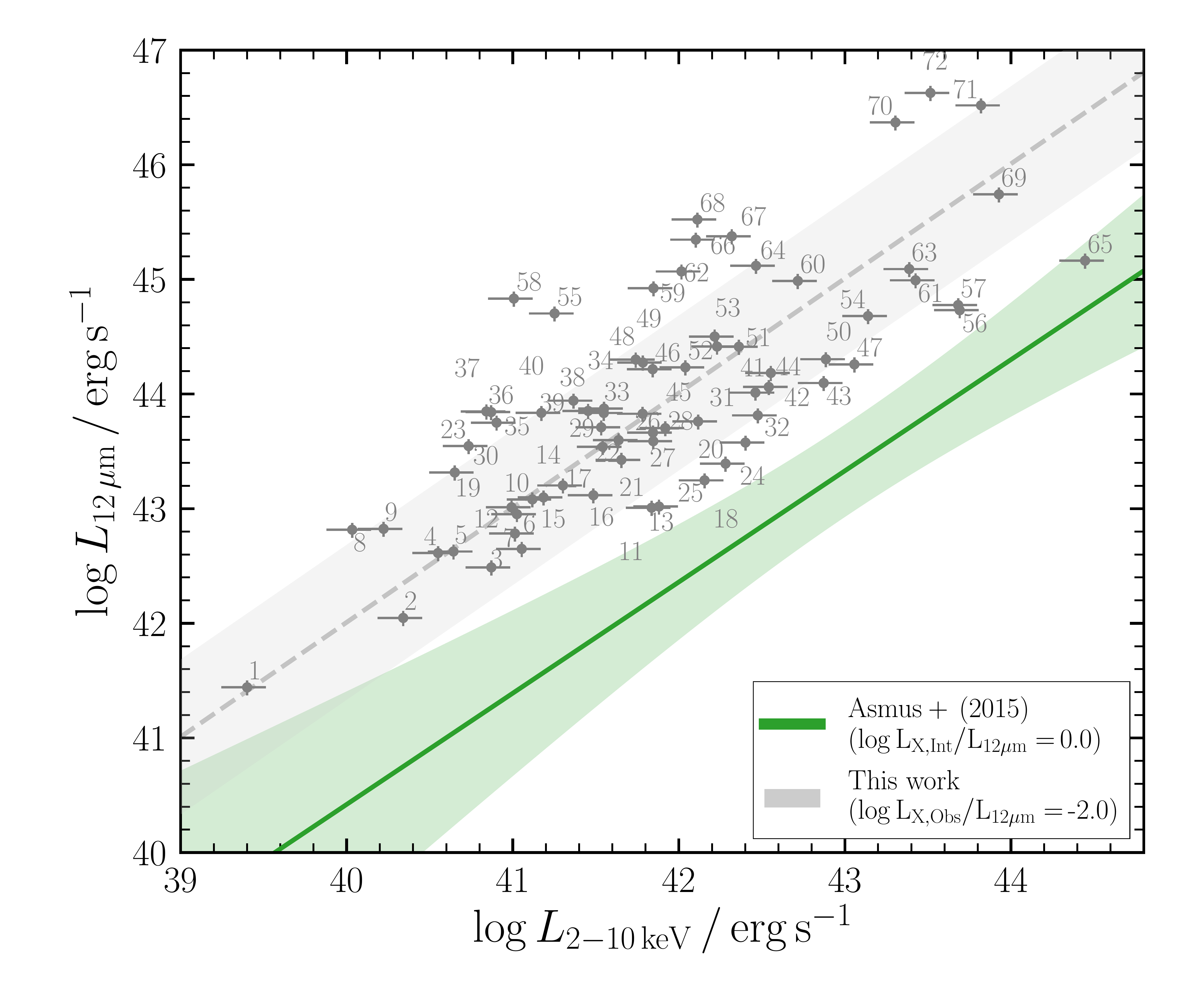}
\caption{\label{fig:selection} Predicted rest-frame $12\,\mu\rm m$ source luminosity (interpolated from the \citet{Mullaney11} infrared AGN spectral template) vs. the rest-frame observed $2-10\,\rm{keV}$ luminosity of our sample of Compton-thick AGN.  The solid (green) line is the best fit correlation from \citet{Asmus15}, together with the 1-$\sigma$ scatter in light green shading.  All sources were assigned a 30\% and 15\% uncertainty to the X-ray and $12\,\mu\rm m$ luminosities, respectively.  The sample shows diminished X-ray emission relative to $12\,\mu\rm m$ emission by a factor of greater than 2 orders of magnitude on average, indicative of Compton-thick obscuration.  The average observed correlation is shown with a grey dotted line and shaded standard deviation.  Labels refer to the ID column in Table \ref{tab:sample}.}
\end{figure*}

The rest-frame observed (i.e. absorbed) $2-10\,\rm{keV}$ luminosity was computed from a fit to the available X-ray spectra within \texttt{xspec} (for objects without a reported observed X-ray flux), and plotted against the rest-frame $12\,\mu\rm m$ luminosity, interpolated from the \citet{Mullaney11} AGN spectral template.  These observed fluxes are plotted in Figure \ref{fig:selection} (grey points), with a 30\% and 15\% uncertainty on the X-ray and $12\,\mu\rm m$ luminosities, respectively.  The original correlation found by \citet{Asmus15} is shown with a solid (green) line for clarity, together with the 1-$\sigma$ scatter.  On average, the sample displays a mean ratio of observed X-ray to mid-infrared flux of $-2.0\pm0.7$ dex, and this is shown over plotted with a dashed (grey) line and shading.  An average deviation of greater than two orders of magnitude from the relation is indicative of Compton-thick levels of obscuration found in previous works.  However, from this relation, 2MASX J09235371-3141305, MCG -02-12-017, NGC 6232, COSMOS 0661 and COSMOS 1517 displayed mid-infrared fluxes that agreed with the observed X-ray flux within the uncertainties found by \citet{Asmus15}.  This could suggest that the observed X-ray flux has a major contribution from the transmitted component, i.e. is only partially obscured and thus were excluded from our Compton-thick sample.

\subsubsection{Star Formation Contamination of $L_{\rm{12\,\mu m}}$}
\label{sec:SF}
To test for infrared star formation contamination, we first used the $12\,\mu \rm m$ observations from \citet{Asmus14}.  This work minimised star-formation contamination in measuring mid-infrared fluxes of local sources by using high-angular resolution ($\lesssim 0\farcs4$) imaging with ground-based 8\,m class telescopes.  Such contamination would not be excluded from \textit{WISE}-based measurements, that were used in our sample for these sources, due to the larger angular resolution (FWHM) of $6\farcs1$, $6\farcs4$, $6\farcs5$ and $12\farcs0$ for W1, W2, W3 and W4, respectively.  16 of our sample of \total{} sources have measured fluxes in \citet{Asmus14}.  The average X-ray to mid-infrared flux ratio for these 16 sources was consistent with the equivalent ratio for the full sample.  To fully account for this in the remainder of our sample without high angular resolution measurements, we conservatively use the average change in flux between \textit{WISE} and \citeauthor{Asmus14} (0.29\,dex) added in quadrature to the original 15\% uncertainty assigned to the template interpolated flux as the lower error bar for all sources lacking a mid-infrared observation from \citet{Asmus14}, giving 0.30\,dex.  For the 16 sources with measured fluxes from \citeauthor{Asmus14}, we use the quoted rest-frame 12\,$\mu$m luminosities and uncertainties therein.

\subsection{Rest-frame Fe\,K$\rm{\alpha}$ line EWs}
\label{subsec:EWs}
Due to the complexity associated with NGC 1068 \citep{Bauer15}, NGC 4945 \citep{Puccetti14} and the Circinus Galaxy \citep{Arevalo14}, our simplified phenomenological model could not provide a reasonable description of the data for these sources.  For this reason, we use the EWs quoted in \citet{Ricci15}, converted to the rest-frame for the corresponding sources.  Additionally, we did not have access to the spectral files for 4 high-redshift sources.  The source of the EWs we use for our analysis are included in Table \ref{tab:sample}, column (12).  In total, we computed the rest-frame neutral Fe\,K$\rm{\alpha}$ fluorescence line EW for 65/\total{} sources, as follows:
\begin{enumerate}
\item Any counts with $E\,<\,3\,\rm{keV}$ in the source rest-frame were ignored for \textit{Chandra} (or $E\,<\,4\,\rm{keV}$ for \textit{NuSTAR}) observations, in order to remove as much soft X-ray contamination from non-primary AGN sources as possible.  Such sources include intrinsic AGN emission scattered into the line of sight, a relativistic jet, X-ray binaries present in the host or photoionised gas.  Furthermore, all counts above 7\,keV in the observed frame were excluded to account for the instrument-based sensitivities of \textit{Chandra}.  The corresponding upper limit for \textit{NuSTAR} was $\sim$\,14--15\,keV in the observed frame, optimising the measurement of the continuum over the most sensitive \textit{NuSTAR} energy range.

\item In the low counts regime, we used Cash-statistics \citep[C-stat]{Cash79} during fitting.  Spectra were either grouped to allow a minimum number of counts, or a minimum signal-to-noise (S/N) ratio per bin, while retaining enough spectral resolution for the Fe\,K$\alpha$ line.  We generally favoured fitting with C-stat unless sources had enough counts or high enough S/N to warrant the use of $\chi^2$ statistics on a correspondingly S/N-binned spectrum.  We experimented with different binning strategies within the sources fitted with C-stat, and found consistent outcomes.

\item Next we fitted each spectrum with a simplified phenomenological model consisting of photoelectric absorption acting on a composite power law plus a narrow Gaussian of $\mathrm{FWHM}\approx\,2\,$eV ($\sigma\,=\,1\,\mathrm{eV}$), modelling the observed continuum plus the narrow core of the Fe\,K$\rm{\alpha}$ fluorescence line.  This model was used only to constrain the shape of the observed spectrum, and the EW of the Fe\,K$\alpha$ line.  If a given source had an observed excess of emission in the softer energy band ($E\lesssim 4\,\mathrm{keV}$) an \texttt{apec} component was additionally included in the model to account for this.  In \texttt{xspec}, this baseline model takes the form:

\begin{gather}
    \textsc{model = gal\_phabs $\times$ (apec + \nonumber}\\
    \textsc{zphabs $\times$ (zpowerlaw[$\Gamma$\,=\,1.4] + \nonumber}\\
    \textsc{zgauss[E$_{\rm L}$ = 6.4\,{\rm keV}]))}
	\label{eq:EW_model}
\end{gather}

\textsc{gal\_phabs} refers to an additional minor contribution to the absorption from the Galaxy.  Items in square brackets refer to fixed parameters.  Although many studies suggest the intrinsic power law of AGN have average photon indices of $\sim$\,1.9, we fit the spectra with a flatter (lower) photon index of 1.4, as this is closer to the value found for the flat ($<10$\,keV) reflection spectra typically observed for Compton-thick AGN, and we required our model to provide a reasonable fit to the observed spectrum.

\item We then computed two-dimensional confidence contours over the \textsc{zpowerlaw} and \textsc{zgaussian} model component normalisations (whilst leaving $N_{\rm H}$ and, if required to describe the soft region of the observed spectrum, the \texttt{apec} normalisation, free).

\item These contours were translated to confidence on the Fe K$\rm{\alpha}$ EW, and plotted as a function of the statistical test difference from the best fit acquired (chi-squared or Cash-statistics depending on the source).  This enables us to determine the minimum, and hence presumed best fit rest-frame EW, together with the 1-$\sigma$ uncertainty.  Irrespective of using chi-squared or Cash-statistic, we use a delta statistic of +2.30 to represent the 1-$\sigma$ (68\%) confidence level for two interesting parameters\footnote{\url{https://heasarc.gsfc.nasa.gov/xanadu/xspec/manual/XSappendixStatistics.html}}.

\item For sources in which the normalisation of the Fe\,K$\alpha$ line could not be constrained in the fit, we use the limit derived by \texttt{xspec} on this parameter to calculate an upper bound on the EW.  For any sources that yielded an unphysical EW\,$>\,$5\,keV, we set the limit to this value.  This is applicable to 3 sources: CDFS 443, CDFS 454 \& COSMOS 2180, with EW\,$\lesssim$\,12\,keV, EW\,$\lesssim48$\,keV and EW\,$\lesssim11$\,keV, respectively.  We defer the reader to the Appendix for the grouped spectrum used for each source.
\end{enumerate}

\subsection{Fitting procedure}
\label{subsec:fitting}

Our final sample consists of \total{} sources, including \lims{} upper limits on the EW.  All sources without quoted luminosities in \citet{Asmus14} were assigned the same lower uncertainty of 0.3\,dex on $12\,\rm{\mu} \rm m$ luminosity specified in Section \ref{subsec:selection}.  We then fitted a linear regression to the EW vs. rest-frame $12\,\rm{\mu} \rm m$ luminosity.  To account for all the uncertainties present in our sample whilst determining a fit, our fitting procedure was as follows\footnote{Similar in method to \citet{Bianchi07}}:

\begin{enumerate}
\item The dataset was bootstrapped by randomly sampling data points from the original whilst allowing repeats.  The new dataset was the same size as the parent sample.

\item Each point in the bootstrapped dataset was randomly resampled depending on the uncertainty of each point, as follows:
\begin{enumerate}
	\item \textit{Non-detections/upper limits:} new points were randomly drawn from a uniform distribution in the interval $[\rm{log}\,100\,\rm{eV}, \rm{log\,limit\,[eV]}]$,
	\item \textit{Detections:} A new value was generated from a Gaussian distribution with standard deviation given by the 1-$\sigma$ error being considered for that point.
\end{enumerate}
To avoid strongly unphysical values from biasing the simulations, we truncated the randomised EWs to between 100\,eV and 5\,keV.
\item A linear least-squares regression was carried out on the Monte Carlo simulated dataset using the \texttt{scipy.linregress} Python package.   The Spearman's Rank Correlation Coefficient ($\rho$) was then found using the \texttt{scipy.spearmanr} package for each fit.

\item Steps (i)\,--\,(iii) were repeated in order to obtain a distribution of gradients, y-intercepts and $\rho$ values for the original dataset.
\end{enumerate}

\section{RESULTS}
\label{sec:results}

Table \ref{tab:sample} includes details of each source used in our final sample, and the Appendix contains the best fit spectrum and EW contour for each source used, as well as the sources ruled out in our analysis.
After carrying out \iterations{} iterations, we obtain a best fit linear regression to the data of:

\begin{gather}
    \rm{log}(\rm{EW_{\rm{Fe\,K\alpha}}/keV})=\nonumber\\
    \hspace*{40pt}-(0.08\pm0.04)\,\rm{log}(L_{12\,\rm{\mu} \rm m}/10^{44}\,\rm{erg\,s}^{-1})+2.87\pm0.05
	\label{eq:best_fit}
\end{gather}

\begin{figure*}
\centering
\includegraphics[width=1.\textwidth]{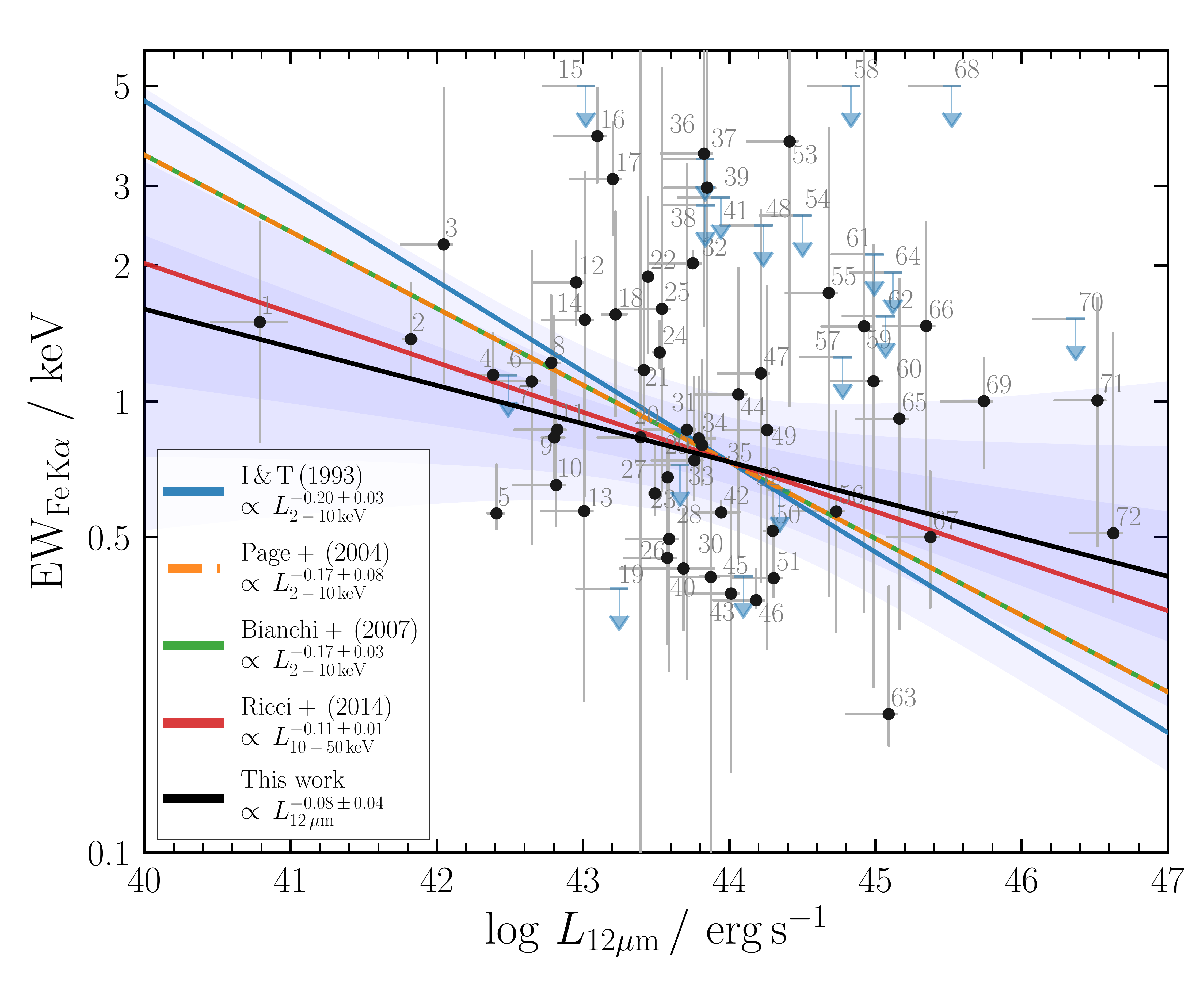}
\caption{\label{fig:fit} Plot of $L_{12\,\mu \rm m}$ vs. (rest-frame) neutral Fe\,K$\rm{\alpha}$ fluorescence line EWs.  Blue arrows represent upper limits.  As a comparison to previous studies into the Iwasawa-Taniguchi effect, we further include the gradients of previous works \citet[I\,\&\,T\,(1993)]{Iwasawa93}, \citet[Page+\,(2004)]{Page04}, \citet[Bianchi+\,(2007)]{Bianchi07} and \citet[Ricci+\,(2014)]{Ricci14}.  These correlations have all been renormalised to match our best fit y-intercept at $10^{44}\,\rm{erg\,s}^{-1}$ for comparison, since we are using the $12\,\mu \rm m$ luminosity, which is different from the intrinsic luminosity proxies used by other Iwasawa-Taniguchi effect studies.  The blue shaded region represents the standard deviation from the mean of our best fit, with lighter shading corresponding to incrementally lower numbers of standard deviation.}
\end{figure*}

Figure \ref{fig:fit} shows all rest-frame $12\,\mu\rm m$ luminosities vs. rest-frame neutral Fe\,K$\rm{\alpha}$ fluorescence line EWs.  Blue arrows represent upper limits.  As a comparison to previous studies into the Iwasawa-Taniguchi effect, we further include the gradients of previous works: \citet{Iwasawa93}, \citet{Page04}, \citet{Bianchi07} and \citet{Ricci14}, normalised to the same y-intercept at $10^{44}\,\rm{erg\,s}^{-1}$.  We make this normalisation since the EWs we report for our sample are measured relative to the observed spectrum, which for Compton-thick obscuration is drastically different to the observed spectrum for unobscured AGN, not to mention our proxy for the bolometric luminosity is different to that previously used by other studies.

\begin{figure}
  \includegraphics[width = 0.5\textwidth]{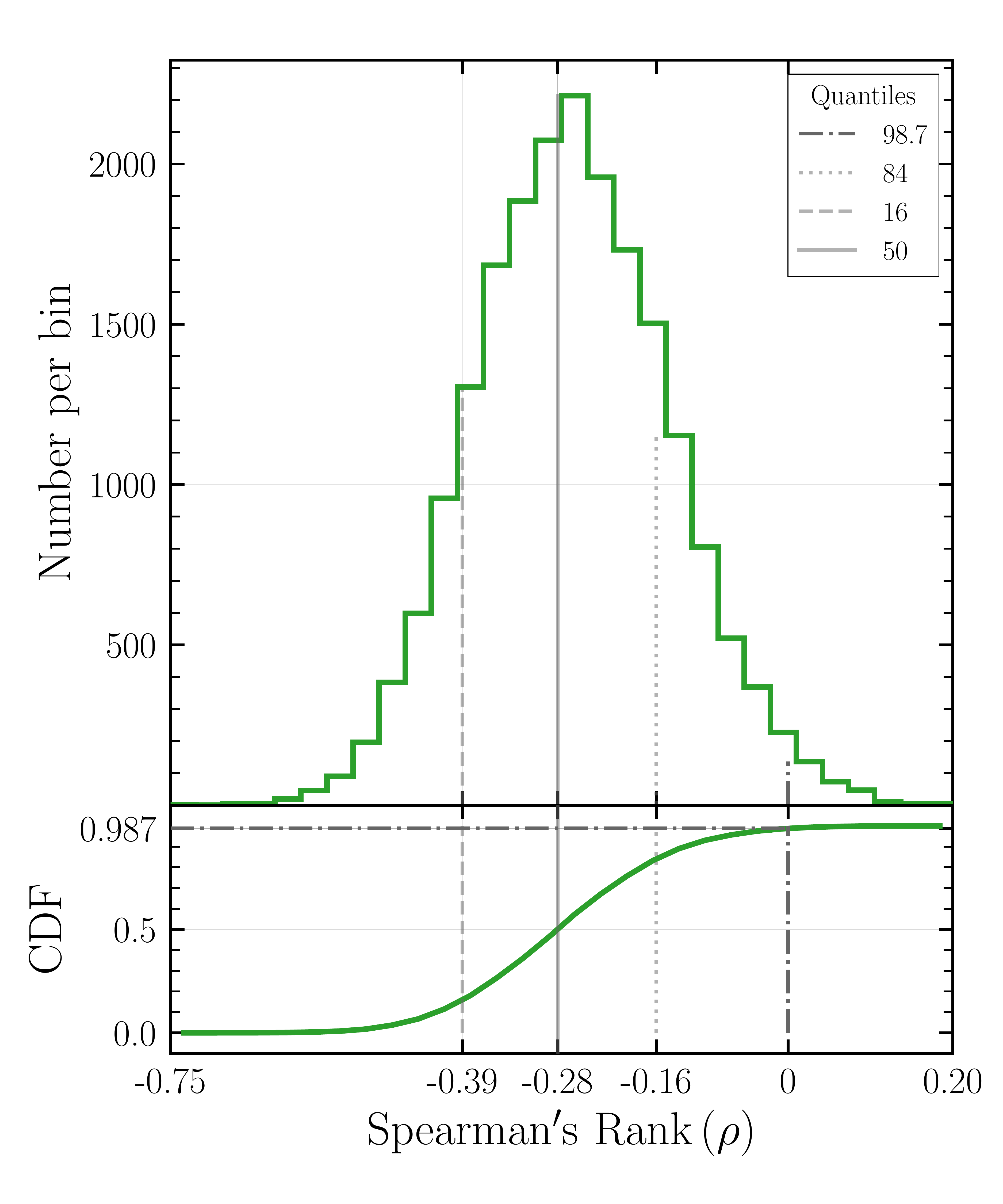}
  \caption{\label{fig:rho_nolims} Distribution of Spearman's Rank Correlation Coefficients ($\rho$) generated from \iterations{} iterations of steps (1) - (3) of our fitting procedure, outlined in Section \ref{subsec:fitting}, for a dataset excluding all upper limits (\nolims{} in total).  The 16th, 50th and 84th quantiles are shown with dashed, solid and dotted lines, respectively.  The dotted (purple) line additionally shows the correlation to be negative to \rhoneg{} confidence.}
\end{figure}

To test the significance of the fit, we computed the Spearman's Rank Correlation Coefficient ($\rho$) of the correlation for our sample, excluding upper limits.  Upper limits were excluded since $\rho$ tests the strength of a monotonic relationship between variables, which can be dramatically effected by the large range of values/orders of variables attainable with the inclusion of limits in our Monte-Carlo based fitting method.  This left \nolims{} sources, and gave a value of \rhonolims{}.  Figure \ref{fig:rho_nolims} shows the corresponding distribution in $\rho$ found, indicating a negative correlation to \rhoneg{} confidence.

Our best fit gradient is fully consistent with \citet{Ricci14} within 1-$\sigma$ errors, who attempted to take into account time-averaging of the spectra for determining EWs - see Section \ref{sec:discussion} for further discussion on this result.  The gradient found here is also flatter than the \citet{Bianchi07} best fit gradient, but consistent within 90\% confidence.  We include the distributions of our best linear fit gradients and y-intercepts in Figures \ref{fig:m} and \ref{fig:c}, respectively for the \iterations{} iterations (including upper limits).

\begin{subfigures}
\begin{figure}
  \includegraphics[width = 0.5\textwidth]{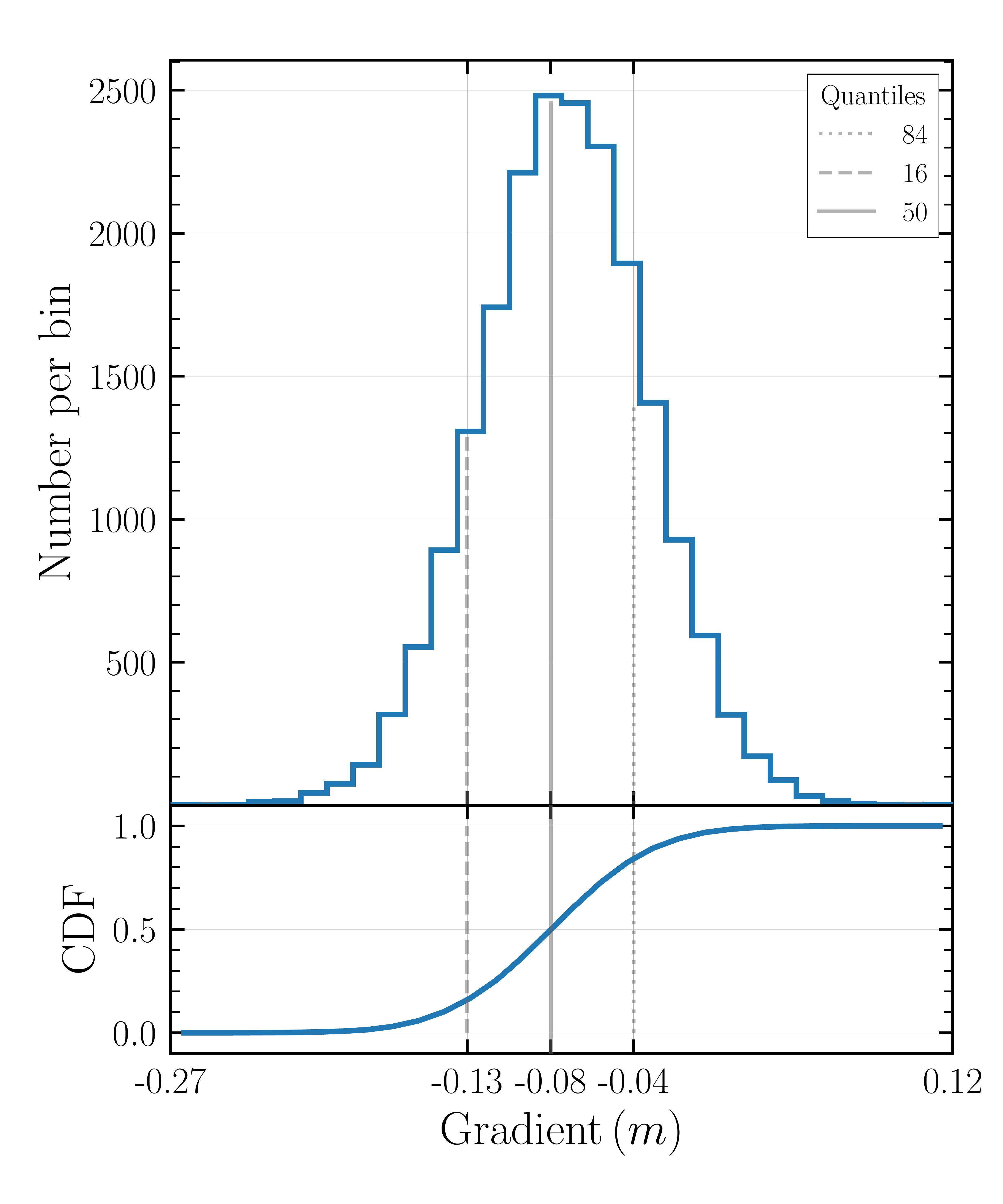}
  \caption{\label{fig:m} Distribution of linear regression gradients generated from \iterations{} iterations of steps (1) - (3) of our fitting procedure (Section \ref{subsec:fitting}).  The 16th, 50th and 84th quantiles are shown with dashed, solid and dotted lines, respectively.}
\end{figure}

\begin{figure}
  \includegraphics[width = 0.5\textwidth]{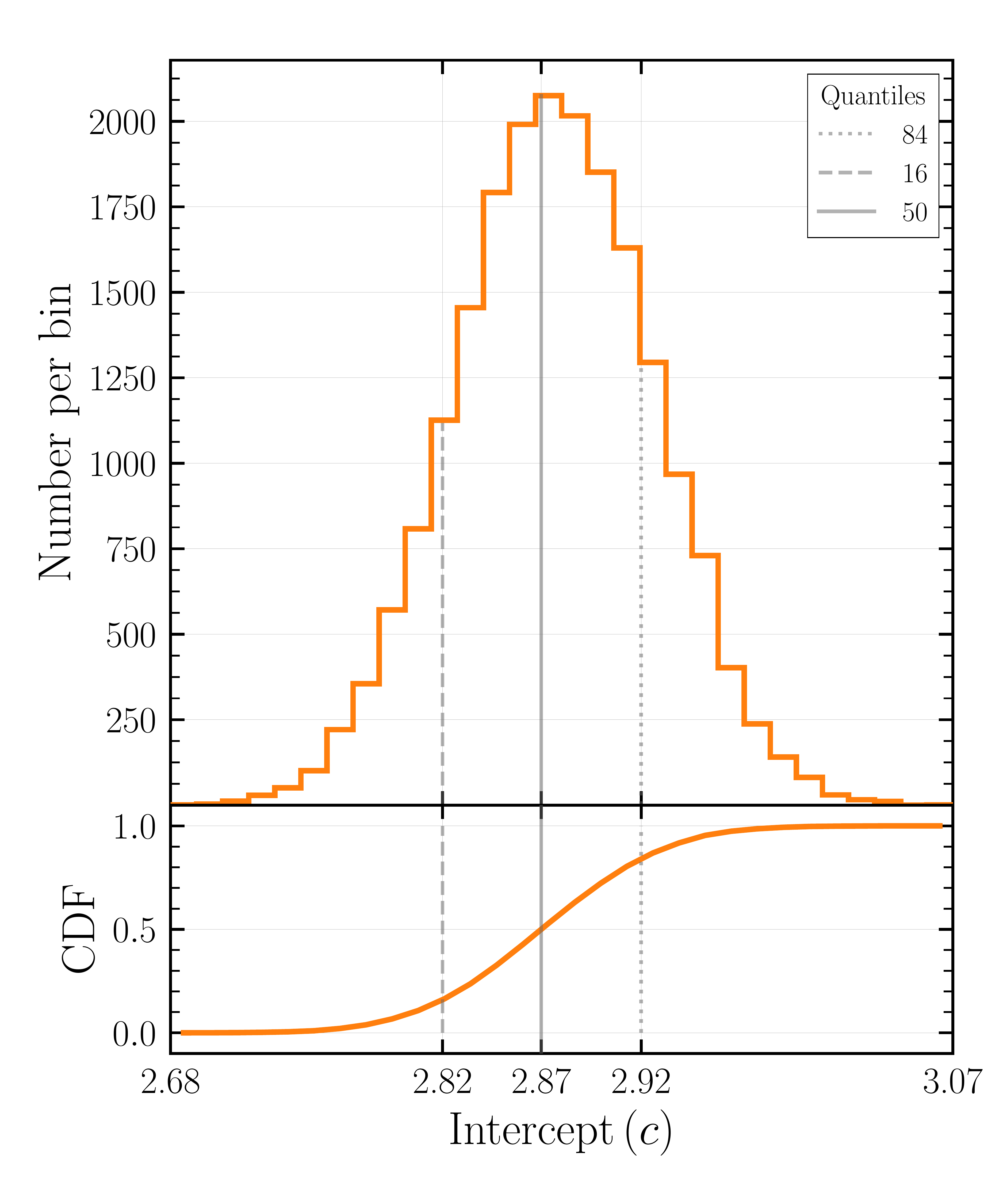}
  \caption{\label{fig:c} Distribution of linear regression y-intercepts generated from \iterations{} iterations of steps (1) - (3) of our fitting procedure, outlined in Section \ref{subsec:fitting}.  The 16th, 50th and 84th quantiles are shown with the same line styles as in Figure \ref{fig:m}.}
\end{figure}
\end{subfigures}

\clearpage
\onecolumn
\begin{landscape}
\LTcapwidth=\columnwidth
\begin{centering}
\begin{longtable}{cccccccccccc}
\hline
 ID &       Identifier &      {\it z} &    RA &  DEC & source ref. &  $\rm{log}\,{\it L}_{\rm{2\,-\,10\,\rm keV}}^{\rm{obs}}$ &  $\rm{log}\,{\it L}_{12\,\mu\rm{m}}$ &  $\rm{EW}$ &     $\rm{+EW}$ &   $\rm{-EW}$ & EW ref. \\
 (1) &       (2) &      (3) &    (4) &  (5) & (6) &  (7) &  (8) &  (9) &     (10) &   (11) & (12) \\
\endfirsthead

\multicolumn{12}{c}%
{{\tablename\ \thetable{} : \textit{cont.}}} \\
\hline
 ID &       Identifier &      {\it z} &    RA &  DEC & source ref. &  $\rm{log}\,{\it L}_{\rm{2\,-\,10\,\rm keV}}^{\rm{obs}}$ &  $\rm{log}\,{\it L}_{12\,\mu\rm{m}}$ &  $\rm{EW}$ &     $\rm{+EW}$ &   $\rm{-EW}$ & EW ref. \\
 (1) &       (2) &      (3) &    (4) &  (5) & (6) &  (7) &  (8) &  (9) &     (10) &   (11) & (12) \\
 \hline
\endhead

\hline
\endlastfoot

\hline
  1 &                 NGC 5194 &  0.002 &  202.4696 &  47.1953 &       Gou12 &    39.4 &      40.8 &    1.50 &   1.01 &  0.69 &    Boo18 \\
  2 &             ESO 005-G004 &  0.006 &   91.4233 & -86.6319 &       Ric15 &    40.6 &      41.8 &    1.37 &   0.47 &  0.23 &    Boo18 \\
  3 &                 NGC 1448 &  0.004 &   56.1329 & -44.6447 &       Ann17 &    40.3 &      42.0 &    2.23 &   2.73 &  1.13 &    Boo18 \\
  4 &                 NGC 5643 &  0.004 &  218.1696 & -44.1744 &       Ric15 &    40.6 &      42.4 &    1.14 &   0.28 &  0.13 &    Boo18 \\
  5 &                 NGC 5728 &  0.009 &  220.5996 & -17.2531 &       Ric15 &    41.5 &      42.4 &    0.56 &   0.16 &  0.05 &    Boo18 \\
  6 &                 CDFS 345 &  0.123 &   53.1028 & -27.9120 &       Bri14 &    40.9 &      42.5 &    1.14 &      0 &    -1 &    Boo18 \\
  7 &                 NGC 4180 &  0.007 &  183.2627 &   7.0388 &       Ric15 &    41.1 &      42.6 &    1.11 &   1.05 &  0.63 &    Boo18 \\
  8 &            ESO 137-G034  &  0.009 &  248.8088 & -58.0800 &       Ric15 &    41.0 &      42.8 &    1.22 &   0.51 &  0.19 &    Boo18 \\
  9 &                 NGC 3393 &  0.012 &  162.0975 & -25.1619 &       Ric15 &    41.1 &      42.8 &    0.83 &   0.72 &  0.16 &    Boo18 \\
 10 &                 NGC 3079 &  0.004 &  150.4908 &  55.6797 &       Ric15 &    40.0 &      42.8 &    0.65 &   0.61 &  0.12 &    Boo18 \\
 11 &                 NGC 4945 &  0.002 &  196.3646 & -49.4683 &       Ric15 &    40.2 &      42.8 &    0.86 &   0.03 &  0.03 &    Ric15 \\
 12 &                 NGC 2273 &  0.006 &  102.5358 &  60.8458 &       Mas16 &    41.0 &      43.0 &    1.83 &   0.44 &  0.36 &    Boo18 \\
 13 &                 NGC 6921 &  0.014 &  307.1202 &  25.7234 &       Ric15 &    41.8 &      43.0 &    0.57 &   0.52 &  0.35 &    Boo18 \\
 14 &               2MFGC02280 &  0.015 &   42.6775 &  54.7049 &       Ric15 &    41.0 &      43.0 &    1.52 &   1.71 &  0.90 &    Boo18 \\
 15 &                 CDFS 443 &  0.895 &   53.1335 & -27.7478 &       Bri14 &    41.9 &      43.0 &    5.00 &      0 &    -1 &    Boo18 \\
 16 &                  IC 2560 &  0.010 &  154.0779 & -33.5639 &       Bal14 &    41.2 &      43.1 &    3.87 &   1.10 &  0.83 &    Boo18 \\
 17 &                 NGC 1320 &  0.009 &   51.2029 &  -3.0422 &       Bal14 &    41.3 &      43.2 &    3.11 &   1.06 &  0.78 &    Boo18 \\
 18 &                 NGC 7130 &  0.016 &  327.0812 & -34.9511 &       Ric15 &    40.9 &      43.2 &    1.56 &   1.08 &  0.63 &    Boo18 \\
 19 &             ESO 464-G016 &  0.036 &  315.5991 & -28.1748 &       Ric15 &    42.2 &      43.2 &    0.38 &      0 &    -1 &    Boo18 \\
 20 &                 CDFS 296 &  0.518 &   53.2734 & -27.8709 &       Bri14 &    42.3 &      43.4 &    0.83 &   5.74 &  0.74 &    Boo18 \\
 21 &                 NGC 7479 &  0.008 &  346.2358 &  12.3228 &       Ric15 &    40.7 &      43.4 &    1.17 &   0.68 &  0.29 &    Boo18 \\
 22 &                  IC 3639 &  0.011 &  190.2200 & -36.7558 &       Boo16 &    40.7 &      43.4 &    1.89 &   0.95 &  0.61 &    Boo18 \\
 23 &                 NGC 1194 &  0.014 &   45.9546 &  -1.1036 &       Ric15 &    41.7 &      43.5 &    0.62 &   0.17 &  0.07 &    Boo18 \\
 24 &                 NGC 3281 &  0.011 &  157.9671 & -34.8536 &       Gou12 &    41.9 &      43.5 &    1.28 &   0.23 &  0.10 &    Boo18 \\
 25 &                 CDFS 114 &  0.310 &   53.0356 & -27.7800 &       Bri14 &    41.5 &      43.5 &    1.60 &   3.89 &  0.97 &    Boo18 \\
 26 &           MCG +08-03-018 &  0.020 &   20.6434 &  50.0550 &       Ric15 &    42.4 &      43.6 &    0.45 &   0.19 &  0.16 &    Boo18 \\
 27 &             ESO 138-G001 &  0.009 &  252.8338 & -59.2347 &       Ric15 &    41.6 &      43.6 &    0.68 &   0.11 &  0.04 &    Boo18 \\
 28 &           MCG +06-16-028 &  0.016 &  108.5162 &  35.2793 &       Ric15 &    41.8 &      43.6 &    0.50 &   0.29 &  0.24 &    Boo18 \\
 29 &             CGCG 164-019 &  0.030 &  221.4033 &  27.0347 &       Ric15 &    41.8 &      43.7 &    0.72 &      0 &    -1 &    Boo18 \\
 30 &                 Arp 299B &  0.010 &  172.1292 &  58.5614 &       Pta15 &    41.8 &      43.7 &    0.43 &   0.15 &  0.12 &    Boo18 \\
 31 &                 CDFS 273 &  0.229 &   53.0825 & -27.6897 &       Bri14 &    41.5 &      43.7 &    0.86 &   2.49 &  0.62 &    Boo18 \\
 32 &          Circinus Galaxy &  0.001 &  213.2912 & -65.3392 &       Ric15 &    40.9 &      43.8 &    2.02 &   0.14 &  0.01 &    Ric15 \\
 33 &            ESO 201-IG004 &  0.036 &   57.5954 & -50.3025 &       Ric15 &    42.1 &      43.8 &    0.74 &   0.40 &  0.14 &    Boo18 \\
 34 &                  NGC 424 &  0.012 &   17.8650 & -38.0833 &       Ric15 &    41.5 &      43.8 &    0.83 &   0.31 &  0.13 &    Boo18 \\
 35 &            NGC 7212NED02 &  0.027 &  331.7583 &  10.2335 &       Ric15 &    42.5 &      43.8 &    0.80 &   0.44 &  0.15 &    Boo18 \\
 36 &                 CDFS 065 &  0.664 &   53.0673 & -27.8282 &       Bri14 &    41.8 &      43.8 &    3.54 &   5.66 &  2.07 &    Boo18 \\
 37 &                 CDFS 421 &  0.738 &   53.0770 & -27.7656 &       Bri14 &    41.5 &      43.8 &    3.44 &      0 &    -1 &    Boo18 \\
 38 &                 CDFS 347 &  0.280 &   53.1458 & -27.9035 &       Bri14 &    41.2 &      43.8 &    2.72 &      0 &    -1 &    Boo18 \\
 39 &                 CDFS 384 &  0.150 &   53.1750 & -27.6639 &       Bri14 &    40.8 &      43.8 &    2.98 &   7.92 &  2.20 &    Boo18 \\
 40 &             ESO 406-G004 &  0.029 &  340.6390 & -37.1853 &       Ric15 &    41.5 &      43.9 &    0.41 &   0.45 &  0.33 &    Boo18 \\
 41 &                 CDFS 063 &  0.670 &   53.0751 & -27.8315 &       Bri14 &    41.4 &      43.9 &    2.83 &      0 &    -1 &    Boo18 \\
 42 &                 NGC 1068 &  0.004 &   40.6696 &  -0.0133 &       Ric15 &    41.3 &      43.9 &    0.57 &   0.04 &  0.01 &    Ric15 \\
 43 &                 NGC 1229 &  0.036 &   47.0451 & -22.9601 &       Ric15 &    42.5 &      44.0 &    0.37 &   0.24 &  0.22 &    Boo18 \\
 44 &                 CDFS 400 &  1.090 &   53.1049 & -27.9138 &       Bri14 &    42.5 &      44.1 &    1.04 &   0.94 &  0.59 &    Boo18 \\
 45 &          IGR J14175-4641 &  0.077 &  214.2652 & -46.6948 &       Ric15 &    42.9 &      44.1 &    0.41 &      0 &    -1 &    Boo18 \\
 46 &             CGCG 420-015 &  0.029 &   73.3571 &   4.0617 &       Ric15 &    42.6 &      44.2 &    0.36 &   0.06 &  0.02 &    Boo18 \\
 47 &                 CDFS 158 &  0.738 &   53.0941 & -27.7406 &       Bri14 &    41.8 &      44.2 &    1.15 &   1.51 &  0.75 &    Boo18 \\
 48 &                AEGIS 567 &  0.536 &  214.8070 &  52.8973 &       Bri14 &    42.0 &      44.2 &    2.45 &      0 &    -1 &    Boo18 \\
 49 &                   Mrk 34 &  0.050 &  158.5358 &  60.0311 &       Gan14 &    43.1 &      44.3 &    0.86 &   0.94 &  0.58 &    Boo18 \\
 50 &                 NGC 7674 &  0.029 &  351.9862 &   8.7792 &       Gan17 &    42.2 &      44.3 &    0.52 &   0.17 &  0.13 &    Boo18 \\
 51 &                 NGC 6240 &  0.024 &  253.2454 &   2.4008 &       Ric15 &    42.9 &      44.3 &    0.41 &   0.15 &  0.04 &    Boo18 \\
 52 &           MCG +10-14-025 &  0.039 &  143.9654 &  61.3529 &       Ric15 &    41.7 &      44.3 &    0.64 &      0 &    -1 &    Boo18 \\
 53 &                 CDFS 459 &  1.609 &   53.1228 & -27.7228 &       Bri14 &    42.4 &      44.4 &    3.76 &  31.54 &  2.79 &    Boo18 \\
 54 &                AEGIS 602 &  0.769 &  214.8420 &  52.9219 &       Bri14 &    42.2 &      44.5 &    2.58 &      0 &    -1 &    Boo18 \\
 55 &                 CDFS 264 &  2.026 &   53.0588 & -27.7084 &       Bri14 &    43.1 &      44.7 &    1.74 &   2.31 &  1.37 &    Boo18 \\
 56 &              XMMID 60152 &  0.579 &  150.3122 &   1.7302 &       Lan15 &    43.7 &      44.7 &    0.57 &   0.38 &  0.26 &    Lan15 \\
 57 &                 CDFS 039 &  3.660 &   53.0785 & -27.8598 &       Bri14 &    43.7 &      44.8 &    1.25 &      0 &    -1 &    Boo18 \\
 58 &                 CDFS 454 &  0.650 &   53.0446 & -27.8019 &       Bri14 &    41.0 &      44.8 &    5.00 &      0 &    -1 &    Boo18 \\
 59 &                 CDFS 448 &  0.680 &   53.0808 & -27.6811 &       Bri14 &    41.8 &      44.9 &    1.47 &   4.86 &  1.12 &    Boo18 \\
 60 &                 CDFS 401 &  1.370 &   52.9604 & -27.8699 &       Bri14 &    42.7 &      45.0 &    1.11 &   1.12 &  0.88 &    Boo18 \\
 61 &              COSMOS 0581 &  1.778 &  150.2910 &   2.0895 &       Bri14 &    43.4 &      45.0 &    2.12 &      0 &    -1 &    Boo18 \\
 62 &              COSMOS 0987 &  0.353 &  149.7929 &   2.1256 &       Bri14 &    42.0 &      45.1 &    1.54 &      0 &    -1 &    Boo18 \\
 63 &  2MASX J03561995-6251391 &  0.108 &   59.0831 & -62.8609 &       Ric15 &    43.4 &      45.1 &    0.20 &   0.19 &  0.03 &    Boo18 \\
 64 &                 CDFS 460 &  2.145 &   53.0976 & -27.7155 &       Bri14 &    42.5 &      45.1 &    1.93 &      0 &    -1 &    Boo18 \\
 65 &              COSMOS 0363 &  2.704 &  150.0459 &   2.2013 &       Bri14 &    44.4 &      45.2 &    0.91 &   0.96 &  0.60 &    Boo18 \\
 66 &              COSMOS 0482 &  0.120 &  150.4250 &   2.0663 &       Bri14 &    42.1 &      45.3 &    1.47 &   1.04 &  0.91 &    Boo18 \\
 67 &               XMMID 2608 &  0.125 &  150.4249 &   2.0660 &       Lan15 &    42.3 &      45.4 &    0.50 &   0.20 &  0.15 &    Lan15 \\
 68 &              COSMOS 2180 &  0.350 &  149.9758 &   2.4615 &       Bri14 &    42.1 &      45.5 &    5.00 &      0 &    -1 &    Boo18 \\
 69 &                BzK 4892  &  2.578 &   53.1488 & -27.8211 &       Fer11 &    43.9 &      45.7 &    1.00 &   0.25 &  0.29 &  Coral16 \\
 70 &                 CDFS 382 &  0.667 &   52.9624 & -27.6879 &       Bri14 &    43.3 &      46.4 &    1.52 &      0 &    -1 &    Boo18 \\
 71 &         IRAS F15307+3252 &  0.930 &  233.1838 &  32.7131 &       H-L17 &    43.8 &      46.5 &    1.00 &   0.70 &  0.53 &    Boo18 \\
 72 &                XMMID 324 &  1.222 &   53.2051 & -27.6806 &       Geo13 &    43.5 &      46.6 &    0.51 &   0.91 &  0.15 &    Geo13 \\
\hline
\\
\caption{Complete list of all sources used in determining the anti-correlation plotted in Figure \ref{fig:fit}, ordered by rest-frame $12\,\mu\rm m$ luminosity.  Information on specific columns is as follows: (1) ID specific to each source, corresponding to the index featured on plots throughout the paper. (2) source name. (3) spectroscopic redshift. (4) right ascension in degrees. (5) declination in degrees; (6) Source reference. (7) Logarithm of the observed rest-frame $2-10\,\rm{keV}$ luminosity, measured in $\rm{erg\,s}^{-1}$; (8) Logarithm of the rest-frame $12\,\mu\rm m$ luminosity predicted from the \citet{Mullaney11} infrared spectral template; (9) Rest-frame neutral Fe\,K$\rm{\alpha}$ fluorescence line EW measured in keV; (10) and (11) upper and lower limits on the rest-frame neutral Fe\,K$\rm{\alpha}$ fluorescence line EW measured in keV (values of 0 and -1 respectively denote an upper limit to EW). (12) reference used for EW value.
References:
\citet[Fer11]{Feruglio11}; \citet[Gou12]{Goulding12}; \citet[Geo13]{Georgantopoulos13}; \citet[Br14]{Brightman14}; \citet[Gan14]{Gandhi14}; \citet[Bal14]{Balokovic14}; \citet[Lan15]{Lanzuisi15}; \citet[Ric15]{Ricci15}; \citet[Pta15]{Ptak15}; \citet[Mas16]{Masini17}, \citet[Cor16]{Corral16}; \citet[Gan17]{Gandhi17}; \citet[H-L17]{Hlavacek-Larrondo17}; \citet[Ann17]{Annuar17}; Boo18: This work.}
\label{tab:sample}.
\end{longtable}
\end{centering}
\end{landscape}
\clearpage
\twocolumn

\section{DISCUSSION}
\label{sec:discussion}

Our results indicate the presence of an Iwasawa-Taniguchi effect for Compton-thick AGN.  This is surprising, since the majority of our sample is presumed to have a noticeable flux contribution from the reflected component in the $\sim$6\,--\,7.9\,keV energy region (e.g. Figure \ref{fig:transmitted_flux} - we will address this further in Section \ref{sec:recedingtorus}), and the Fe K$\alpha$ EW is not expected to vary relative to the underlying Compton-scattered reflection continuum.  Here we discuss the significance of our result, as well as possible physical interpretations if confirmed on larger samples.

\subsection{Significance of result}
\label{sec:significance}

As a comparison to previous works, Table \ref{tab:it_significances} includes the anti-correlation significance metric quoted for the studies into the Iwasawa-Taniguchi effect that we include in Figure \ref{fig:fit}, as well as the obscuration type of the sources included in the corresponding samples.

\begin{table*}
\begin{center}
\begin{tabular}{rcccc}
\hline
\hline
\noalign{\smallskip}
Reference               & Metric & Value     & No. of sources & Obscuration class\\ 
\noalign{\smallskip}
\multicolumn{1}{r}{(1)} & (2)       & (3)       & (4)            & (5)        \\ 
\noalign{\smallskip}
\hline
\noalign{\smallskip}
\noalign{\smallskip}
This work               & Prob.& \rhoneg{}          & \nolims{}             & Compton-thick   \\  
\noalign{\smallskip}
\citealp{Ricci14}       & Prob.& 99\%          & 47             & Seyfert 2   \\  
\noalign{\smallskip}
\citealp{Bianchi07}     & Prob.    & 99.6\%          & 157            & Unobscured \\                    		
\noalign{\smallskip}
\citealp{Jiang06}       & $\rho$    & -0.47          & 101            & Unobscured \\  
\noalign{\smallskip}
\citealp{Page04}        & Prob.& $>$\,99.98\%           & 53             & Unobscured \\                    		
\noalign{\smallskip}
\citealp{Iwasawa93}     & Corr. coeff.          & $>$\,0.8          & 37             & Unobscured \\      
\noalign{\smallskip}
\noalign{\smallskip}
\hline
\noalign{\smallskip}
\noalign{\smallskip}
\end{tabular}
\caption{\label{tab:it_significances}Summary of the anti-correlation probability significances found for the studies into the Iwasawa-Taniguchi effect included in Figure \ref{fig:fit}. The table lists the (1) study reference, (2) the correlation metric used to quantify the significance of the resulting anti-correlation: `Prob.' - probability of an anti-correlation; `$\rho$' - Spearman's Rank correlation coefficient; `Corr. coeff.' - correlation coefficient, (3) the value of the correlation metric found by the work, (4) the number of objects in each sample, and (5) the obscuration classes of the sources included in the corresponding work.}
\end{center}
\end{table*}

\subsubsection{AGN dominance}
\begin{figure*}
\centering
  \includegraphics[width = 0.82\textwidth]{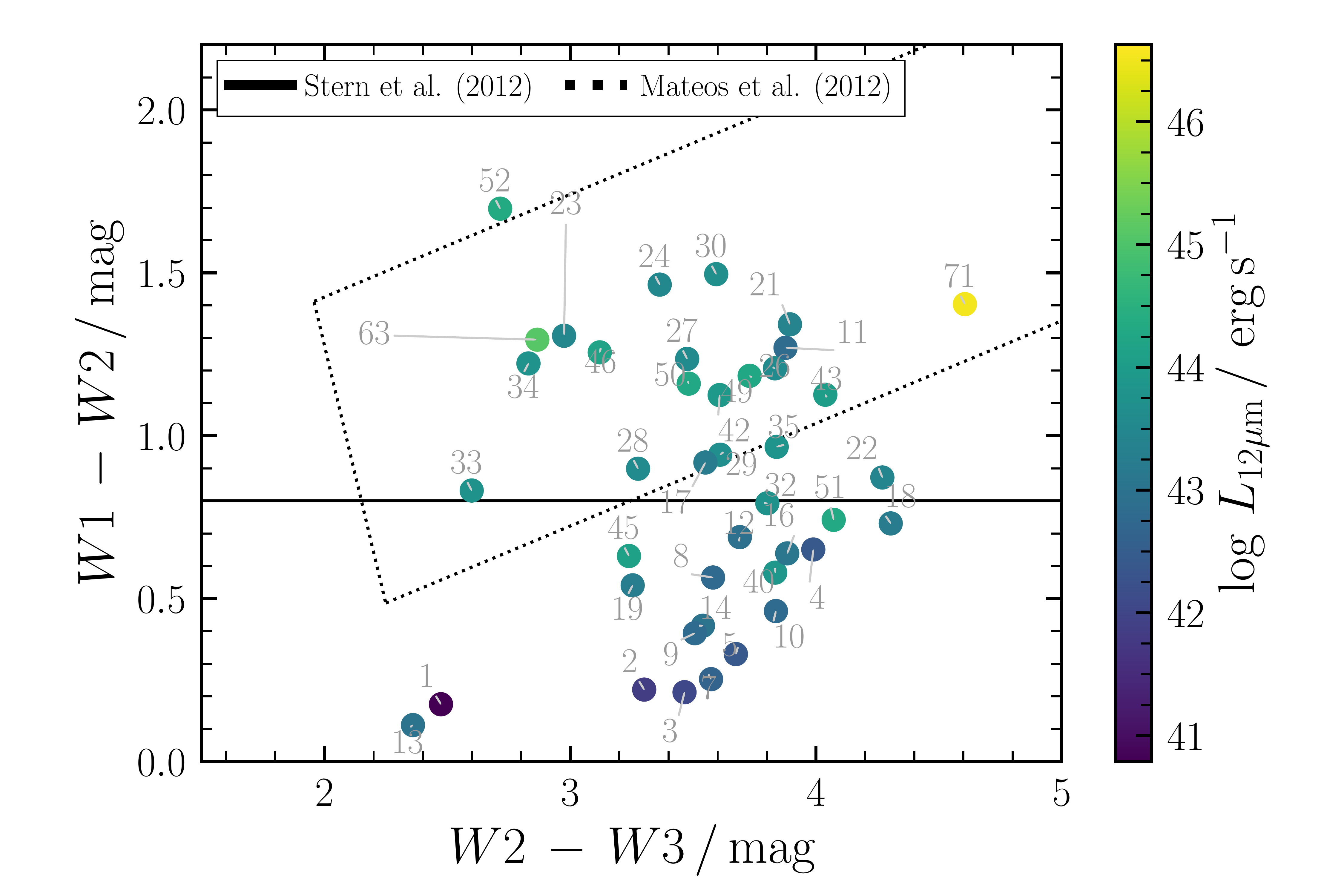}
  \includegraphics[width = 0.82\textwidth]{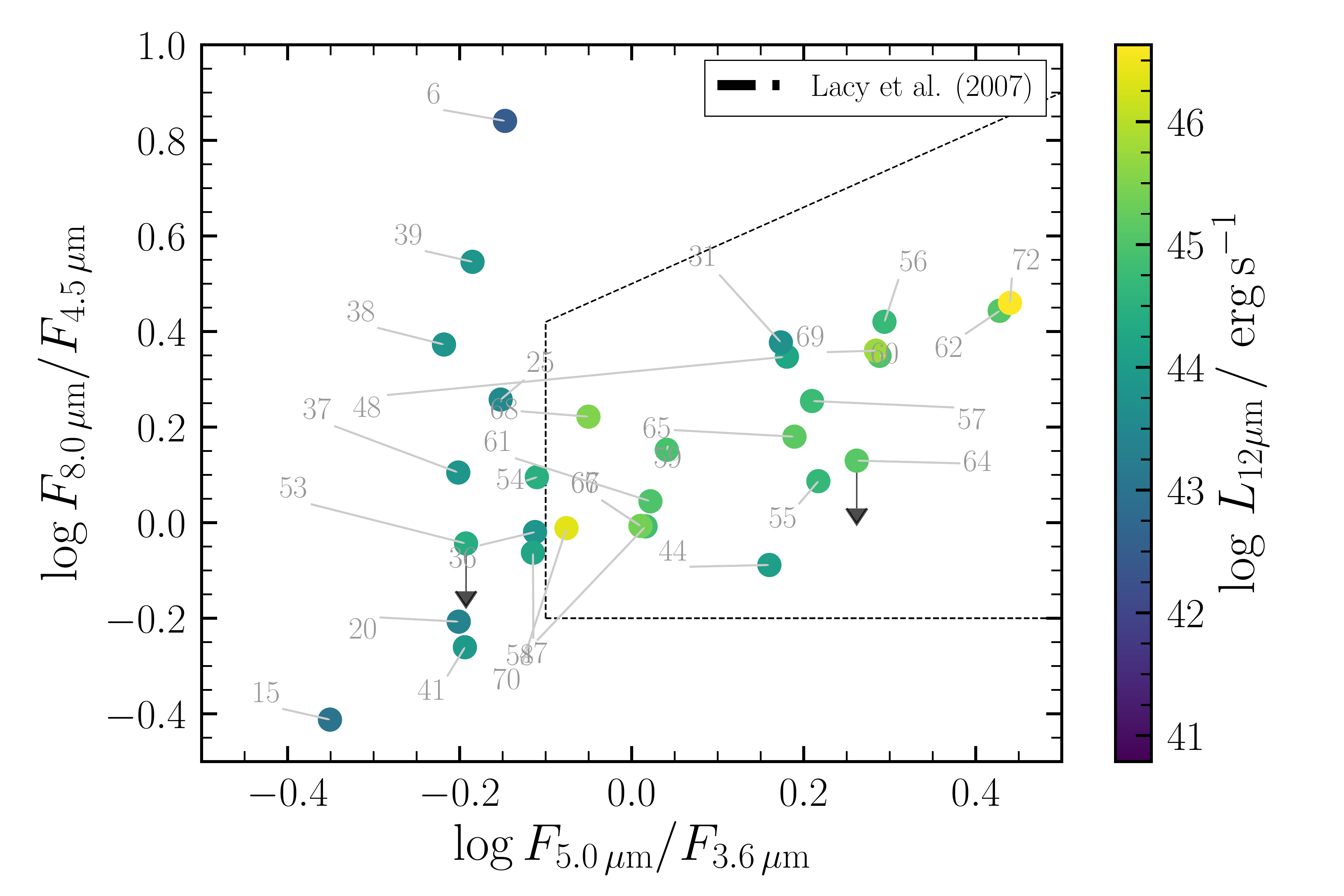}
  \caption{\label{fig:threshold} Colour selection criteria for \textit{WISE} (top panel) and \textit{Spitzer} (bottom panel) flux densities.  The \textit{WISE} thresholds are from \citet{Stern12} and \citet{Mateos12}, and the \textit{Spitzer} colour wedge is from \citet{Lacy07}.  In total, \MIRdominated{}/\total{} sources lie outside these criteria, but after testing the possible effects of star formation on these sources, we still require a significant anti-correlation to fit the data - see Section \ref{sec:SF} for details.}
\end{figure*}

In addition to estimating the effect of star formation contamination using high angular resolution observations (see Section \ref{sec:SF}), we sought to test the AGN dominance of our interpolated infrared luminosities.  We use the colour criteria of \citet{Stern12}, \citet{Mateos12} and \citet{Lacy07} for observed flux densities (used to renormalise the \citealt{Mullaney11} AGN template and interpolate a rest-frame 12$\mu$m flux) from \textit{WISE} and \textit{Spitzer}, respectively.  However, \citet{Stern12} do note that the efficiency of such mid-infrared colour selections of AGN increases strongly with X-ray luminosity.  As such, lower luminosity sources, e.g. with $L_{\rm{2\,-\,10\,keV}}\lesssim 10^{43}\,\rm{erg\,s}^{-1}$, in our sample may not display AGN-like mid-infrared colours, and thus lie outside the wedge and cut thresholds.  Alternatively, the sources lying outside the selection criteria may not be intrinsically weak - for example, \citet{Gandhi15a} notes that bluer $W1-W2$ colours could arise from strong host star formation contamination or anisotropic/weak reprocessed torus emission.  As such, sources satisfying any one of these three colour thresholds are most likely to not display star formation contamination, and be AGN-dominated in the mid-infrared.  The flux densities are plotted in Figure \ref{fig:threshold}.  In total, \MIRdominated{}/\total{} sources satisfy either \citet{Stern12} and \citet{Mateos12} combined (for \text{WISE}-based observations) or \citet{Lacy07} (for \textit{Spitzer}-based observations).  This potentially indicates some form of star formation contamination (or another form of contaminant) present in the sources that do not satisfy these criteria.  However, running a fit to only the \MIRdominated{} predicted AGN-dominated sources in our sample results in a gradient of \MIRdominatedFIT{} between EW and $L_{12\,\rm{\mu} \rm m}$, fully consistent with the main fit presented in Section \ref{sec:results}, albeit with larger uncertainty.

As a further test of contamination in our sample, we carried out a fit only to sources displaying a deficit in observed X-ray to $12\,\mu\rm m$ luminosity of greater than two orders of magnitude (see Figure \ref{fig:selection}).  This again returned a consistent result with our main fit of \ratioLTtwoFIT{} for \ratioLTtwo{} sources.  These tests suggest that any star formation contamination does not dominate the trend that we observe.

\subsubsection{Binning}

Lastly, we carried out a fit to the sample binned by $12\,\mu \rm m$ luminosity.  A maximum binning of \srcperbin{} sources optimised total number of bins together with sources per bin.  We approximated all upper limits as the average between the limit and log\,100\,eV, then assigned a 1-$\sigma$ error to the new point that encompassed log\,100\,eV to the limit.  The corresponding binned EW error for each bin was then given by:

\begin{gather}
		\sigma_{\rm EW}\,=\,\frac{\sqrt{\sum  \rm{\delta EW}^{2}}}{N}
\end{gather}

\begin{figure*}
\centering
\includegraphics[width=.8\textwidth]{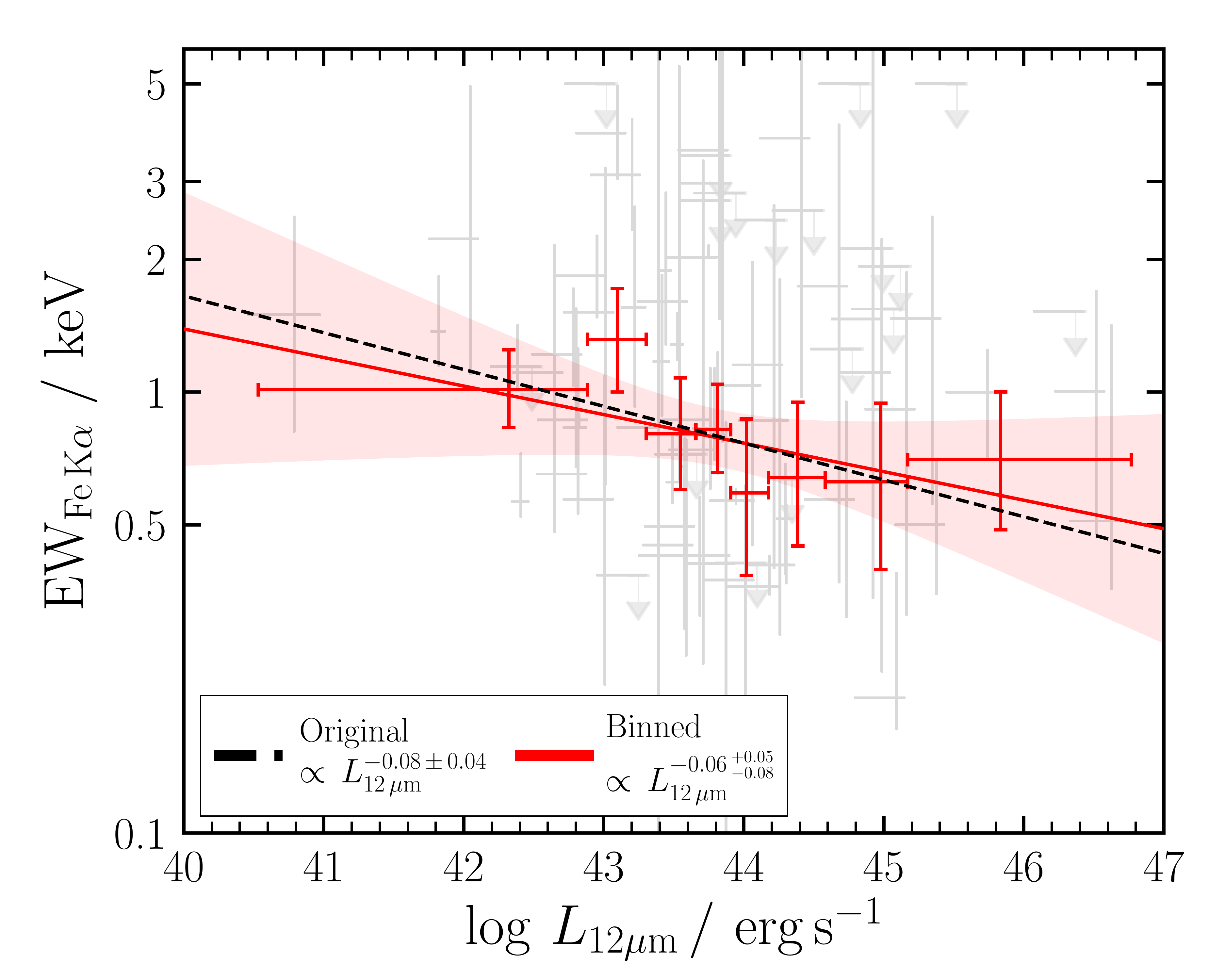}
\caption{\label{fig:bin_fit} The best fit to the binned sample, with a maximum of \srcperbin{} sources per bin (see the text for details of the binning method used).  The red shaded region shows the standard deviation from the mean of our fit, which is fully consistent with the original best fit found for the unbinned data (shown with a black dotted line).  This has been renormalised to the same y-intercept as the binned fit to allow an easier comparison between gradients.  Background grey points show the source data that was binned.}
\end{figure*}

Here \textit{N} refers to the number of sources in each bin.  The best fit gradient we get to the binned data is $\mathrm{log}(\mathrm{EW}_{\mathrm{Fe\,K}\alpha})\,\propto\,\binFIT{}\,\mathrm{log}(L_{12\,\rm{\mu} \rm m})$, and is plotted in Figure \ref{fig:bin_fit} with a red solid line and one standard deviation shading.  The best fit to the original sample is shown with a dashed (black) line, which is fully consistent with the binned gradient.  The binned data has been renormalised to have the same y-intercept as the original result for easier visual comparison of gradients.  The background grey points show the original data.

\subsection{Physical interpretation}
\label{sec:interpretation}

\subsubsection{Covering factor dependence}
\label{sec:recedingtorus}

The EW of the Fe\,K$\alpha$ line measured solely relative to the reprocessed continuum is very rarely $<1\,$keV (see Figure 8 of \citealt{Murphy09}).  One way to achieve an observed EW significantly less than 1\,keV is via leaked intrinsic AGN emission contributing some flux to the observed spectrum in the iron line region.  An example of this would be in transmission-dominated Compton-thick AGN, or a `clumpy' torus.  Typical X-ray spectral model predictions (see Figure \ref{fig:transmitted_flux}) for AGN show that for column densities, $N_{\rm H}\gtrsim1.5\times10^{24}\,\mathrm{cm}^{-2}$, the observed reflected flux is $\gtrsim10\,$times more than the transmitted flux.  Below this column density, a borderline Compton-thin/thick AGN could have a reasonable contribution in flux from the transmitted component, and hence a variable continuum with intrinsic luminosity. One way to obtain less reflected flux with increasing column would be a decreased covering factor, which has been dubbed the `receding torus', and was suggested to explain the Iwasawa-Taniguchi effect for unobscured and mildly obscured sources previously (e.g. \citealt{Page04,Ricci13a}, see Section \ref{sec:intro} of this paper).  However, \citet{Lawrence10} discuss that the apparent decrease of obscured AGN fraction with bolometric luminosity is much less significant in IR/radio samples than with X-ray samples, suggesting that the receding torus model may not exist in nature.

\begin{figure*}
\centering
  \includegraphics[angle=0,width=0.75\textwidth,trim={0 0 15cm 0}]{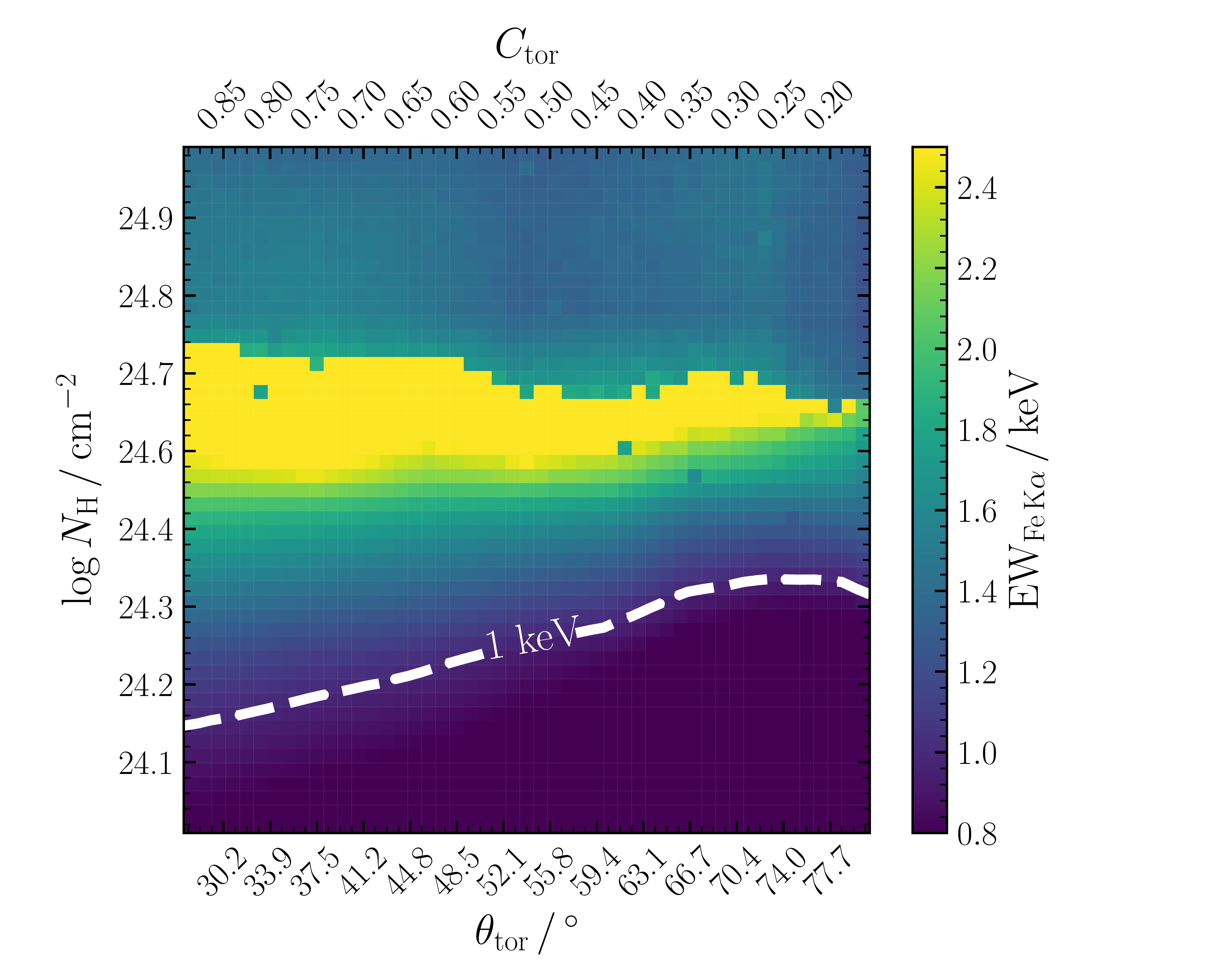}
  \caption{\label{fig:CF_EW} A colour map of simulated neutral iron K$\alpha$ EW for a series of line of sight column densities ($N_{\rm H}$) and torus covering factors ($C_{\rm tor}=\rm{cos}\,\theta_{\rm tor}$) in the \texttt{borus02} \citep{Balokovic18} model.  For these simulations, the line of sight column density and equatorial column density were tied together.  Spectra were simulated using the \textit{NuSTAR} response matrices.  The simulated spectra were then fit in the 6\,--\,7\,keV energy region with a \textsc{powerlaw}\,+\,\textsc{Gaussian} to derive a predicted EW.  The contour shows the boundary at which all EW are predicted to be $>$\,1\,keV.  All EWs predicted to be $<$\,0.8\,keV or $>$\,2.5\,keV were capped (shown with the same minimum and maximum colour, respectively).  Due to our selection of sources with $\rm{log}\,\textit N_{\rm H}>24.18$ ($N_{\rm H}>1.5\times10^{24}\rm{cm}^{-2}$) to 90\% confidence from literature values, we cannot rule out sources as having at least a partial EW dependence with covering factor, as can be seen by this region of the colour map.}
\end{figure*}

In Figure \ref{fig:CF_EW}, we show a colour map of simulated EWs predicted from the \texttt{borus02} model for an $\sim$\,edge on (84$^\circ$) viewing angle, with varying column densities and covering factors.  All spectra were simulated using \texttt{fakeit} from within \texttt{xspec} with the \textit{NuSTAR} simulation files provided by the \textit{NuSTAR} team\footnote{\url{https://www.nustar.caltech.edu/page/response_files}}, and then the resulting spectrum was re-fit in the 6\,--\,7\,keV energy region by a \textsc{powerlaw}\,+\,\textsc{Gaussian} model.  The EW of a narrow (FWHM\,$\sim$\,2\,eV) Gaussian was then derived at fixed line centroid ($E_{\rm L}=6.4$\,keV, using the \texttt{eqwidth} command in \texttt{xspec}.  We also overplot the limiting contour at which all EWs are $>$\,1\,keV, which we take as a proxy for reflection-dominance.  Interestingly, for lower covering factors (higher opening angles), the column density can be high ($\rm{log}\,N_{\rm H}>24.3$ in some cases), and still feature a spectrum with presumably leaked transmitted emission.  Since our original $N_{\rm H}$ selection was ${\rm log}\,N_{\rm H}>24.18$ ($N_{\rm H}>1.5\times10^{24}\rm{cm}^{-2}$) to 90\% confidence from literature values, from this plot we cannot rule out that the higher luminosity sources (with assumed lower covering factors) would feature some sort of leaked transmitted flux contributing to the continuum around 6.4\,keV and decreasing observed neutral Fe K$\alpha$ EW.  Furthermore, if the luminosity-dependent covering factor explanation is correct, then we are currently lacking such reflection-dominated Compton-thick AGN at high luminosities since the Fe K$\alpha$ EW is predicted to always be greater than 1\,keV for $N_{\rm H}\gtrsim1.5\times10^{24}\,\rm{cm}^{-2}$.

As an additional test, we selected sources with literature best fit \textit{lower} 90\% uncertainty on the column density to be $N_{\mathrm{H}}>1.5\times10^{24}\,\mathrm{cm}^{-2}$.  This returned \REFLdom{} sources, with only two sources at higher redshift (COSMOS 0363; $z=2.704$ and BzK 4892; $z=2.578$).  Although the corresponding best fit to this sample returns a gradient of \REFLdomFIT{}, which is entirely consistent with the result for the full sample, we lack enough robustly reflection-dominated sources at higher redshifts to draw precise conclusions for a transmitted component altering the narrow Fe\,K$\alpha$ line.

An alternative way to detect a considerable contribution from the transmitted component in Compton-thick AGN would be via changing-look AGN variability.  In this scenario, the Compton-thick obscurer is clumpy, enabling clouds of differing column to traverse the line of sight, potentially resulting in leaked intrinsic emission.  Such extreme eclipsing events from Compton-thin to Compton-thick levels of obscuration have been observed previously \citep{Risaliti07}, but are rare.  As such, this is unlikely to be responsible for the diminished Fe\,K$\alpha$ lines observed in all higher luminosity sources of the sample where we see the greatest decrease in EW, but may play a non-negligible role.

\subsubsection{Ionisation}
\label{sec:ionisation}
Finally, we consider the effects of ionisation on Compton-thick X-ray spectra.  As mentioned in Section \ref{sec:intro}, many obscured candidate AGN not only show diminished neutral Fe K$\alpha$ line EWs, but also increased ionised Fe\,K$\alpha$ EWs with intrinsic luminosity \citep{Iwasawa09}.  Indeed, a correlation between spectral slope (a proxy for the accretion efficiency of AGN) with the Fe\,K$\alpha$ line energy was found by \citet{Dewangan02}.  From Figure \ref{fig:fit}, one would expect this effect to be most prevalent for intrinsically bright (i.e. high $L_{\rm{bol}}$ and/or high Eddington ratio) systems, which may be more \textit{intrinsically} UV-luminous relative to X-rays.

\begin{figure}
  \includegraphics[angle=0,width=1.1\columnwidth]{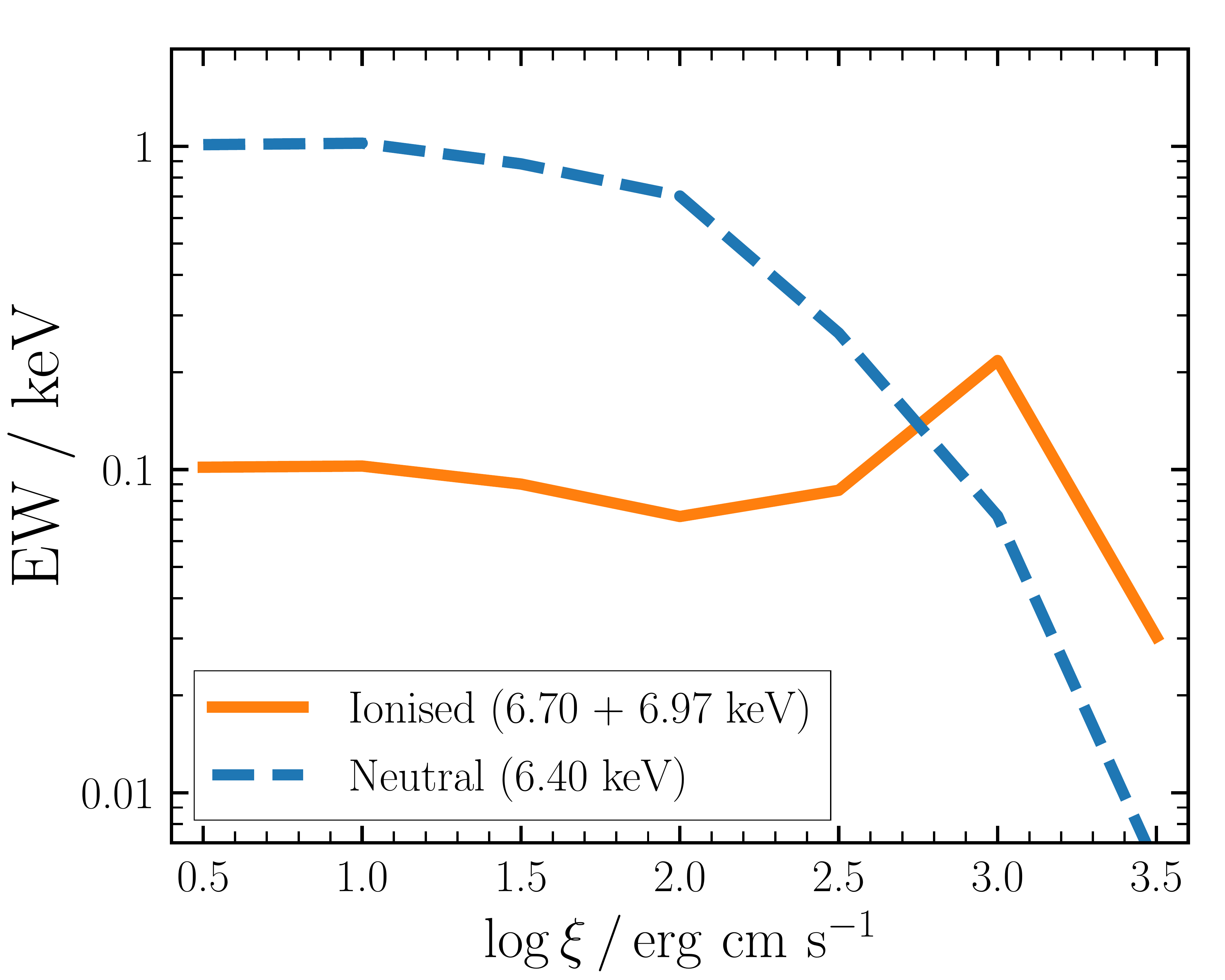}
  \caption{\label{fig:xillver_sim} \texttt{xillver} Simulated EWs of the ionised iron lines at 6.70 and 6.97\,keV (solid orange line) together with neutral Fe K$\alpha$ line EW (dotted blue line) as a function of ionisation parameter.  The EWs were calculated by integrating the flux in the continuum between 5.9\,--\,7.2\,keV, subtracted from the flux in the lines (integrated $\pm\,0.05$\,keV of each predicted line centroid), and then divided by the interpolated continuum at the line centroid.  The spectra were simulated using the \texttt{xillver} spectral model, assuming an inclination of $18.2^\circ$ ($\sim$ face-on, i.e. maximum reflection).}
\end{figure}

To robustly test this would require an ionised toroidal X-ray reprocessing model, which is currently unavailable.  For this reason, we use the \texttt{xillver} \citep{Garcia13} disc reflection model, which calculates a spectrum from the accretion disc surrounding AGN including reflection and also ionised emission lines.  Figure \ref{fig:xillver_sim} illustrates the approximate EWs of the ionised 6.70 and 6.97\,keV iron emission lines and neutral Fe K$\alpha$ line as a function of ionisation parameter.  This is defined as $\xi=4\,\pi\,F_x/n_e$ \citep{Garcia13}, where $F_x$ is the net flux in the 1\,--\,1000\,Ry energy region, and $n_e$ is the electron number density.  From Figure \ref{fig:xillver_sim}, one can infer that the dominance of ionised lines increases with respect to the neutral ones for high values of $\xi$.  Thus, ionisation could be a potential explanation for our results.

\subsubsection{Other possibilities}
\label{sec:other_possibilities}
Another possibility for a depleted neutral iron line in a Compton-thick candidate AGN would be dilution of the reflection spectrum by scattered primary emission from the AGN that is reprocessed by a diffuse ionised `mirror' in a line of sight direction of lower column density.  Such a component would scale with intrinsic luminosity and thus contribute to the Iwasawa-Taniguchi effect.  In fact, a considerable scattered fraction of intrinsic emission was found to explain the observed X-ray spectrum of the local Compton-thick AGN NGC 7674 by \citet{Gandhi17}.  The authors find a fraction of $\sim2-10\%$ to 90\% confidence could explain the low observed EW of the neutral Fe\,K$\alpha$ line from this Compton-thick source.

\begin{figure}
\centering
  \includegraphics[width = 0.5\textwidth]{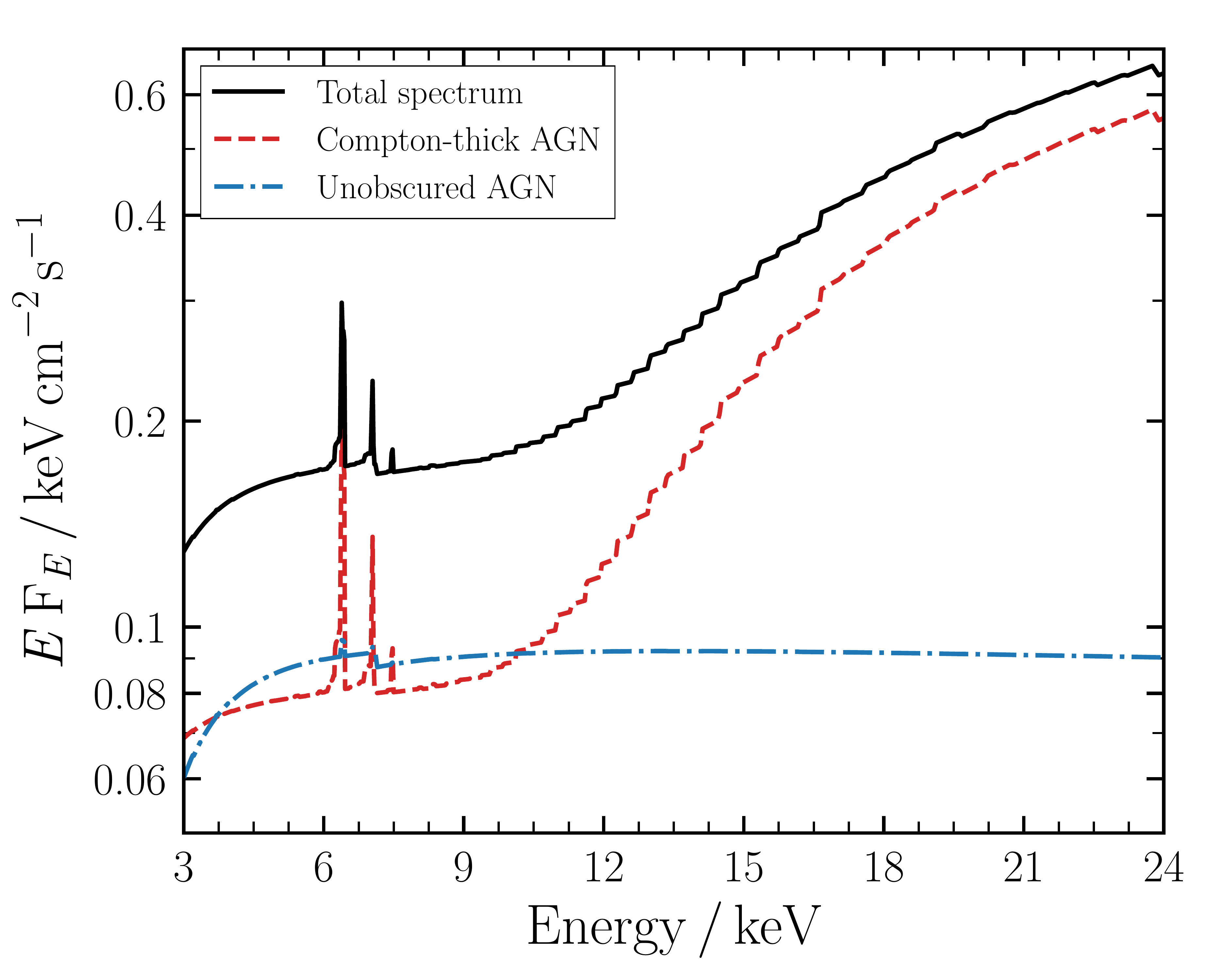}
  \caption{\label{fig:dualAGN} Dual-AGN schematic to explain the lower observed Fe K$\alpha$ EWs observed in luminous and ultraluminous infrared galaxies ($L_{8\,-\,1000\,\mu \rm m}>10^{11}\,\rm{L}_{\rm \odot}$ and $L_{8\,-\,1000\,\mu \rm m}>10^{12}\,\rm{L}_{\rm \odot}$, respectively).  If an AGN were observed post-merger, the two supermassive black holes could exist temporarily separately, but spatially unresolved, leading to an observed (solid black line) composite spectra from two AGN.  The individual unobscured (dot-dash, blue) and obscured (dashed, red) predicted AGN spectra are also plotted.  Depending on the relative contributions in observed flux from either component, the total spectrum could have characteristics of a mildly obscured AGN, despite containing a heavily obscured source.  Both the AGN spectra were simulated using the \texttt{borus02} model.}
\end{figure}

A second tentative explanation for an obscured Iwasawa-Taniguchi effect comes from dual AGN candidate systems, in which a spatially unresolved, less-obscured AGN is present in combination with a heavily obscured source.  Supermassive black hole evolution simulations (from, e.g. \citealt{Hopkins08}) predict luminous quasars to originate from gas-rich mergers.  Immediately post-merger, these sources are predicted to be deeply embedded in the large dust and gas reservoirs that are rapidly being accreted, which absorb optical to X-ray emission and reprocess this at infrared wavelengths.  Depending on the spatial separation of the merging supermassive black holes, the extracted X-ray spectrum could actually be the combined contribution from two components of differing obscuration levels.  \citet{Koss16c} used \textit{NuSTAR} to spatially resolve the emission from the dual AGN in NGC 6921.  The authors found the two AGN components to be Compton-thick, but were able to separately study each independently in the $>$\,10\,keV waveband.  If one component of a dual AGN were less obscured, but contributed a considerable proportion of the total flux contribution, this could result in a diminished Fe K$\alpha$ complex in some post-merger candidates such as hyperluminous ($L_{8\,-\,1000\,\mu \rm m}>10^{13}\,\rm{L}_{\rm \odot}$) infrared galaxies, Dust Obscured Galaxies \citep[DOGs]{Dey08} and Hot DOGs \citep{Wu12}.  For example, recent works have postulated the presence of dual AGN in NGC 7674 \citep{Kharb17} and Mrk 273 \citep{Iwasawa17b}.  Furthermore, \citet{Vito18} recently studied the X-ray emission from 20 Hot DOGs, and found typical predicted X-ray line of sight column densities of $N_{\mathrm{H}}\sim1-1.5\times10^{24}\,\mathrm{cm}^{-2}$.  This is illustrated in Figure \ref{fig:dualAGN}, in which we plot the composite spectrum in black (solid line), formed by combining the spectrum from a reflection-dominated AGN ($\mathrm{log}\,N_{\mathrm{H}}/\mathrm{cm}^{-2}=24.5$; red dashed line) and an unobscured AGN ($\mathrm{log}\,N_{\mathrm{H}}/\mathrm{cm}^{-2}=22$; blue dot-dashed line).  Higher angular resolution X-ray instruments would be required to separate the two components and stringently test this hypothesis.  We further note that both a strong scattered component as well as dual AGN would struggle to explain the observed prominence of ionised iron emission lines often observed in the X-ray spectra of infrared-luminous systems \citep{Teng14,Gilli14,Farrah16}.

A final possibility arises from the effects of dust grains on X-ray photons.  Typical X-ray reprocessing spectral models consist of ray-tracing through a dust-free gas, but not all consider the possible effects that dust grains have on the observed X-ray spectrum in detail.  In fact, \citet{Draine03} has shown that $\sim$\,90\% of the incident power at energy $E=6.4$\,keV on dust grains scatter with angle, $\theta_{\rm s}<0.05^{\circ}$, relative to the incident photon direction.  \citet{Gohil15} further found that such a large anisotropic emission associated with dust grains, as opposed to the isotropic emission of hot gas typically invoked in X-ray reprocessing torus models could enhance the Fe\,K$\alpha$ line EW relative to the underlying reflection continuum by up to factors of $\sim$\,8 for Compton-thick gas.  More luminous AGN would be expected to have a larger dust sublimation radius, and thus have an altered Fe K$\alpha$ EW relative to the less luminous sources, presumably with smaller dust sublimation radii.  Such a scenario effectively decouples the Fe\,K$\alpha$ line from the underlying reflection continuum, and there is already tentative evidence suggesting that these two components may arise from physically separate regions within the torus dust sublimation zone \citep{Gandhi15b}.

\subsection{Implications}
\label{sec:implications}

\subsubsection{Redshift Evolution of Compton-thick AGN}
\label{sec:redshift}

As stated earlier, multiple works predict the obscured fraction of AGN to decrease with increasing luminosity and/or Eddington fraction \citep{Ueda11,Merloni14,Georgakakis17}.  In addition, since the number of luminous AGN is predicted to increase with redshift, this would imply a redshift evolution of obscuration amongst AGN.  However, the anti-correlation we report could lead to a correction to X-ray inferred column densities, that were derived based on fitting an observed iron line.  From Figure \ref{fig:fit}, this correction factor would be largest for the most luminous sources.  Depending on the relative contributions at different luminosities, this could then alter the obscured fraction dependence with luminosity.  Some evidence has indicated a weak or no evolution of the obscured AGN fraction.  For example, \citet{Vito14} studied a sample of 141 X-ray selected AGN at $3<z\leq5$ and found no evidence for an anti-correlation between obscured fraction and luminosity, despite suggesting that this may be due to the non-detection of the lower luminosity obscured sources at higher redshift.  In contrast, \citet{Mateos17} only found a weak luminosity dependence of the type 2 AGN fraction for covering factors derived from infrared clumpy torus modelling for $z\leq1$, and \citet{Buchner15} further found a constant Compton-thick fraction with redshift or accretion luminosity for a sample of $\sim$\,2000 AGN.

\begin{figure}
  \includegraphics[width = 0.5\textwidth]{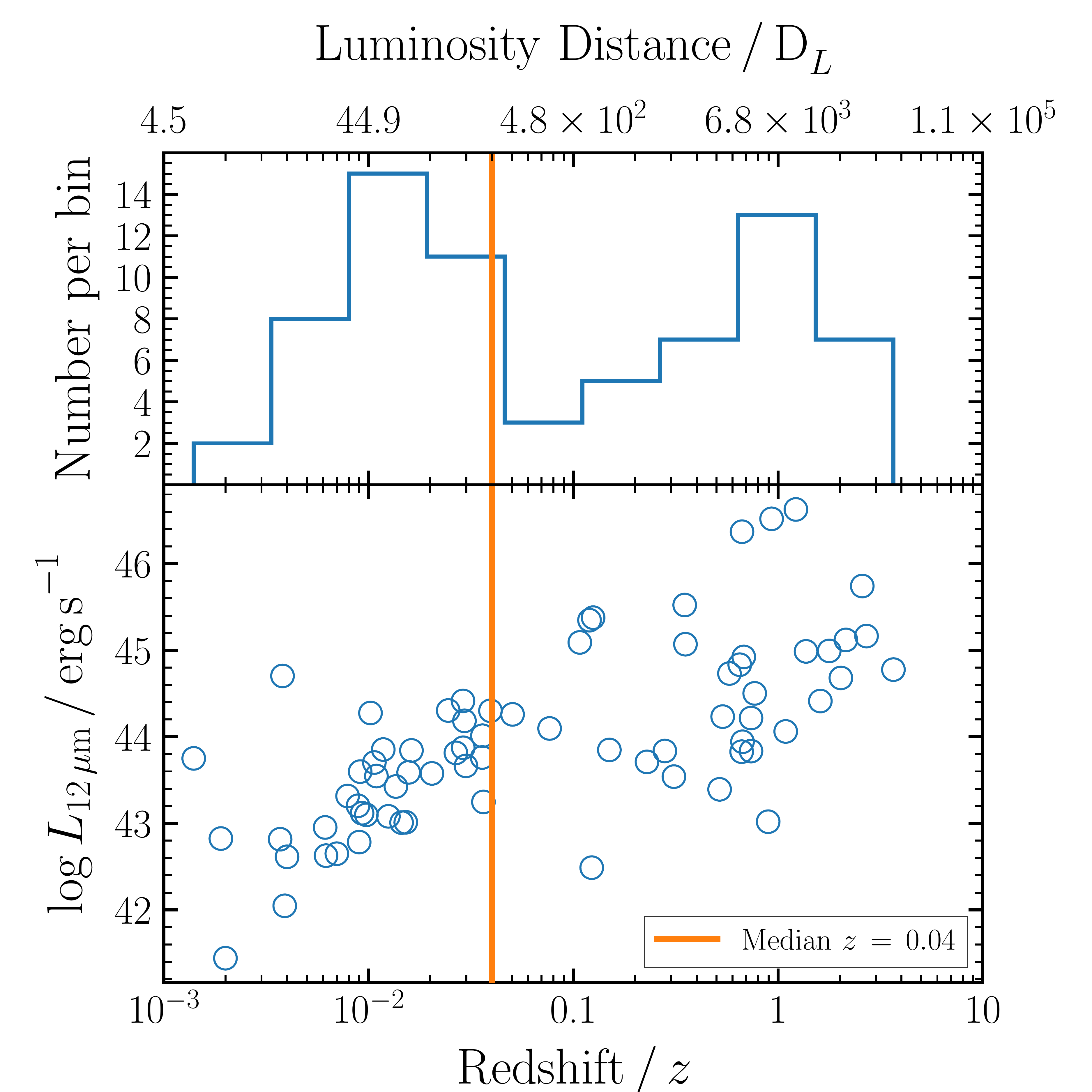}
  \caption{\label{fig:z_dist} Redshift distribution of our sample, according to the rest-frame 12\,$\mu \rm m$ luminosity.  The median redshift was chosen to split the sample into two redshift bins, in order to provide comparable numbers of sources in either bin.}
\end{figure}

To test if our result is biased by redshift, we further separated our sample into two redshift bins, below and above the median redshift of \medred{}.  Figure \ref{fig:z_dist} shows the redshift distribution of our sample with the median redshift shown with an orange (vertical) line.  Carrying out a fit to either redshift bin independently yielded consistent gradients of \LTmedzFIT{} and \GTmedzFIT{}, for the low and high redshift bins respectively, albeit with large scatter.

\subsubsection{The Growth Rate of AGN}
\label{sec:other_implications}

Current X-ray reprocessing torus models do not account for the possible effects of dust grains, and/or reflector ionisation on the observed reflection spectrum.  To zeroth order, such a model could interpret a less prominent neutral Fe\,K$\alpha$ line as evidence for a lower obscuring column than the true value for intrinsically bright, heavily obscured objects if our results are confirmed using larger sample studies.  In the most extreme case, a Compton-thick system could be predicted to be only mildly obscured.  From Figure \ref{fig:selection}, this would mean underpredicting the intrinsic X-ray luminosity, and hence the growth rate of such systems, potentially by factors of around two orders of magnitude.

\subsection{The Future}

\begin{figure}
\begin{center}
\includegraphics[angle=0,width=\columnwidth]{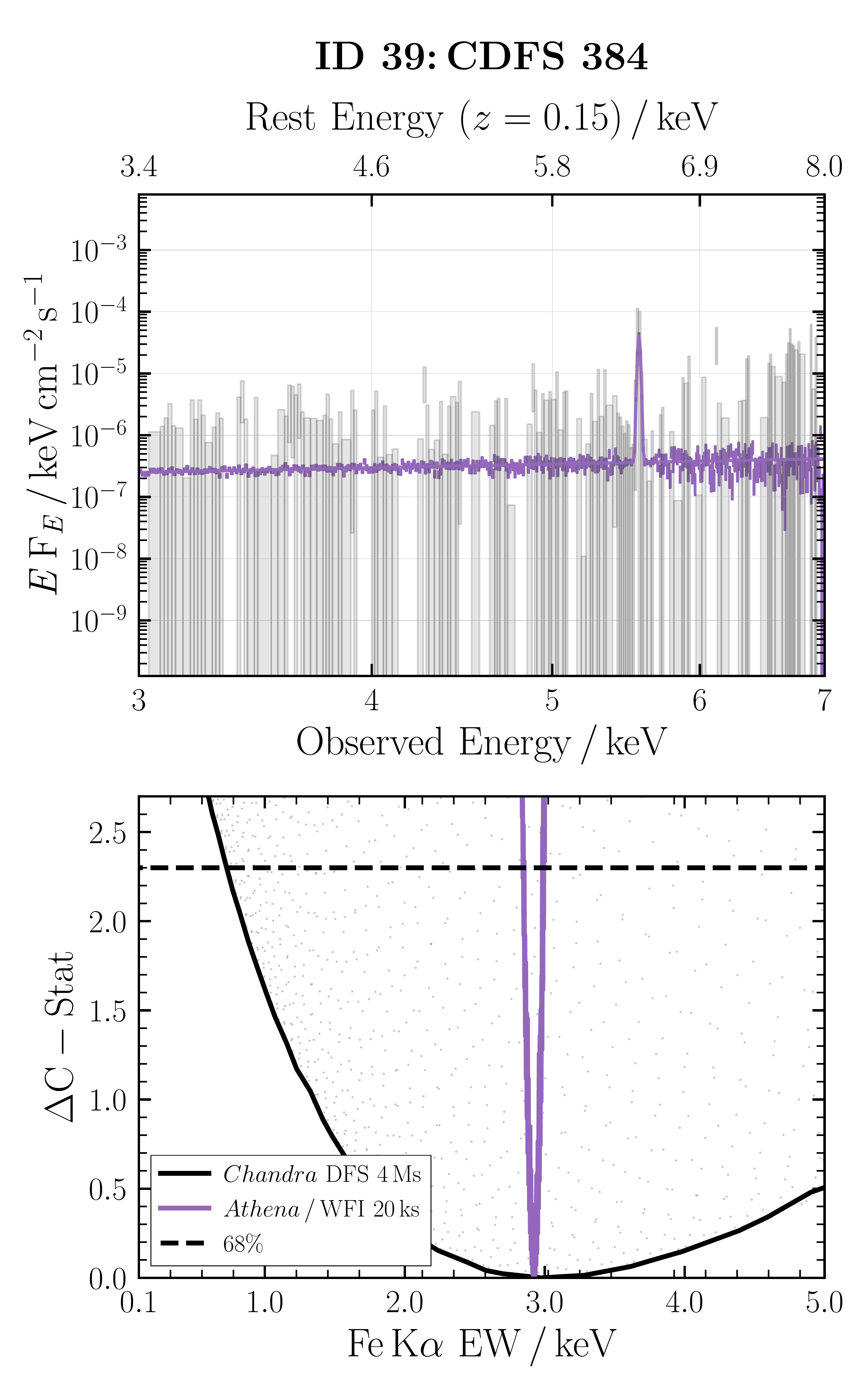}\\
\caption{\label{fig:cdfs384_athena} Top panel: Observed \textit{Chandra} Deep Field South (CDFS) 4\,Ms spectrum for CDFS 384 together with simulated \textit{Athena}/WFI spectrum for a 20\,ks observation, shown in black and purple, respectively.  Lower panel: Reported EW contour for CDFS 384 used in the paper, together with the improved contour attained with the simulated \textit{Athena}/WFI spectrum shown above.  A dramatic improvement to the S/N of the observed spectrum is clearly attainable with \textit{Athena}, not to mention the improved EW confidence region.  Such observations of Compton-thick AGN are to be carried out as part of the \textit{Athena} mission, with the aim to study obscured accretion and galaxy formation with the WFI instrument.}
\end{center}
\end{figure}

Clearly one of the major sources of uncertainty in the relation we report here is on the EW of the Fe\,K$\alpha$ line.  However, many future X-ray missions can improve on this uncertainty, such as the \textit{X-ray Astronomy Recovery Mission (XARM)}\footnote{https://heasarc.gsfc.nasa.gov/docs/xarm/}, or \textit{Athena} \citep{Nandra13}.  \textit{XARM} will enable high spectral resolution studies of the iron line (e.g. \citealt{hitomi17c}) and be able to test the ionisation scenario directly.  Additionally, \textit{Athena} will probe high redshift Compton-thick AGN sensitively.  Figure \ref{fig:cdfs384_athena} shows the possibilities with the \textit{Athena} Wide Field Imager (WFI), with a 20\,ks simulated spectrum shown in purple together with the original 4\,Ms \textit{Chandra} Deep Field South observed spectrum for CDFS 384 from our sample, shown in black. Clearly the signal to noise is dramatically enhanced at the Fe\,K$\alpha$ (rest-frame 6.4\,keV) line as well as the neighbouring continuum, enabling a huge improvement on the calculated EW contour in the lower panel of the figure.  The simulated spectrum was calculated from the original best fit model to the observed \textit{Chandra} spectrum.  What this figure clearly shows, however, is that the confidence range on the EW of such obscured objects will be powerfully improved with the advent of such high-sensitivity instruments in the future.

\section{Summary}
\label{sec:summary}
Here we have carried out the first study into the Iwasawa-Taniguchi effect for Compton-thick AGN.  Our key findings are enumerated below:

\begin{enumerate}
\item We select from the literature a sample of \total{} Compton-thick candidate AGN, covering a redshift range of $z \sim 0.0014-3.7$.  The candidates were confirmed via an offset between predicted intrinsic and observed X-ray luminosity, given the rest-frame 12\,$\mu$m luminosity interpolated from the \citet{Mullaney11} AGN infrared spectral template.

\item We find an anti-correlation between the rest-frame equivalent width of the narrow core of the neutral Fe\,K$\alpha$ fluorescence emission line and the mid-infrared $12\,\mu \rm m$ continuum luminosity, which we use as a proxy for the bolometric AGN luminosity.  From the Spearman's Rank, we find the anti-correlation to be significant to \rhoneg{} confidence.

\item We discuss four possible interpretations of such an anti-correlation:
\begin{enumerate}
\item A luminosity-dependent covering factor (Section \ref{sec:recedingtorus}).
\item Luminosity-dependent ionisation state of the circumnuclear reprocessing material (Section \ref{sec:ionisation}).
\item Other possibilities including the contribution from two AGN in a dual system (Section \ref{sec:other_possibilities}).
\end{enumerate}

\item Possible implications of the Compton-thick Iwasawa-Taniguchi effect include:
\begin{enumerate}
\item An increased number density of Compton-thick AGN at higher redshifts due to predicted higher intrinsic luminosities (Section \ref{sec:redshift}).
\item Current X-ray reprocessing models do not account for this effect, and as such may incorrectly interpret a weak Fe\,K$\alpha$ line as a signature of Compton-thin reprocessing, leading to an under-estimation of the true intrinsic luminosity and hence growth rate of X-ray-obscured AGN.
\item If a luminosity-dependent covering factor can explain the Iwasawa-Taniguchi effect, it would imply that we are still lacking a population of truly reflection-dominated, luminous Compton-thick AGN, since for $N_{\rm H}\gtrsim1.5\times 10^{24}\,\rm{cm}^{-2}$, the EW of the Fe K$\alpha$ line are predicted to always be $>$\,1\,keV, contrary to what we find.
\end{enumerate}

\item This work further illustrates why the Fe\,K$\alpha$ line alone cannot be directly used to accurately determine the line of sight column density to a source.  Future dedicated studies of Compton-thick AGN over broad redshift ranges are required to be able to confirm this effect, which could further hold the answers to understanding the physical geometry and evolution of obscuration surrounding AGN.

\end{enumerate}

\section*{Acknowledgements}

We thank the anonymous referee for invaluable comments on the paper.

P.\,B. and P.\,G. (grant reference ST/J003697/2) thank the STFC for support.  In addition, the authors thank R. Gilli and C. Circosta for providing the 4 and 7\,Ms spectra of LESS J0033229.4-275619 used in Figure \ref{fig:ctagn}.

This work was supported in part by the Black Hole Initiative at Harvard University, which is funded by a grant from the John Templeton Foundation.

M.\,B. acknowledges support from NASA Headquarters under the NASA Earth and Space Science Fellowship Program, grant NNX14AQ07H.

We acknowledge financial support from FONDECYT 1141218 (C.\,R.), Basal-CATA PFB--06/2007 (C.\,R.), the China-CONICYT fund (C.\,R.). This work is partly sponsored by the Chinese Academy of Sciences (CAS), through a grant to the CAS South America Center for Astronomy (CASSACA) in Santiago, Chile.

The scientific results reported in this article are based on observations made by the \textit{Chandra} X-ray Observatory.

This research has made use of data, software and/or web tools obtained from the High Energy Astrophysics Science Archive Research Center (HEASARC), a service of the Astrophysics Science Division at NASA/GSFC and of the Smithsonian Astrophysical Observatory's High Energy Astrophysics Division.

This research has made use of the NASA/IPAC Extragalactic Database (NED), which is operated by the Jet Propulsion Laboratory, California Institute of Technology, under contract with the National Aeronautics and Space Administration.

This work made use of data from the {\it NuSTAR} mission, a project led by the California Institute of Technology, managed by the Jet Propulsion Laboratory, and funded by the National Aeronautics and Space Administration. We thank the {\it NuSTAR} Operations, Software and Calibration teams for support with the execution and analysis of these observations.  This research has made use of the {\it NuSTAR} Data Analysis Software (NuSTARDAS) jointly developed by the ASI Science Data Center (ASDC, Italy) and the California Institute of Technology (USA).

This work is based [in part] on observations made with the \textit{Spitzer} Space Telescope, which is operated by the Jet Propulsion Laboratory, California Institute of Technology under a contract with NASA.

This publication makes use of data products from the Wide-field Infrared Survey Explorer, which is a joint project of the University of California, Los Angeles, and the Jet Propulsion Laboratory/California Institute of Technology, funded by the National Aeronautics and Space Administration.

This work made use of the \texttt{NumPy} \citep{NumPy11}, \texttt{Matplotlib} \citep{Hunter07}, \texttt{SciPy}\citep{Scipy01}, \texttt{pandas} \citep{Pandas10}, \texttt{Astropy} \citep{Astropy13} and \texttt{adjustText}\footnote{https://github.com/Phlya/adjustText} \texttt{Python} packages.

P.\,B. would also like to thank S.\,H{\"o}nig, C.\,Knigge, J.\,Matthews, M.\,Middleton, M.\,Smith, A.\,Beri, A.\,Hill, J.\,Buchner and others for vital scientific discussions into the data analysis and interpretations of the Compton-thick Iwasawa-Taniguchi effect.




\bibliographystyle{mnras}

\bibliography{/Users/pboorman/Dropbox/1_work/bibliography}


\appendix
\,
\,
\,
\noindent All spectra presented here are plotted with energies in the source observed frame on the lower axis, with the source rest-frame energy shown on the upper axis.  Sources with an additional \texttt{apec} component included in the spectral fit are shown with a corresponding label in their legend.

The grouping used is annotated on each plot and has one of two possibilities:
\begin{enumerate}
  \item Binning by a minimum number of counts per bin.
  \item Binning to have a minimum S/N ratio in each bin.
\end{enumerate}

All sources were fitted with a simplified phenomenological model consisting of photoelectric absorption acting on a composite powerlaw ($\Gamma$, the photon index of the powerlaw was assigned to 1.4 for all cases) plus a narrow Gaussian of $\mathrm{FWHM}\approx\,2\,$eV ($\sigma\,=\,1\,\mathrm{eV}$), modelling the observed continuum and narrow core of the Fe\,K$\rm{\alpha}$ fluorescence line, respectively.  See Section 3.3 for further details on the spectral model adopted.  The corresponding confidence contours shown (where applicable) in the top right panel illustrate a delta statistic of +2.30 to represent the 1-$\sigma$ (68\%) confidence level for two interesting parameters\footnote{\url{https://heasarc.gsfc.nasa.gov/xanadu/xspec/manual/XSappendixStatistics.html}}.

All spectra shown feature the spectral fit to the data and the \textsc{del} for the fit in the top and bottom panels, respectively.  \textsc{del} is defined as the (data\,--\,model)/error.

\section{Sources Excluded}
\label{app:ruled_out}
\textit{NuSTAR} data was not publicly available for 19/55 low redshift sources from the \textit{Neil Gehrels Swift}/BAT sample of \citet{Ricci15}, and so were excluded from this work.  See Section 2.2 for more information.  This excluded ESO 565-G019, which is in the \citet{Gandhi14} bona-fide Compton-thick AGN sample, and has been studied individually in \citet{Gandhi15c} with \textit{Suzaku} data.

In addition, our own analysis of the archival archival XMM-\textit{Newton} EPIC/PN spectrum as compared to the more recent \textit{NuSTAR} FPMA \& FPMB spectra strongly indicated a changing-look AGN scenario for NGC 4102 and NGC 4939.  These sources were thus excluded since changing-look AGN could adhere to variable obscuration effects.

\begin{landscape}
\begin{figure}
\begin{multicols}{3}
\includegraphics[angle=0,width=\columnwidth]{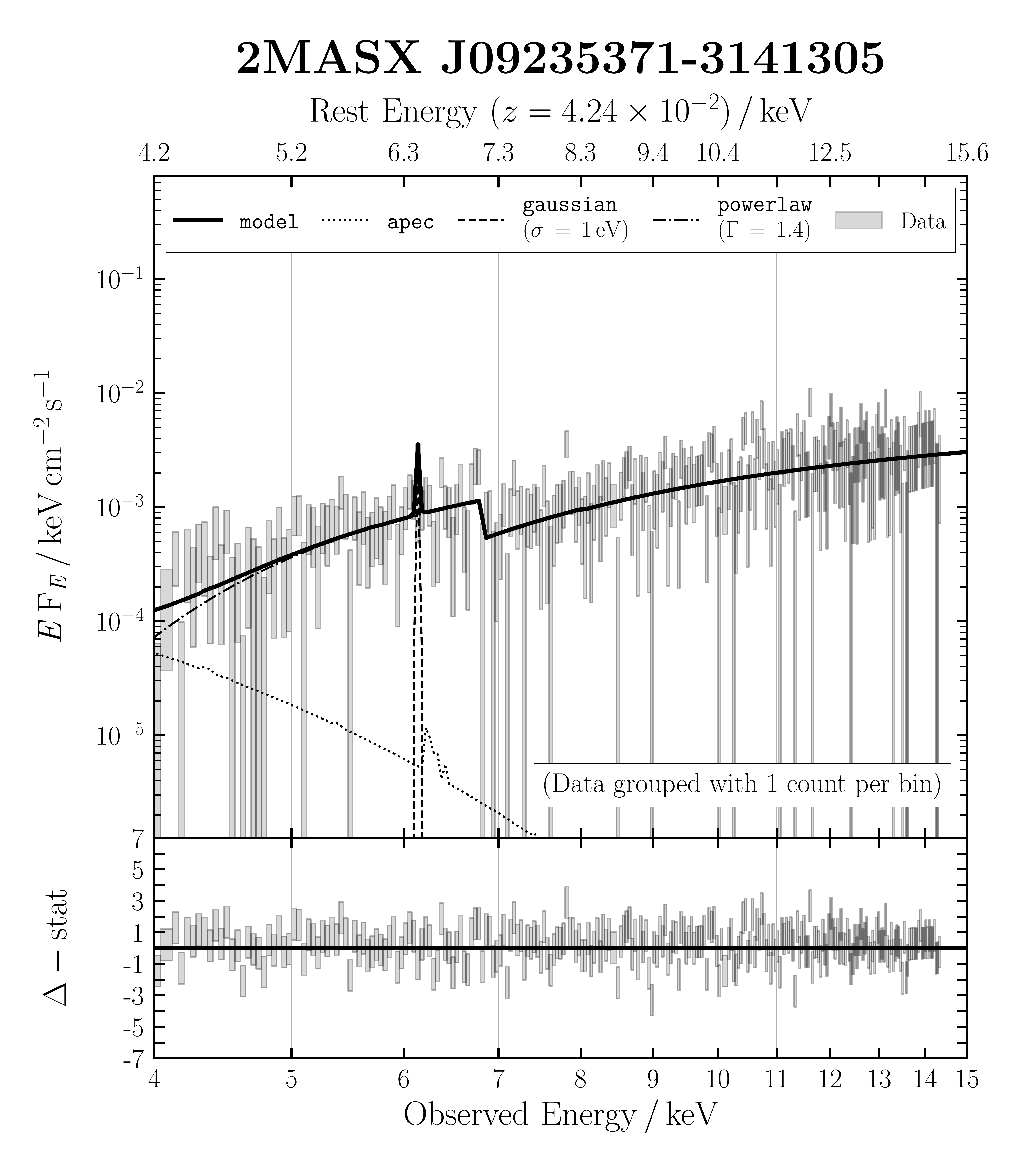}\\
\vspace{0.15cm}
\includegraphics[angle=0,width=\columnwidth]{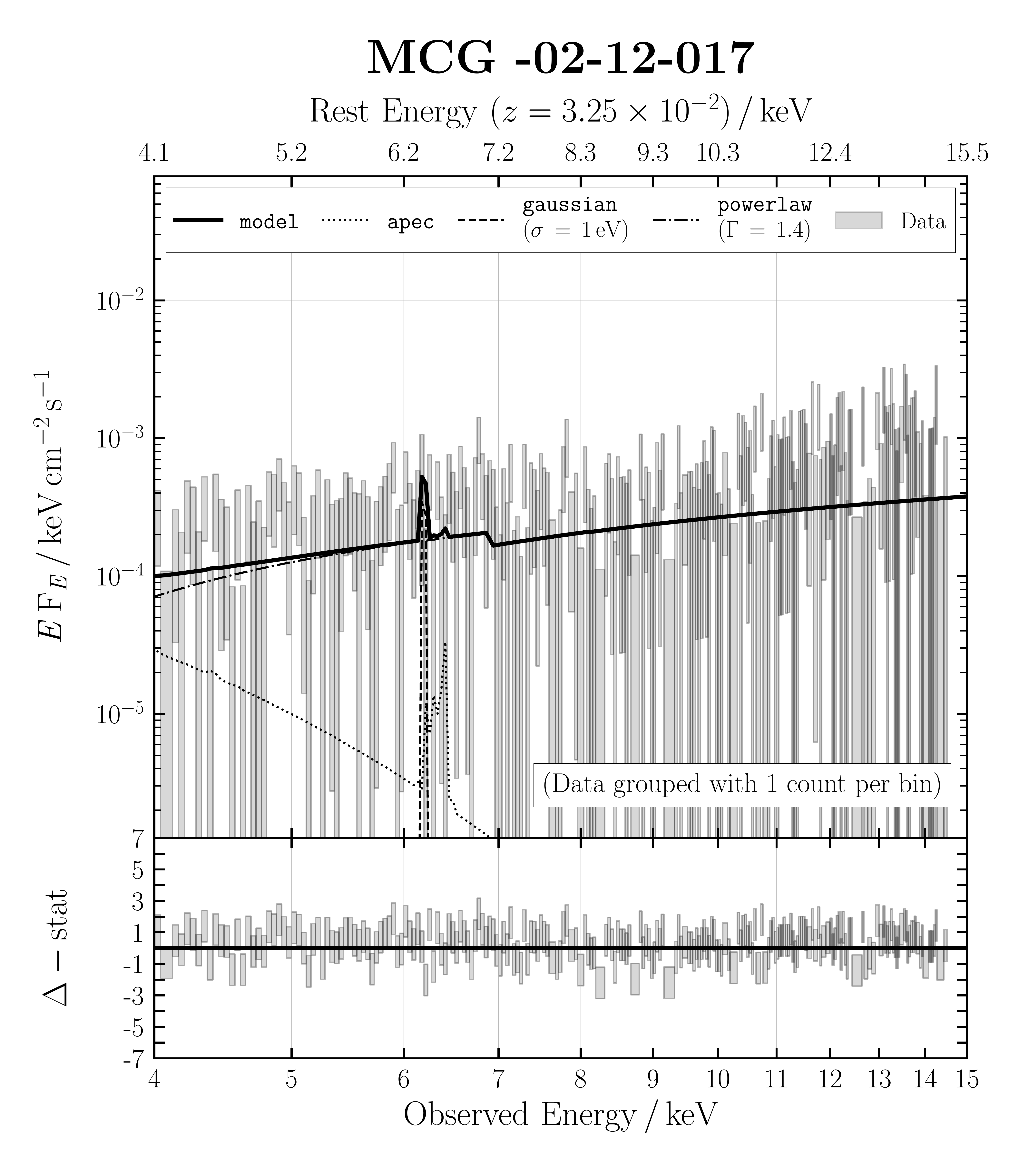}\\
\vspace{0.15cm}
\includegraphics[angle=0,width=\columnwidth]{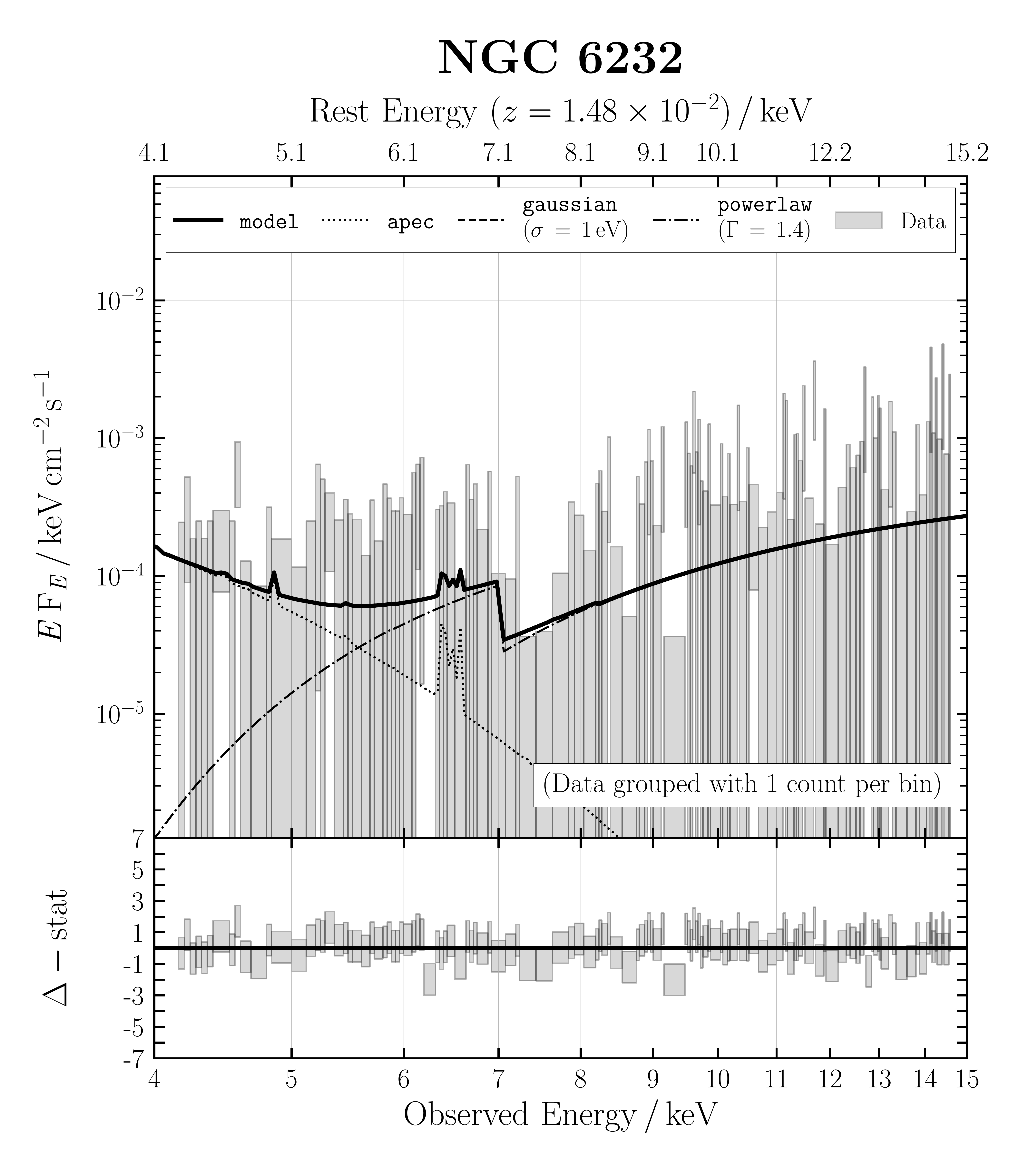}\\
\vspace{0.15cm}
\includegraphics[angle=0,width=\columnwidth]{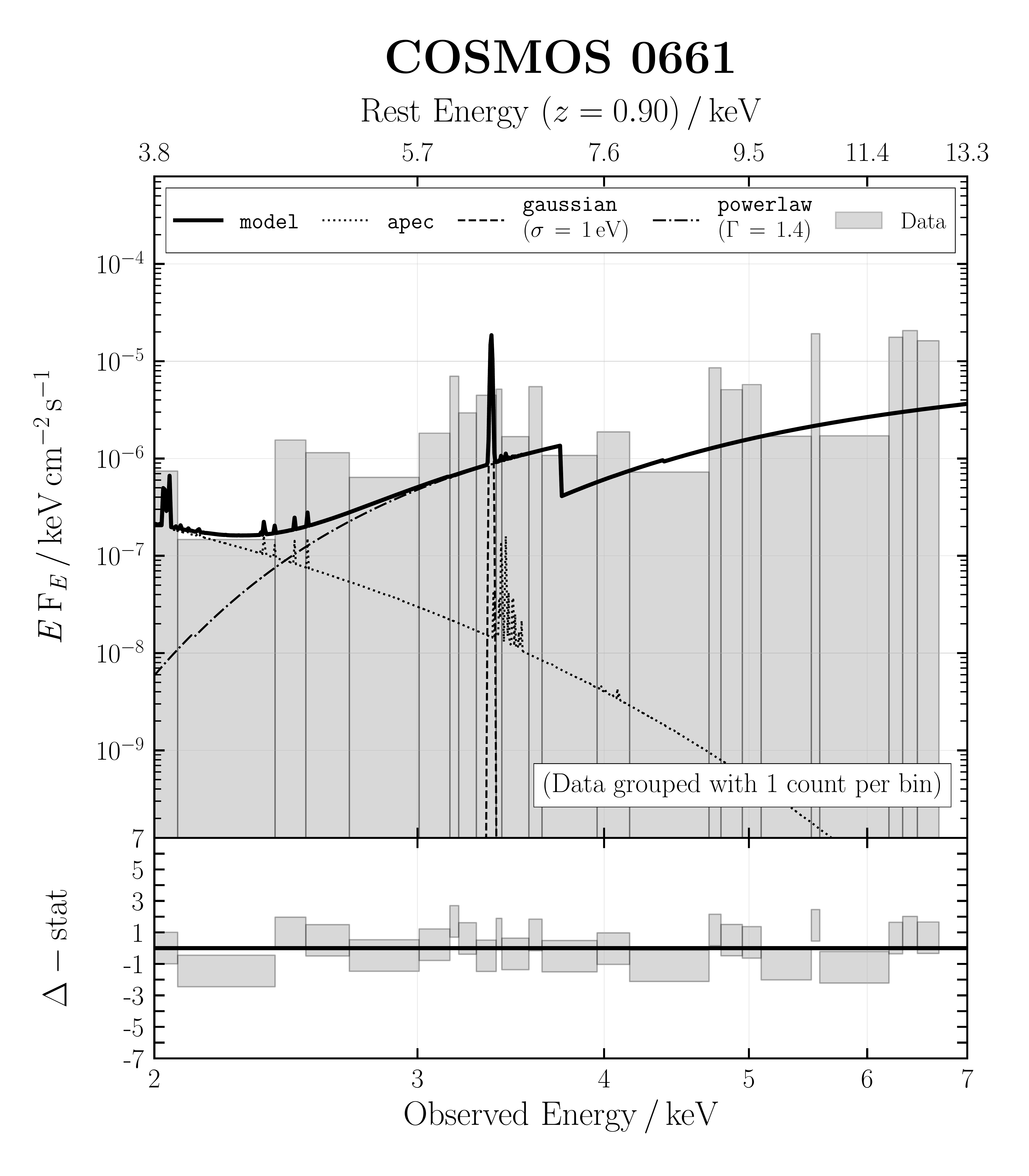}\\
\vspace{0.15cm}
\includegraphics[angle=0,width=\columnwidth]{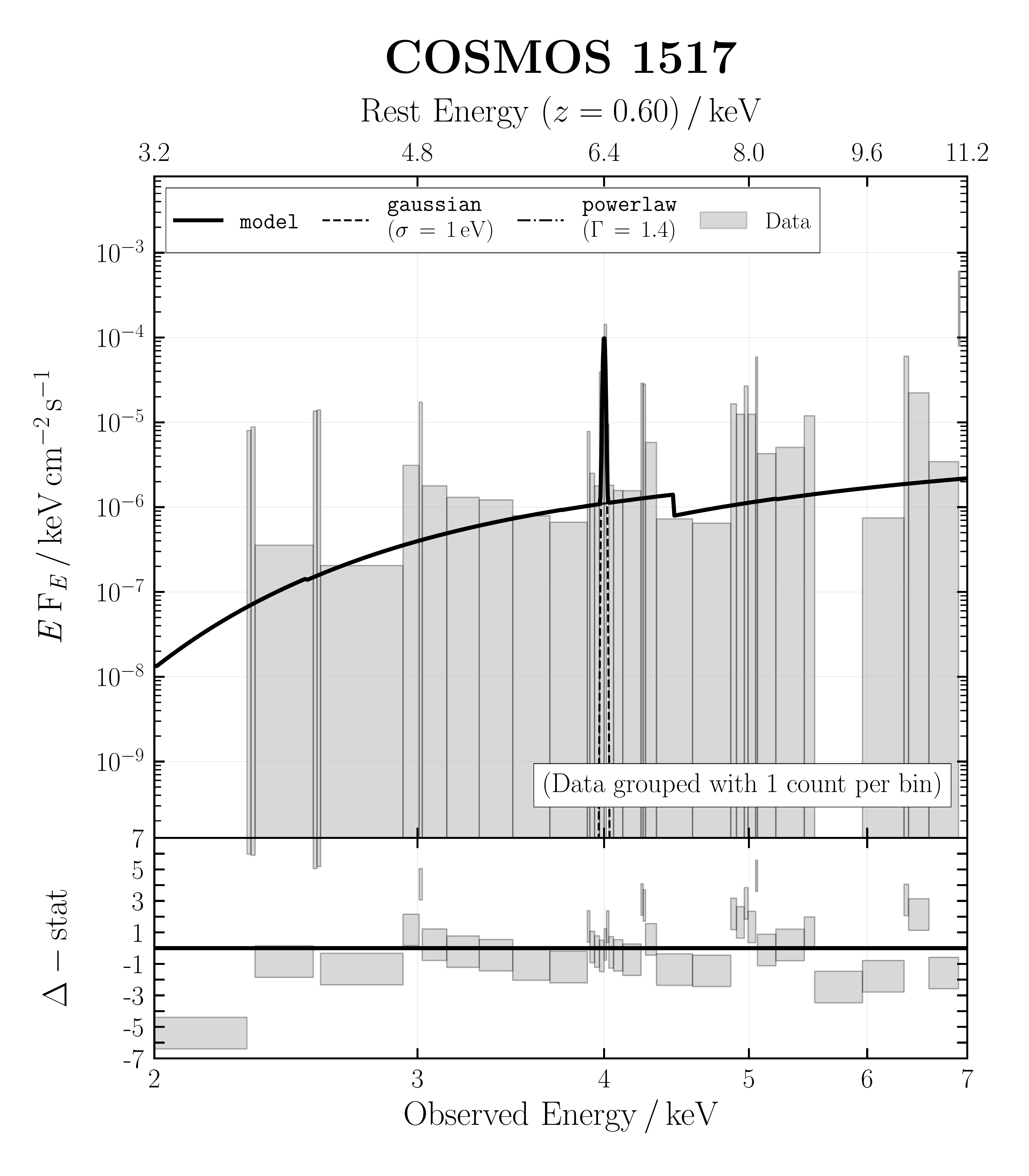}\\
\vspace{0.15cm}
\end{multicols}
\caption{\label{fig:excluded} Spectra of the 5 sources ruled out in our analysis due to an agreement with the \citet{Asmus15} correlation between observed X-ray and mid-infrared luminosity, which can infer less than Compton-thick obscuration.  For details of the spectrum, see the description at the start of the Appendix.  The top and bottom panels show the spectral fit to the data and the \textsc{del} for the fit, respectively.  \textsc{del} is defined as the (data\,--\,model)/error.}
\end{figure}
\end{landscape}

Finally, 5 sources had observed rest-frame 2\,-\,10\,keV fluxes in agreement with the interpolated rest-frame 12\,$\mu$m flux, predicted from the relation presented in \citet{Asmus15}.  These 5 sources were ruled out from our sample, and their spectra are shown in Figure \ref{fig:excluded}.

\section{Sources included}
\label{app:EW}

Here we include individual spectra and equivalent width (EW) contours for the sources we derive EWs for ourselves.  The sources are ordered in ascending 12$\mu$m luminosity, as in Table 1 of the paper.  We used the limit derived from best-fit parameters for 3 sources that the contour method did not provide a reasonable constraint for.  These sources are: COSMOS0581, COSMOS 0987 and CDFS 460.  Furthermore, due to an unphysical EW determined for CDFS 443, CDFS 454 and COSMOS 2180, we fixed the EW for these sources to be $<\,$5\,keV.

The upper right panel for each source figure indicates the contour plot for the EW, with the grid best-fit values shown as faint grey points.  Statistical details of the spectral fit are tabulated in the bottom right panel of each source figure.  All uncertainties shown from the intersection of the horizontal black line with the solid line contour correspond to the 68\% confidence level for two interesting parameters.

\clearpage
\onecolumn

\begin{figure}
\begin{center}
\includegraphics[angle=0,width=\columnwidth]{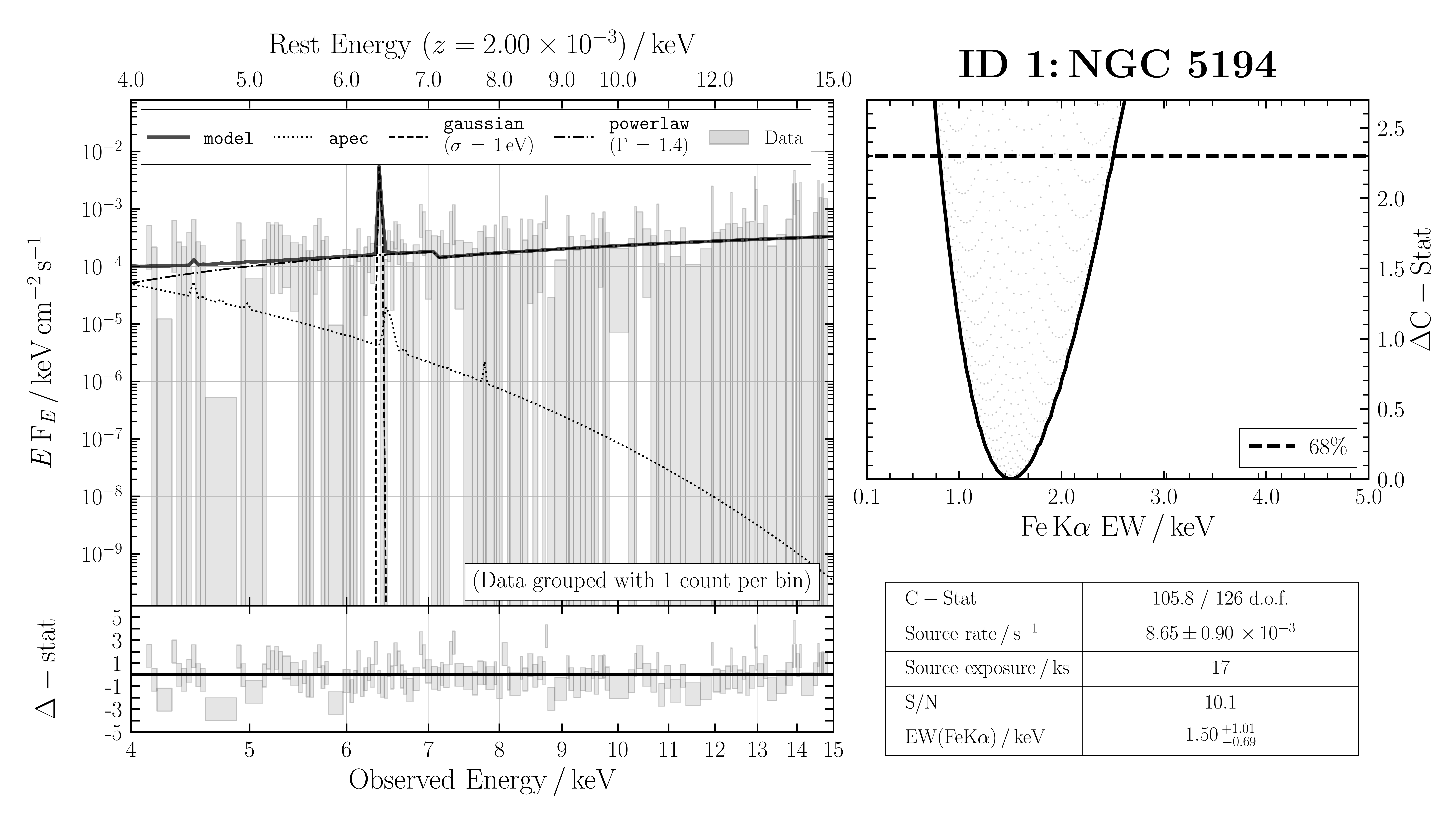}\\
\caption{\label{fig:NGC5194} ID 1: NGC 5194}
\end{center}
\end{figure}

\begin{figure}
\begin{center}
\includegraphics[angle=0,width=\columnwidth]{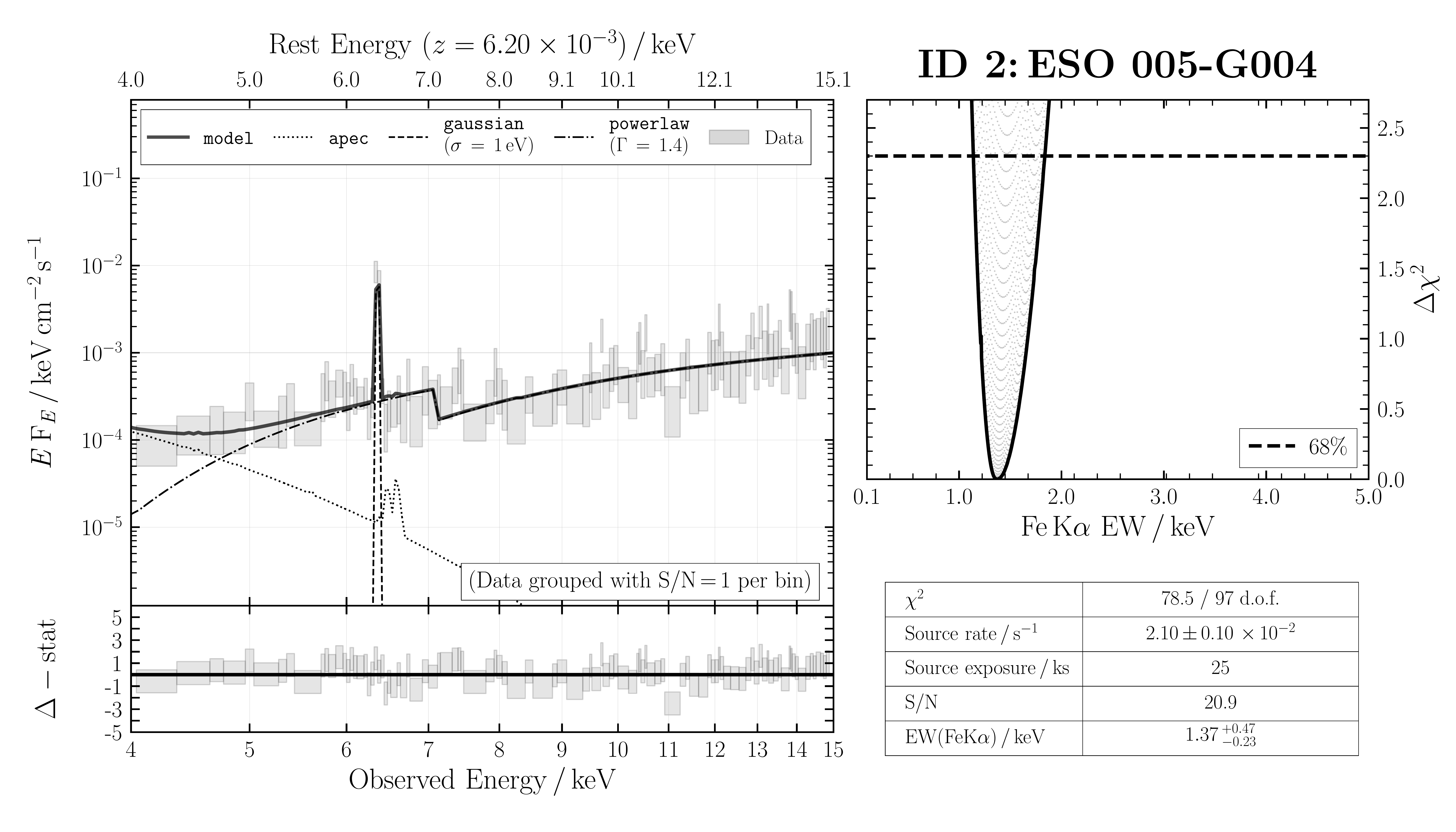}\\
\caption{\label{fig:ESO005-G004} ID 2: ESO 005-G004}
\end{center}
\end{figure}

\begin{figure}
\begin{center}
\includegraphics[angle=0,width=\columnwidth]{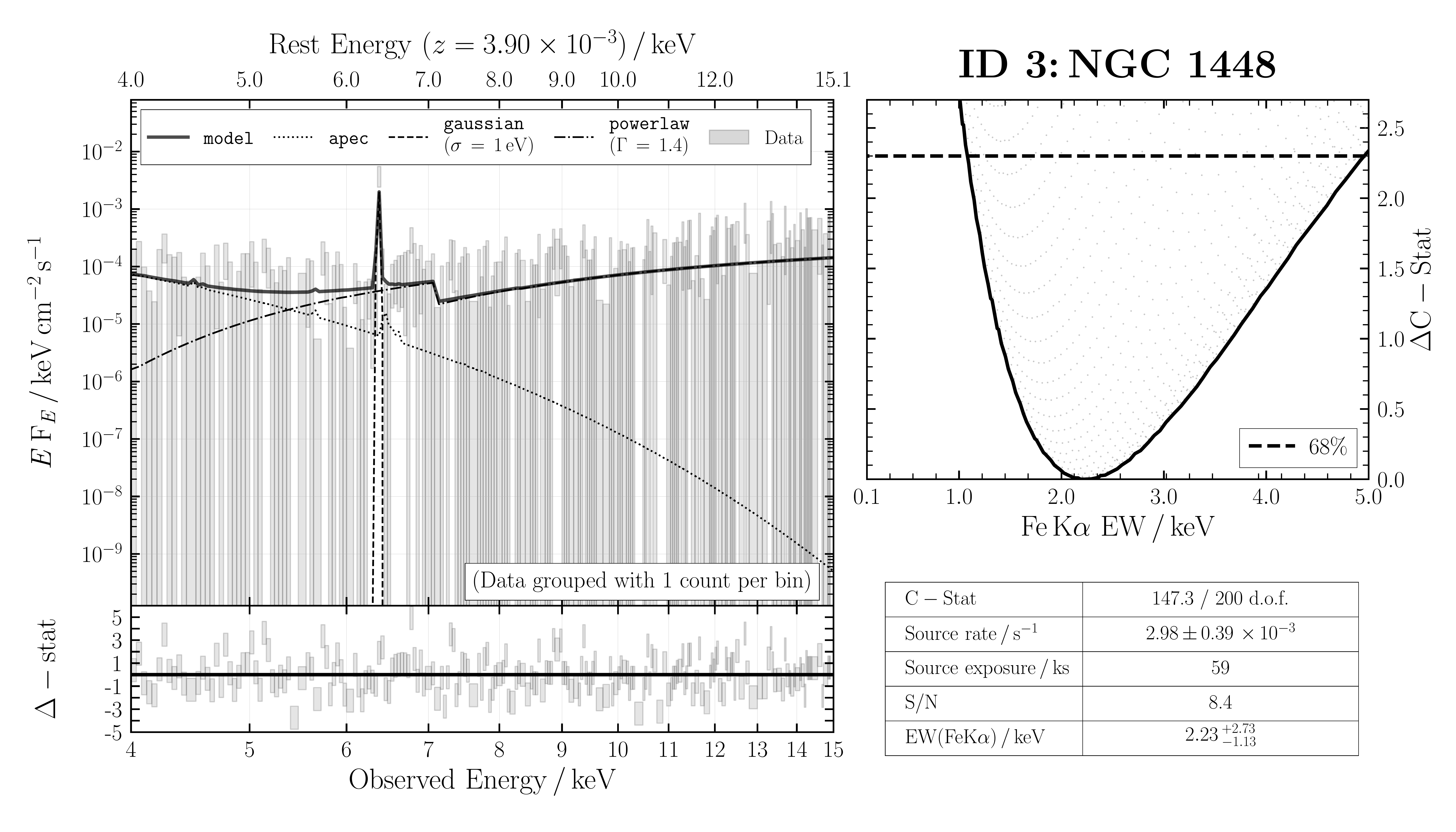}\\
\caption{\label{fig:NGC1448} ID 3: NGC 1448}
\end{center}
\end{figure}

\begin{figure}
\begin{center}
\includegraphics[angle=0,width=\columnwidth]{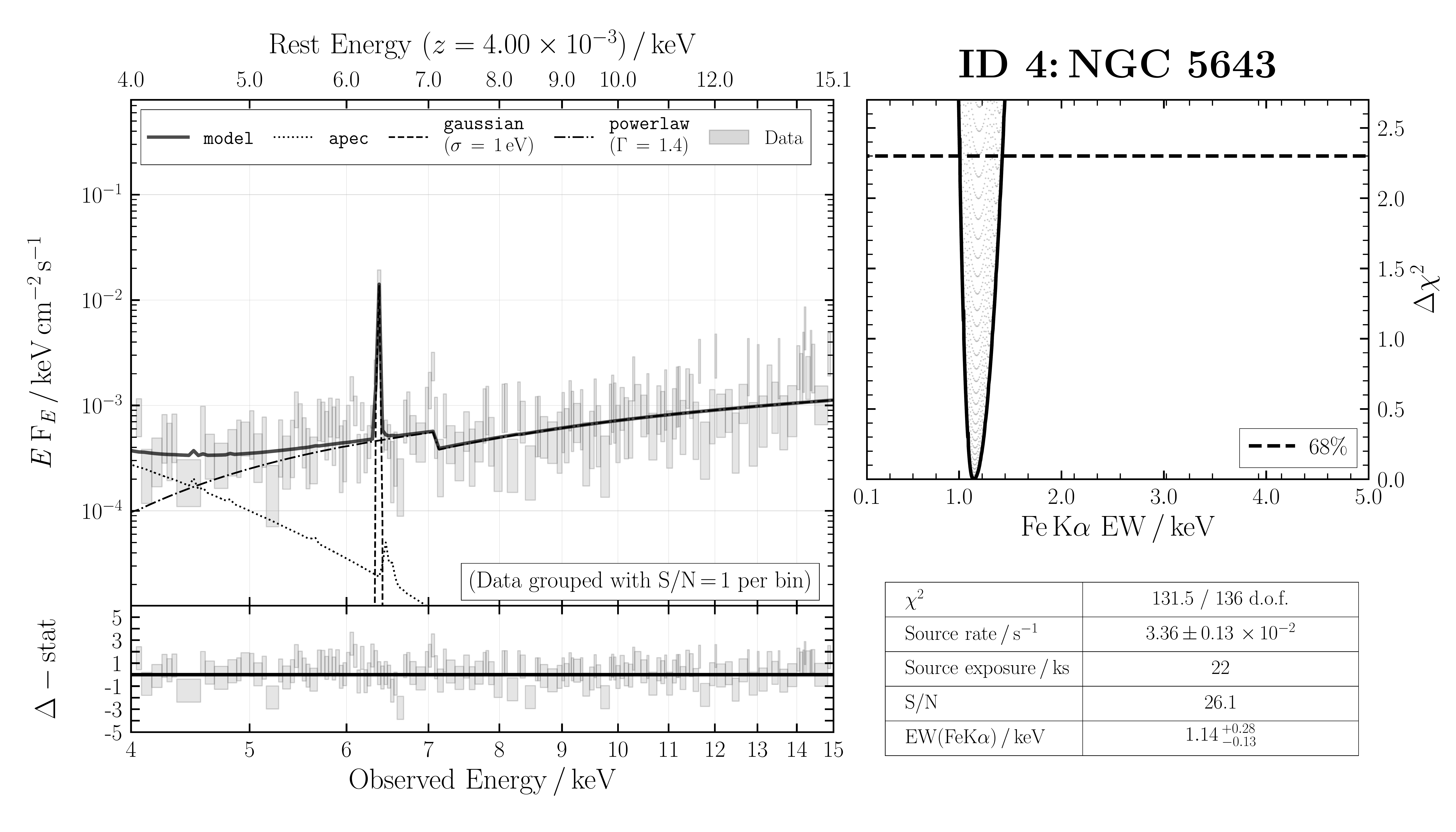}\\
\caption{\label{fig:NGC5643} ID 4: NGC 5643}
\end{center}
\end{figure}

\begin{figure}
\begin{center}
\includegraphics[angle=0,width=\columnwidth]{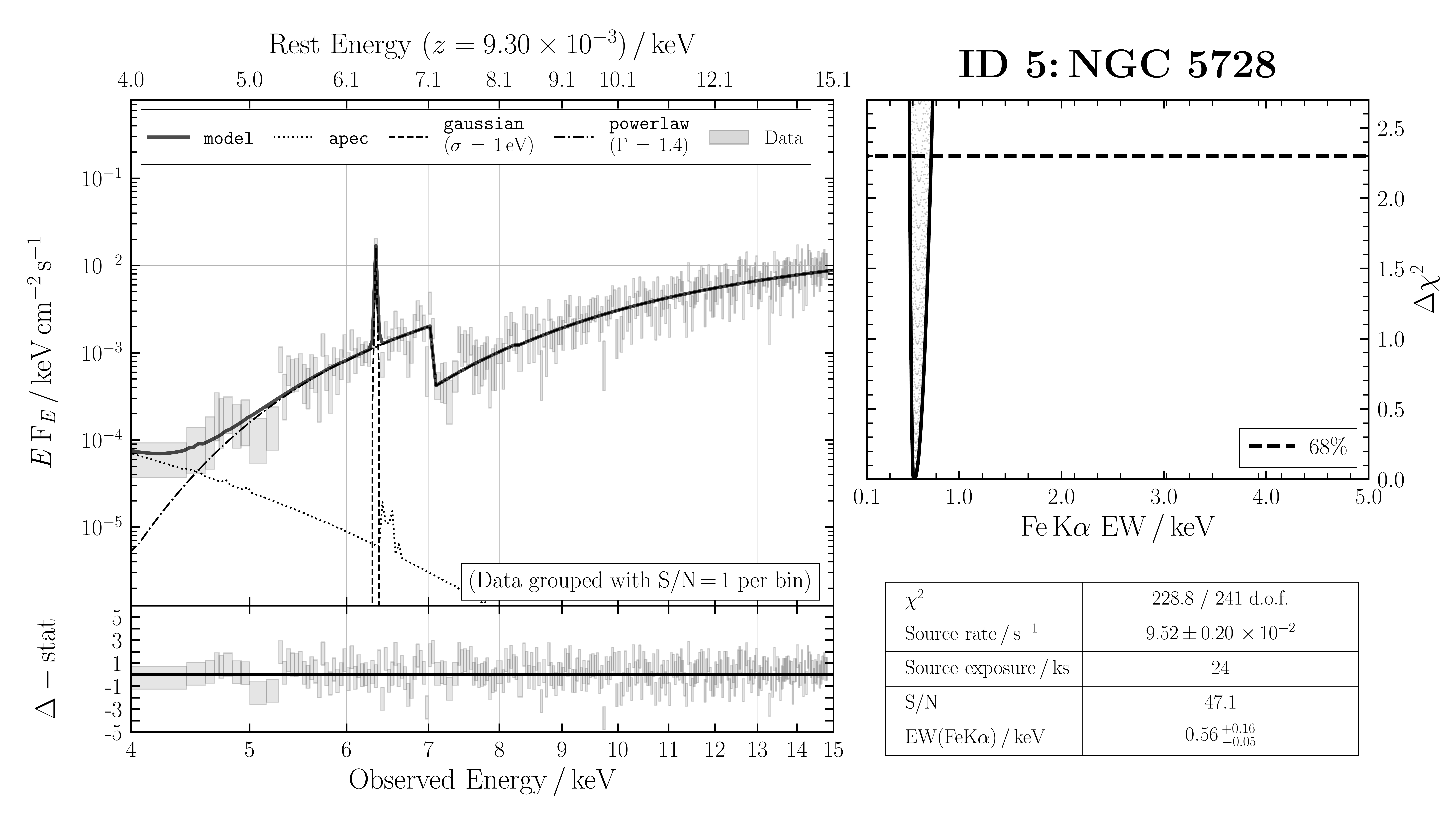}\\
\caption{\label{fig:NGC5728} ID 5: NGC 5728}
\end{center}
\end{figure}

\begin{figure}
\begin{center}
\includegraphics[angle=0,width=\columnwidth]{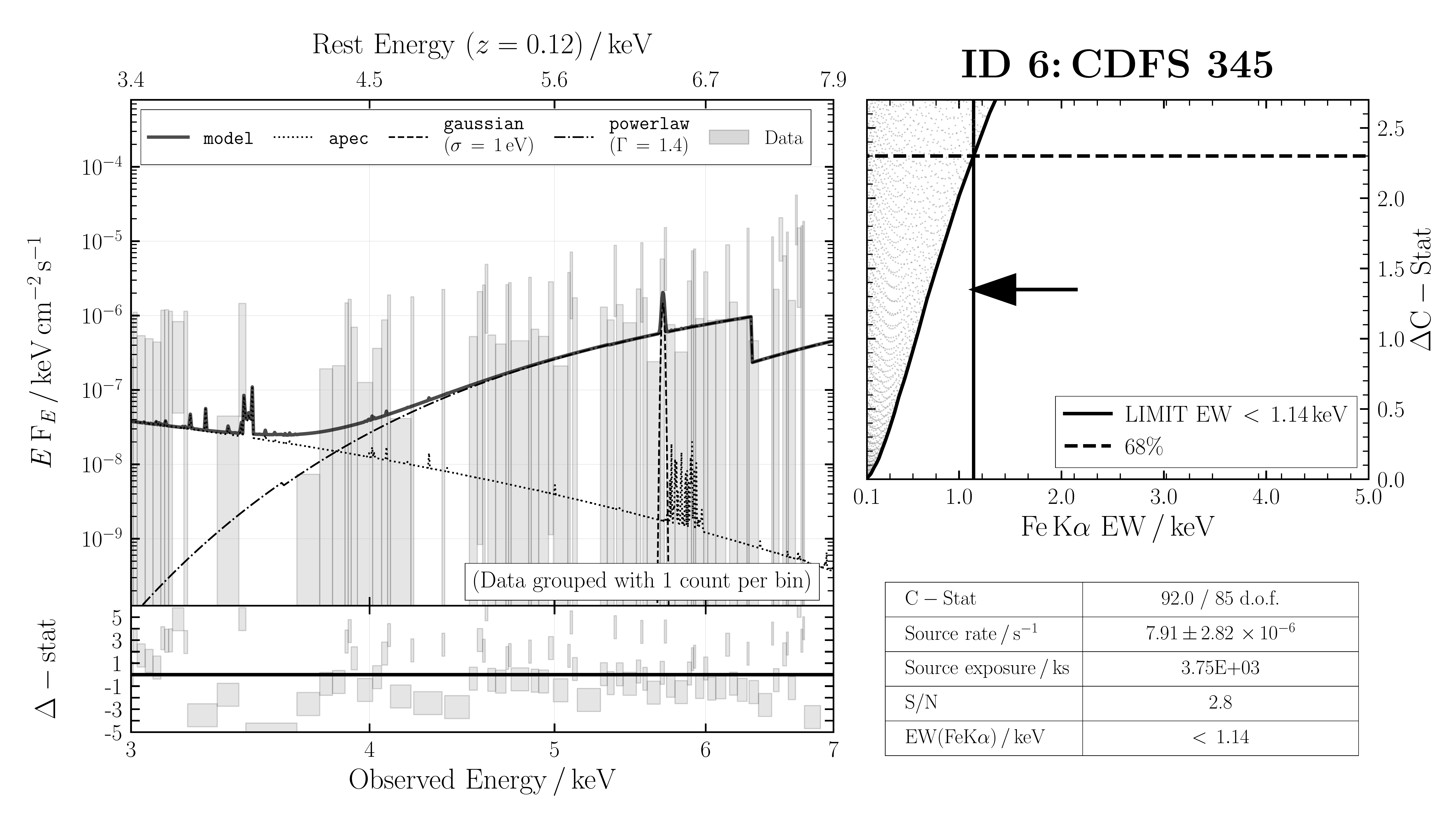}\\
\caption{\label{fig:CDFS345} ID 6: CDFS 345}
\end{center}
\end{figure}

\begin{figure}
\begin{center}
\includegraphics[angle=0,width=\columnwidth]{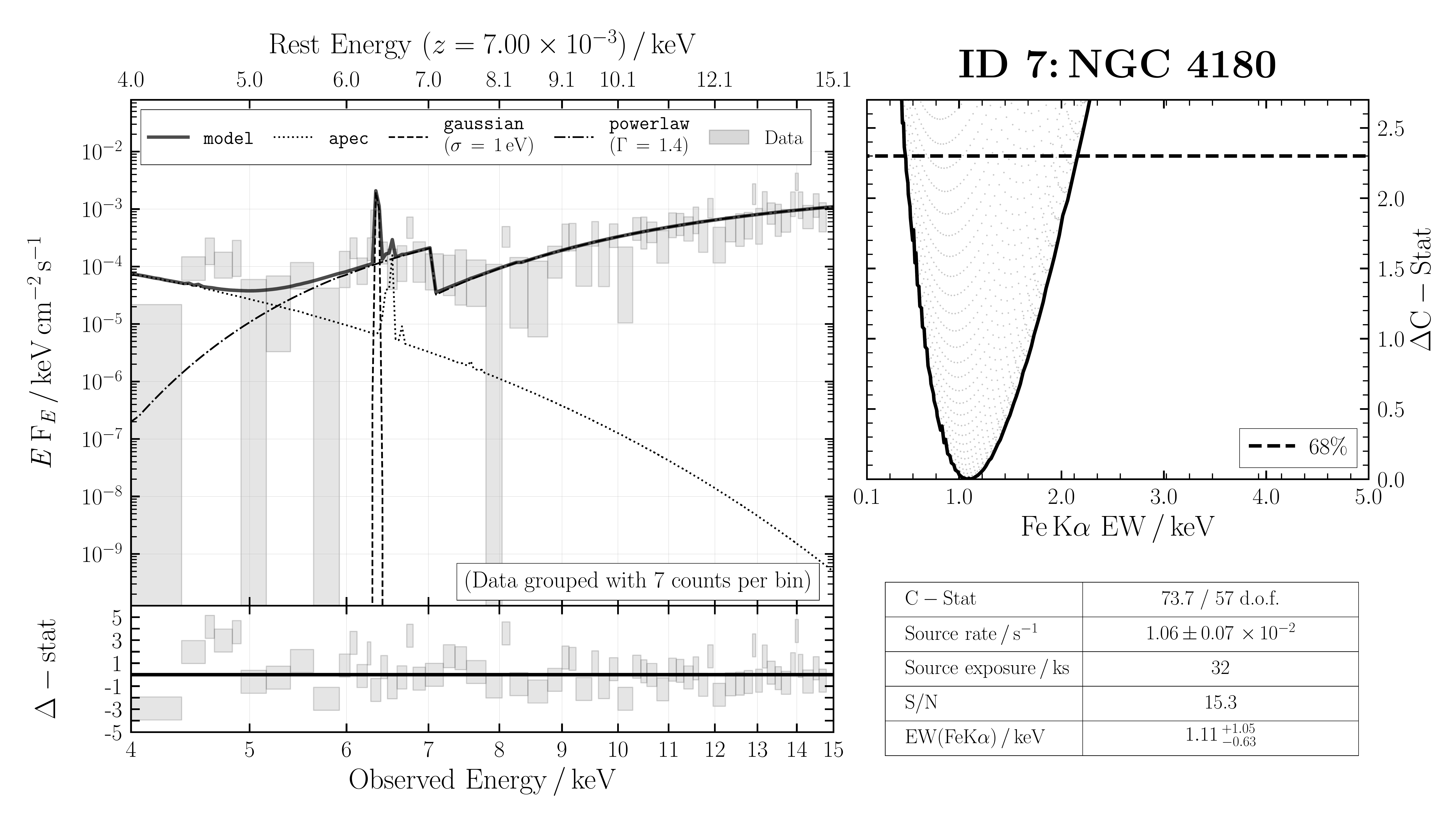}\\
\caption{\label{fig:NGC4180} ID 7: NGC 4180}
\end{center}
\end{figure}

\begin{figure}
\begin{center}
\includegraphics[angle=0,width=\columnwidth]{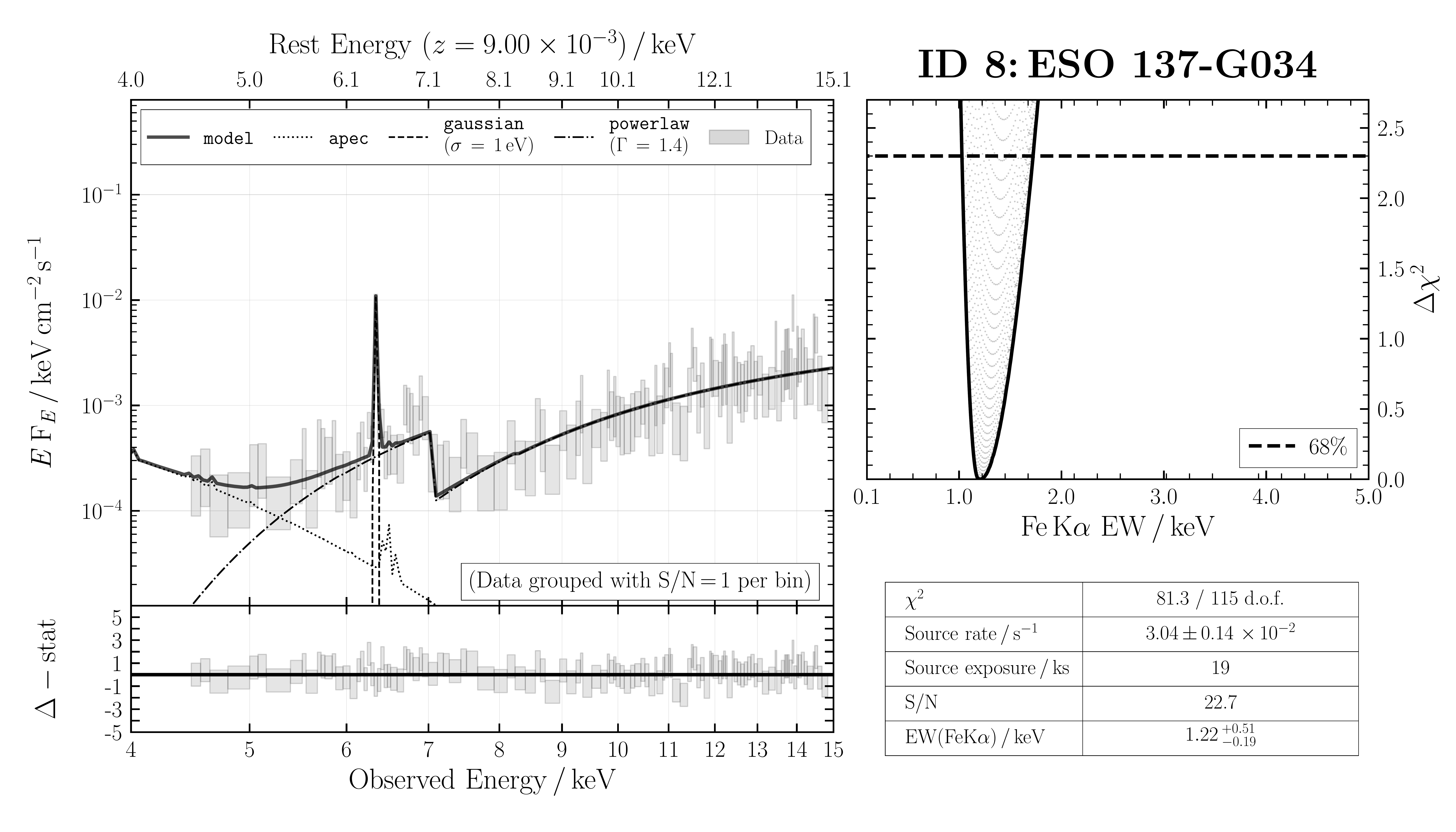}\\
\caption{\label{fig:ESO137-G034} ID 8: ESO 137-G034 }
\end{center}
\end{figure}

\begin{figure}
\begin{center}
\includegraphics[angle=0,width=\columnwidth]{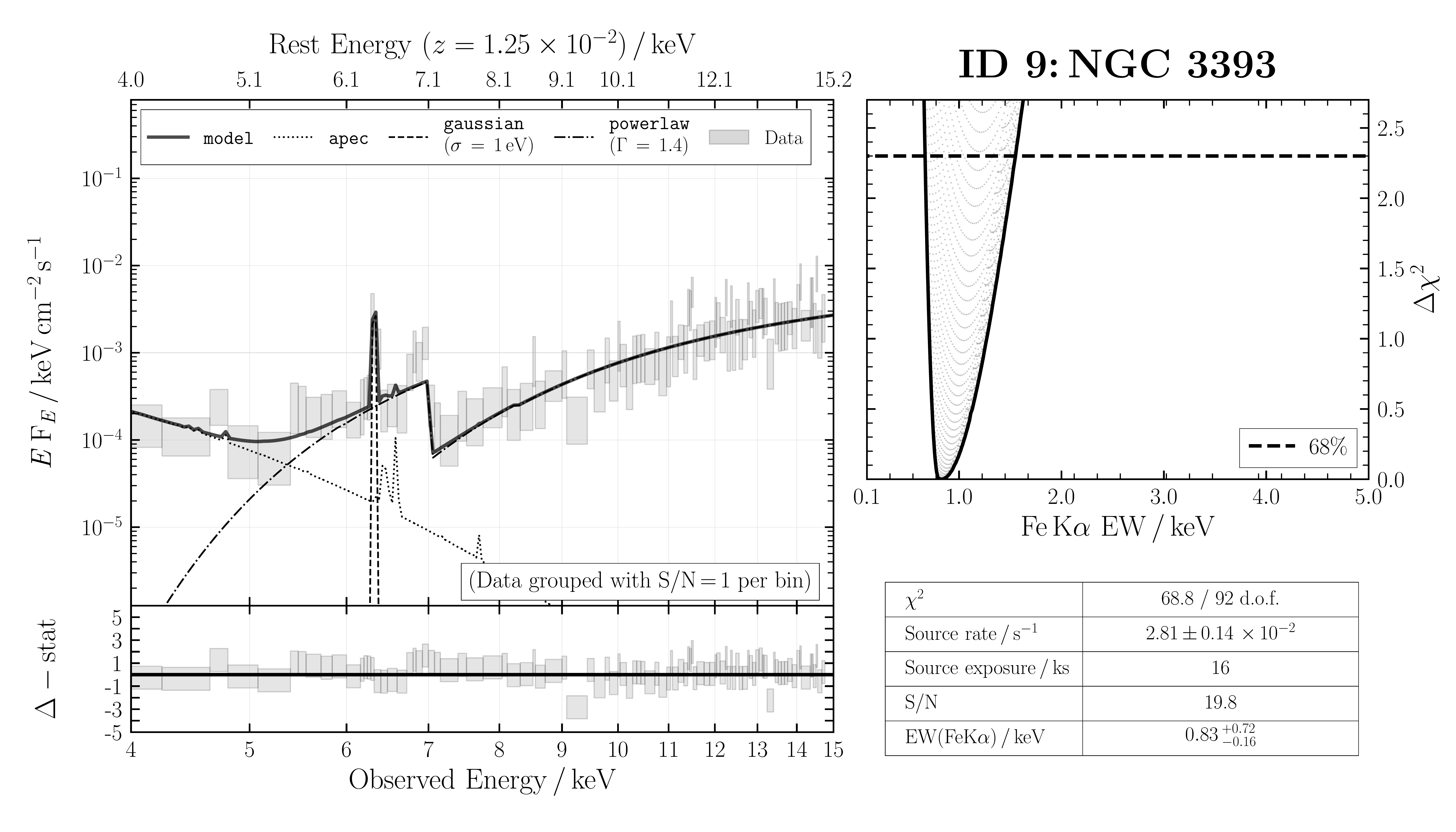}\\
\caption{\label{fig:NGC3393} ID 9: NGC 3393}
\end{center}
\end{figure}

\begin{figure}
\begin{center}
\includegraphics[angle=0,width=\columnwidth]{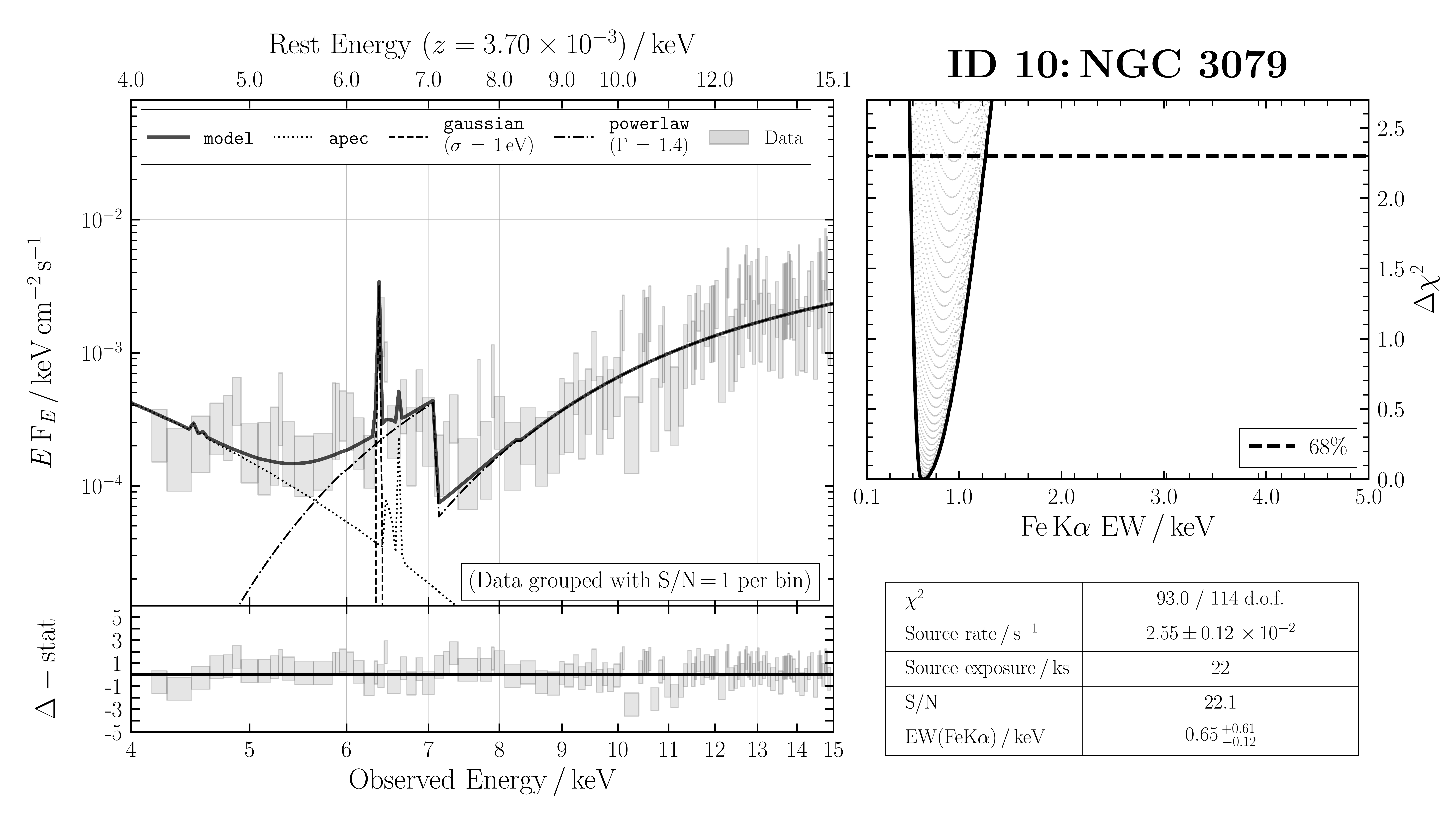}\\
\caption{\label{fig:NGC3079} ID 10: NGC 3079}
\end{center}
\end{figure}

\begin{figure}
\begin{center}
\includegraphics[angle=0,width=\columnwidth]{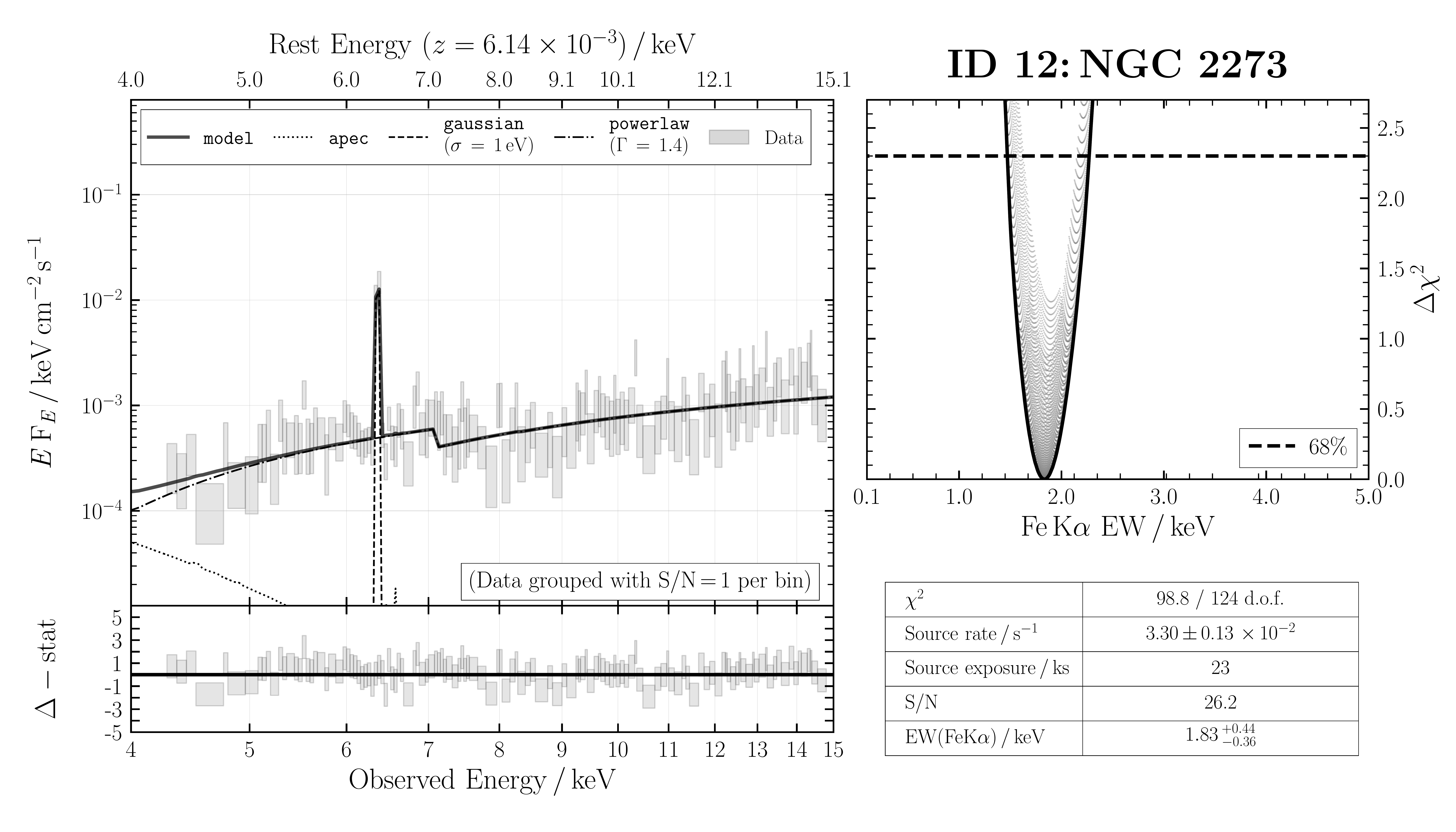}\\
\caption{\label{fig:NGC2273} ID 12: NGC 2273}
\end{center}
\end{figure}

\begin{figure}
\begin{center}
\includegraphics[angle=0,width=\columnwidth]{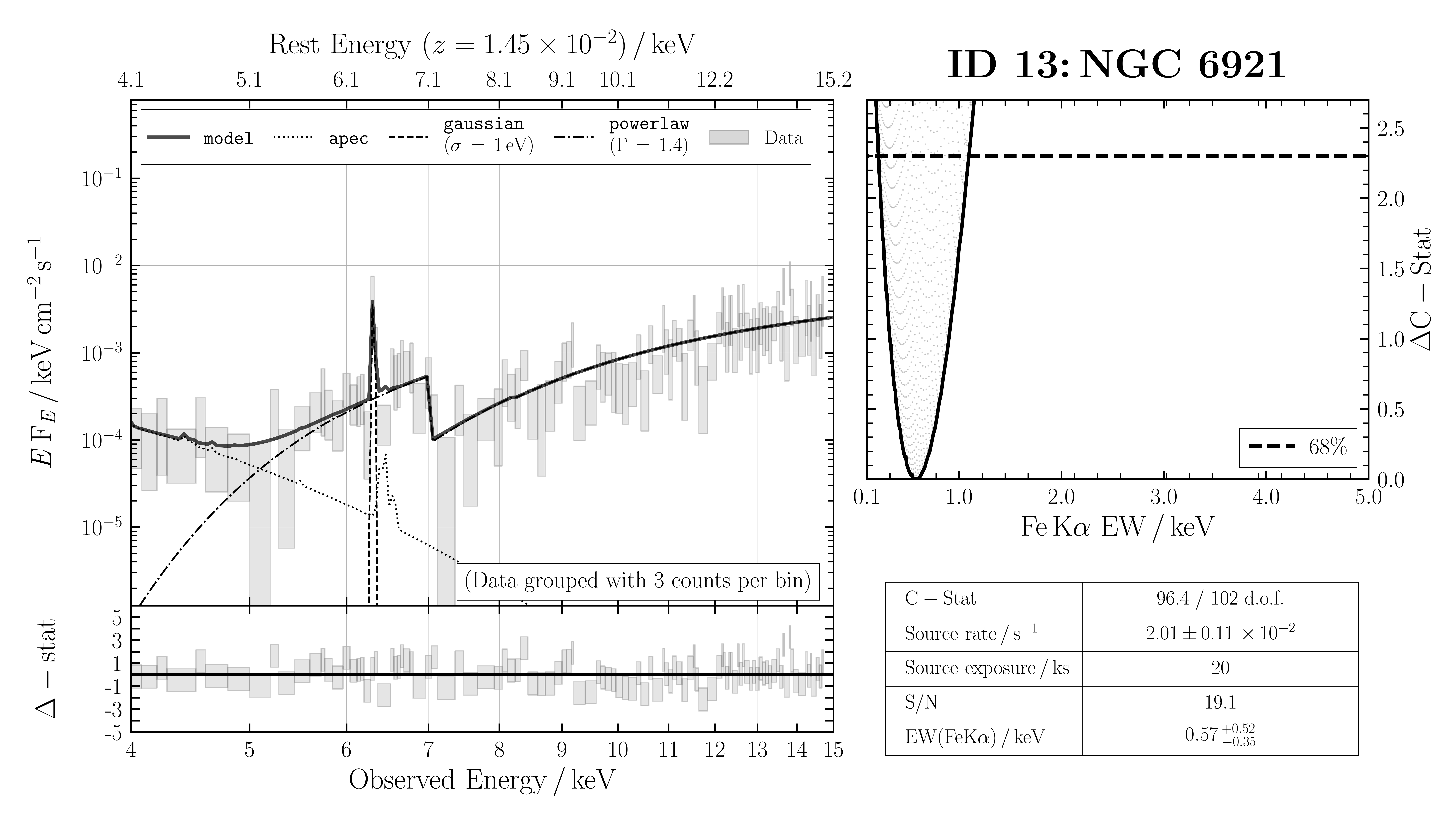}\\
\caption{\label{fig:NGC6921} ID 13: NGC 6921}
\end{center}
\end{figure}

\begin{figure}
\begin{center}
\includegraphics[angle=0,width=\columnwidth]{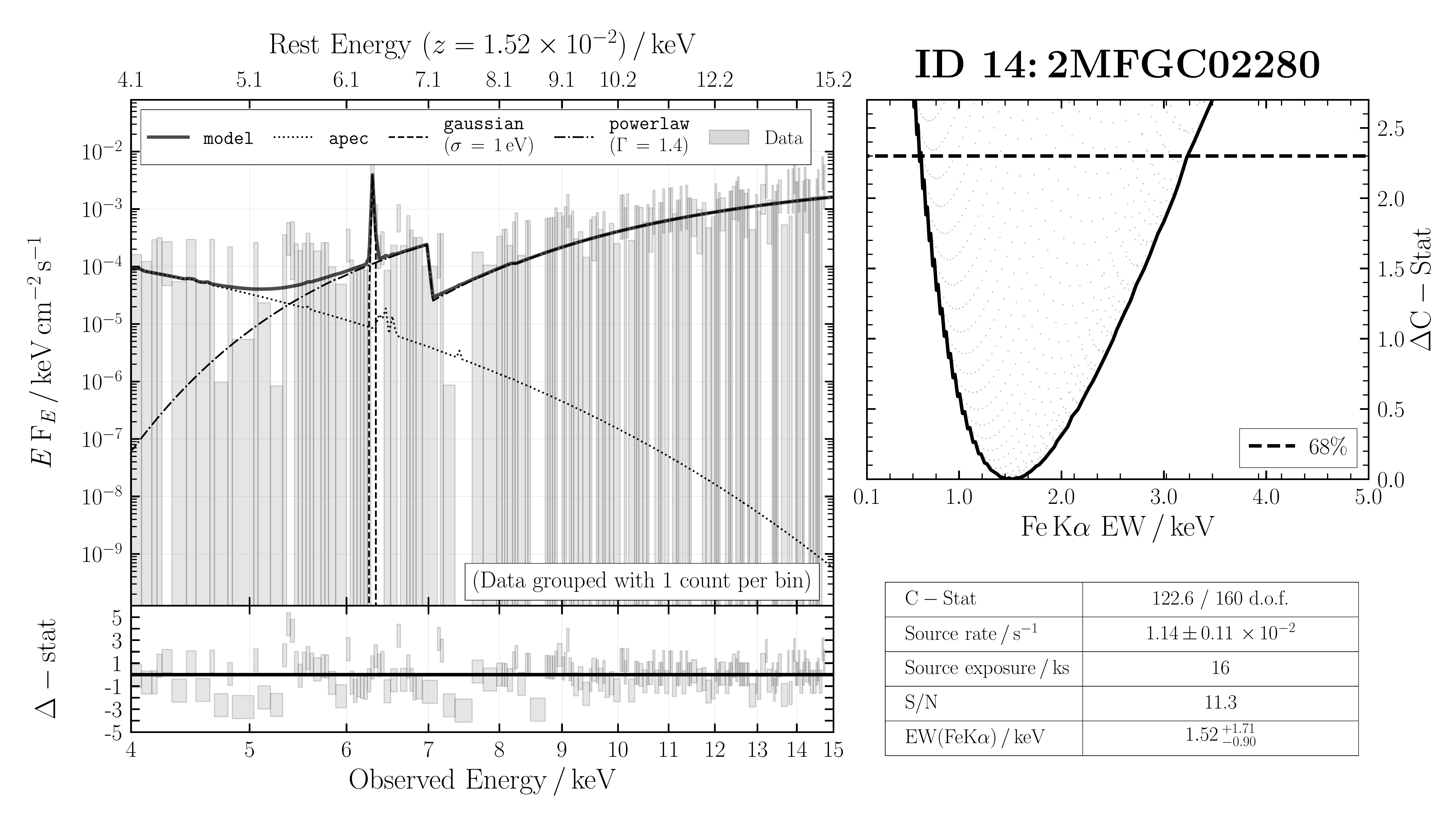}\\
\caption{\label{fig:2MFGC02280} ID 14: 2MFGC02280}
\end{center}
\end{figure}

\begin{figure}
\begin{center}
\includegraphics[angle=0,width=\columnwidth]{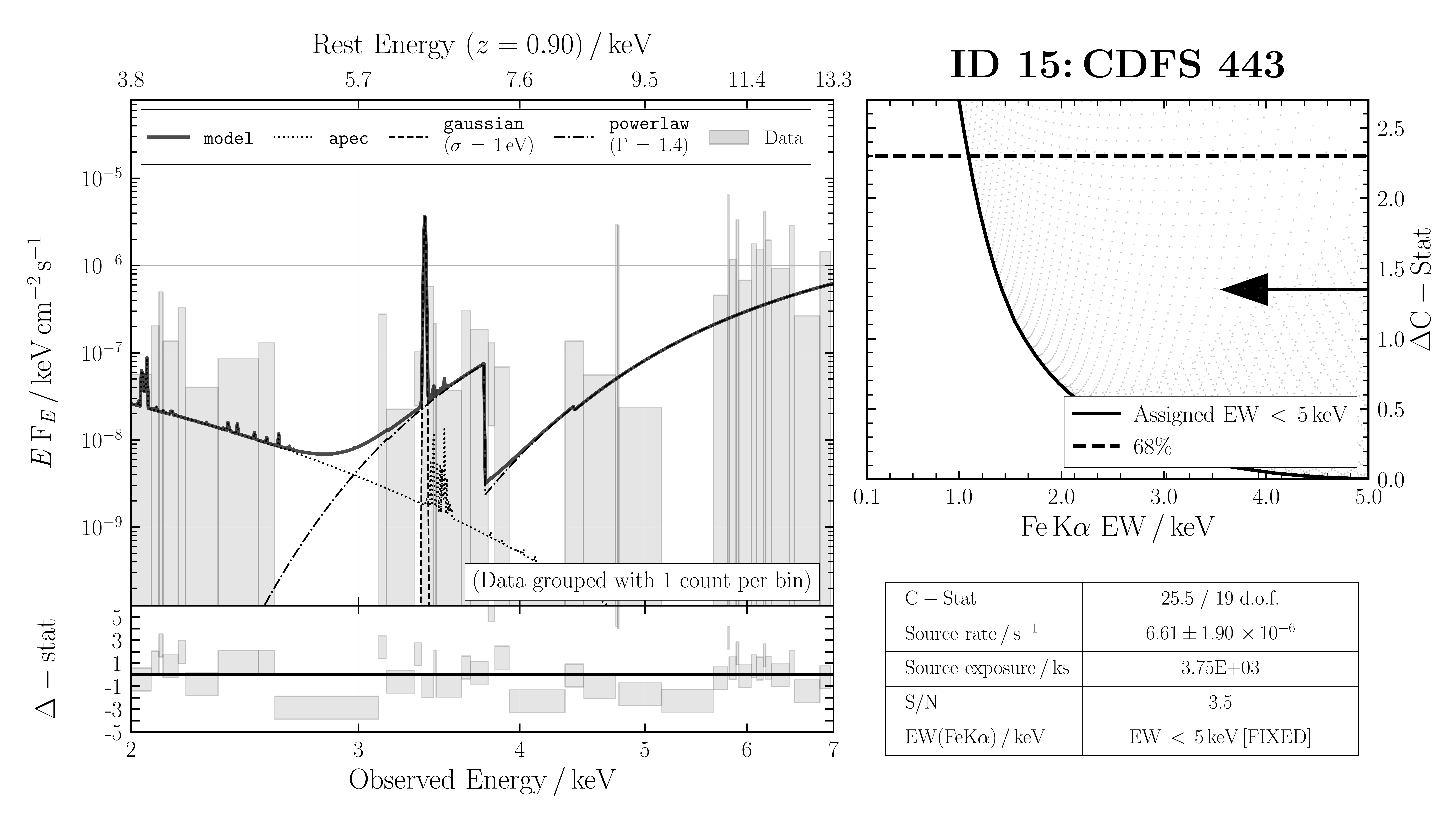}\\
\caption{\label{fig:CDFS443} ID 15: CDFS 443}
\end{center}
\end{figure}

\begin{figure}
\begin{center}
\includegraphics[angle=0,width=\columnwidth]{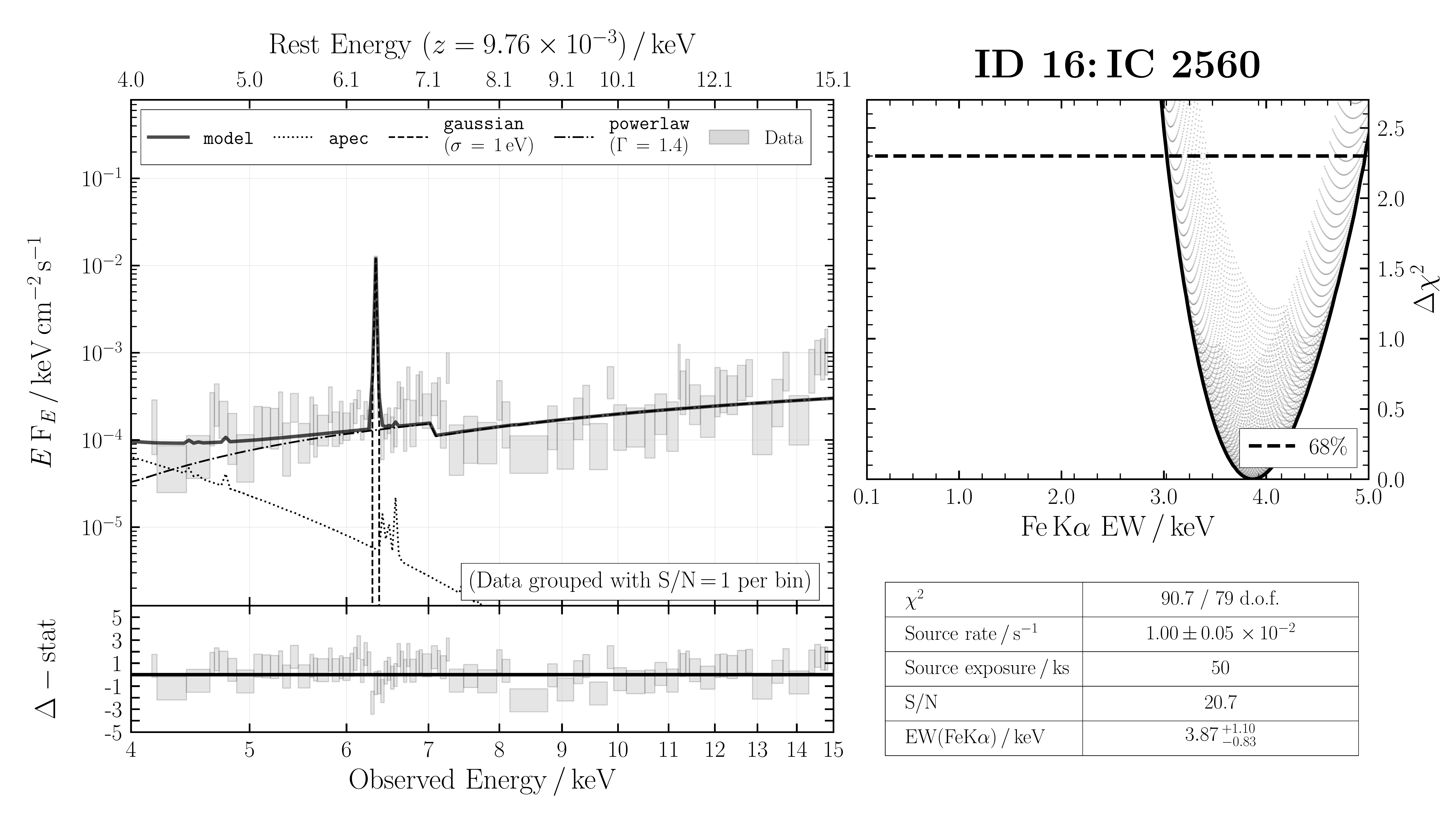}\\
\caption{\label{fig:IC2560} ID 16: IC 2560}
\end{center}
\end{figure}

\begin{figure}
\begin{center}
\includegraphics[angle=0,width=\columnwidth]{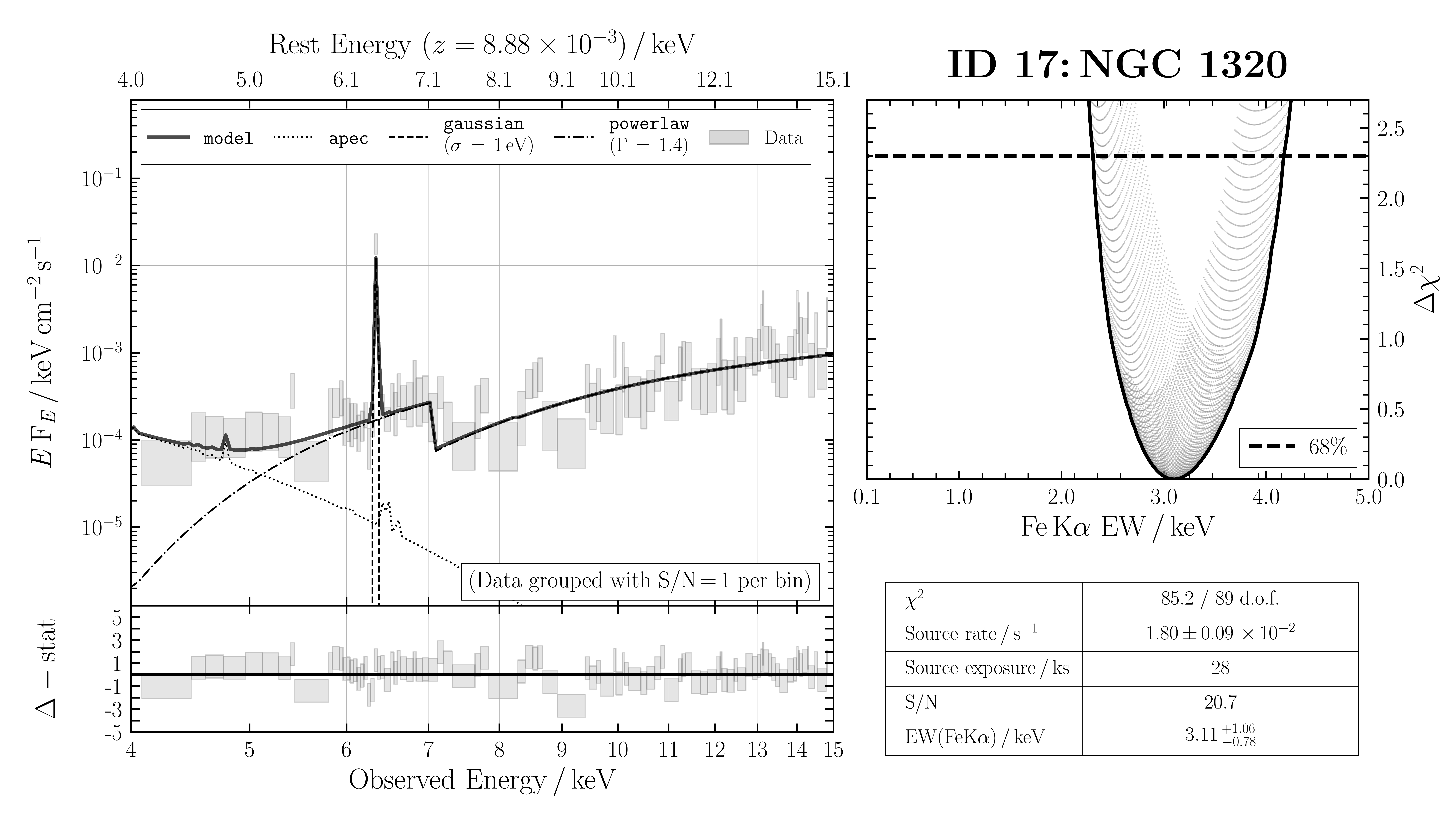}\\
\caption{\label{fig:NGC1320} ID 17: NGC 1320}
\end{center}
\end{figure}

\begin{figure}
\begin{center}
\includegraphics[angle=0,width=\columnwidth]{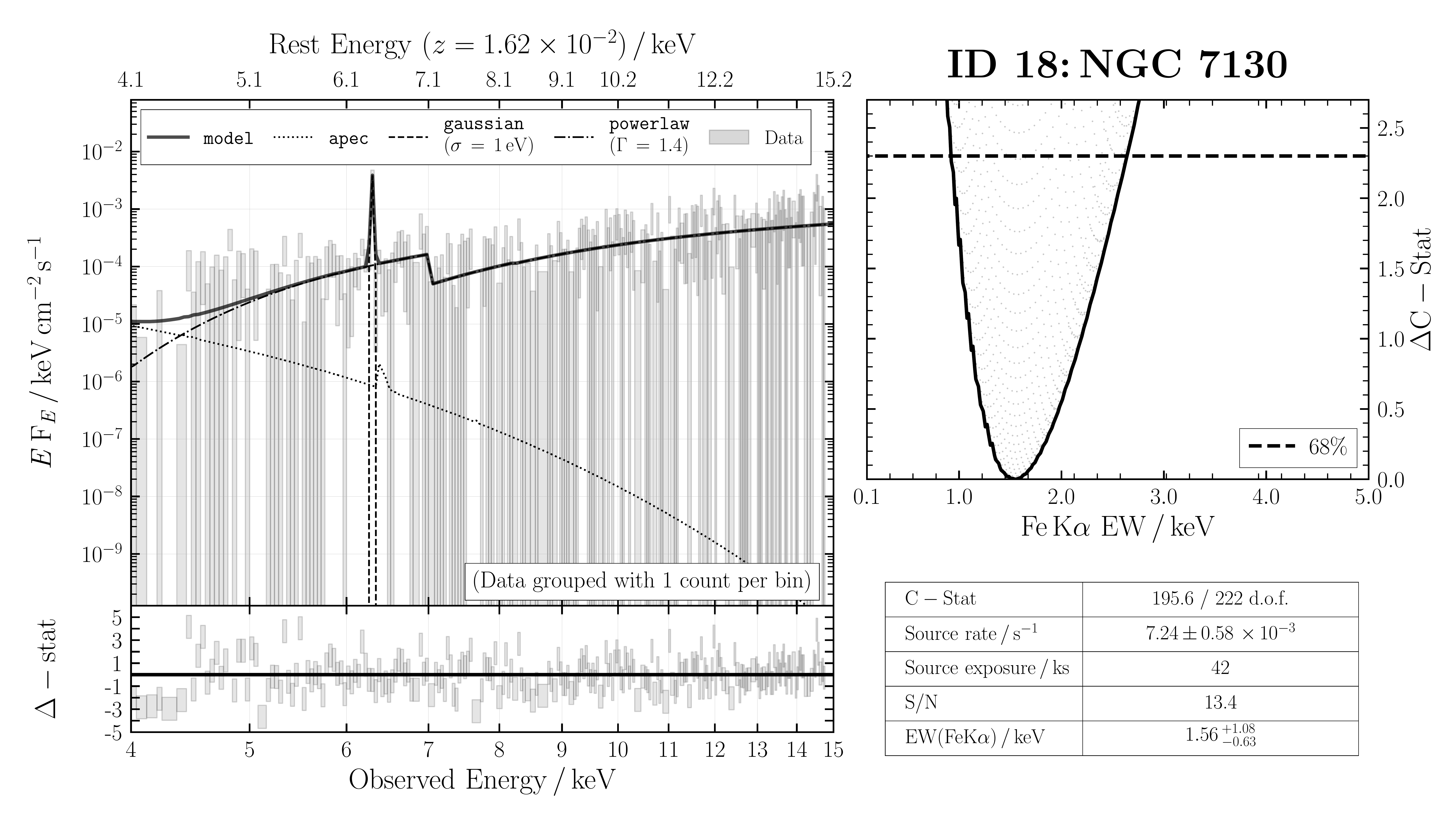}\\
\caption{\label{fig:NGC7130} ID 18: NGC 7130}
\end{center}
\end{figure}

\begin{figure}
\begin{center}
\includegraphics[angle=0,width=\columnwidth]{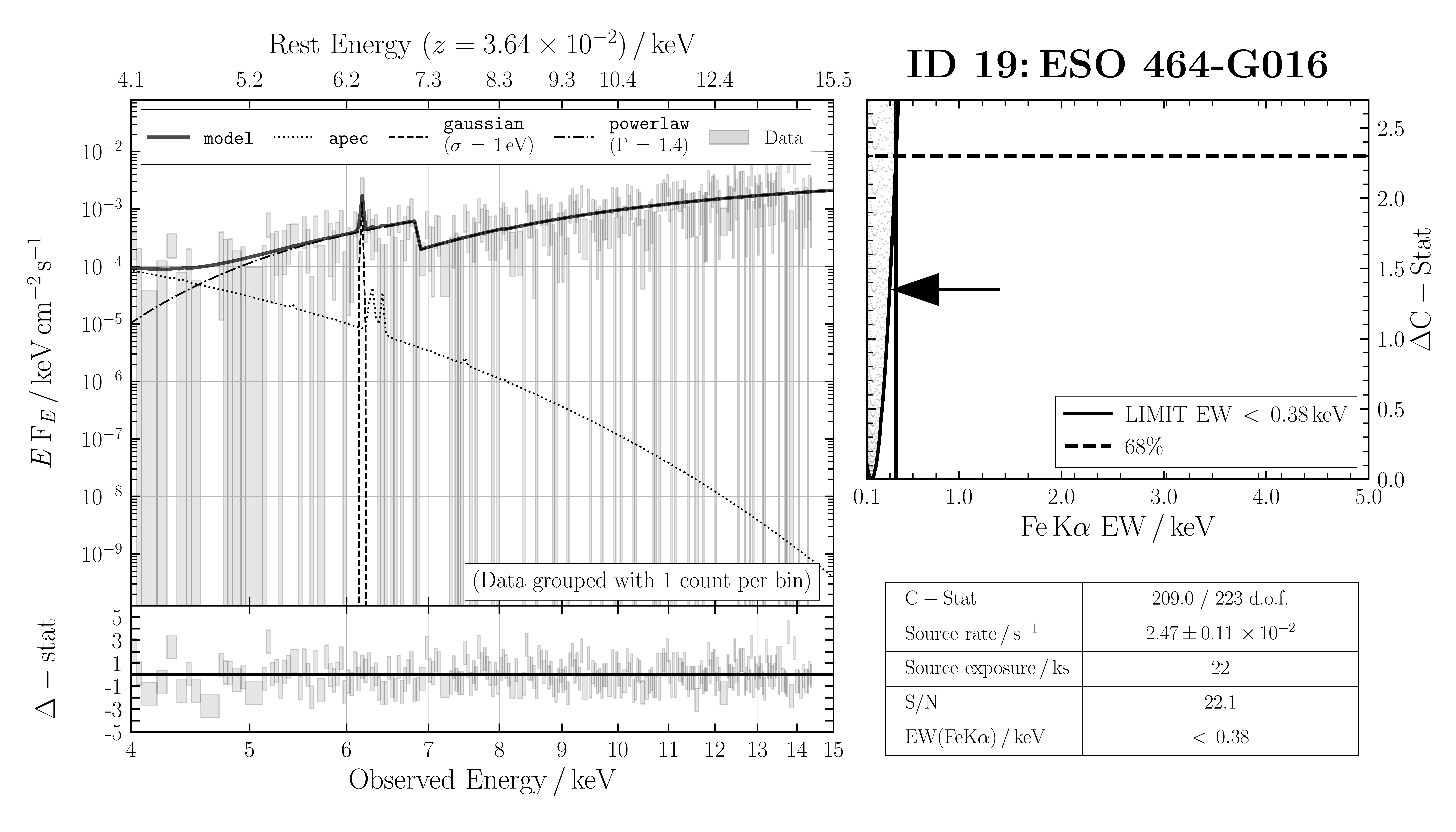}\\
\caption{\label{fig:ESO464-G016} ID 19: ESO 464-G016}
\end{center}
\end{figure}

\begin{figure}
\begin{center}
\includegraphics[angle=0,width=\columnwidth]{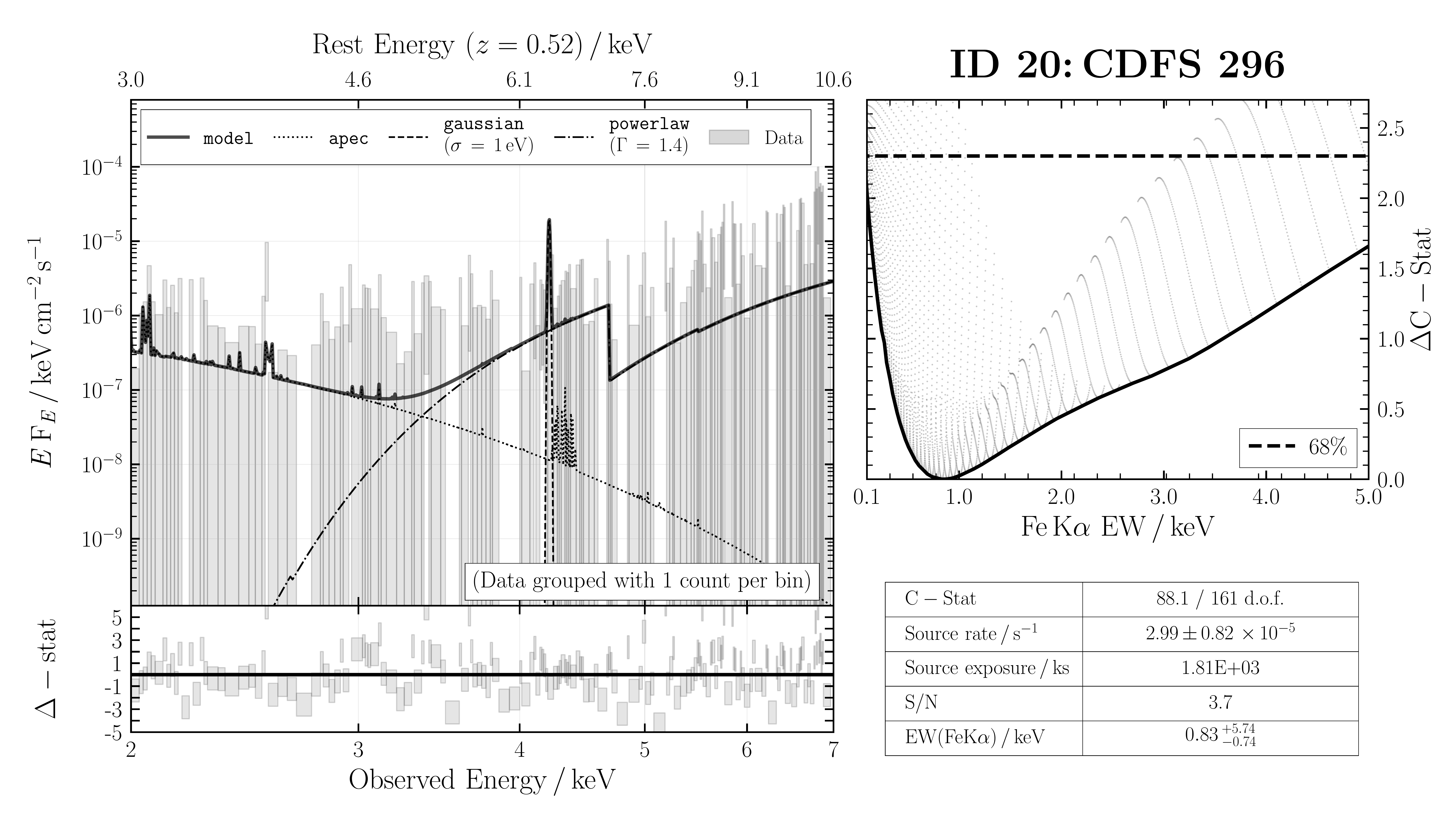}\\
\caption{\label{fig:CDFS296} ID 20: CDFS 296}
\end{center}
\end{figure}

\begin{figure}
\begin{center}
\includegraphics[angle=0,width=\columnwidth]{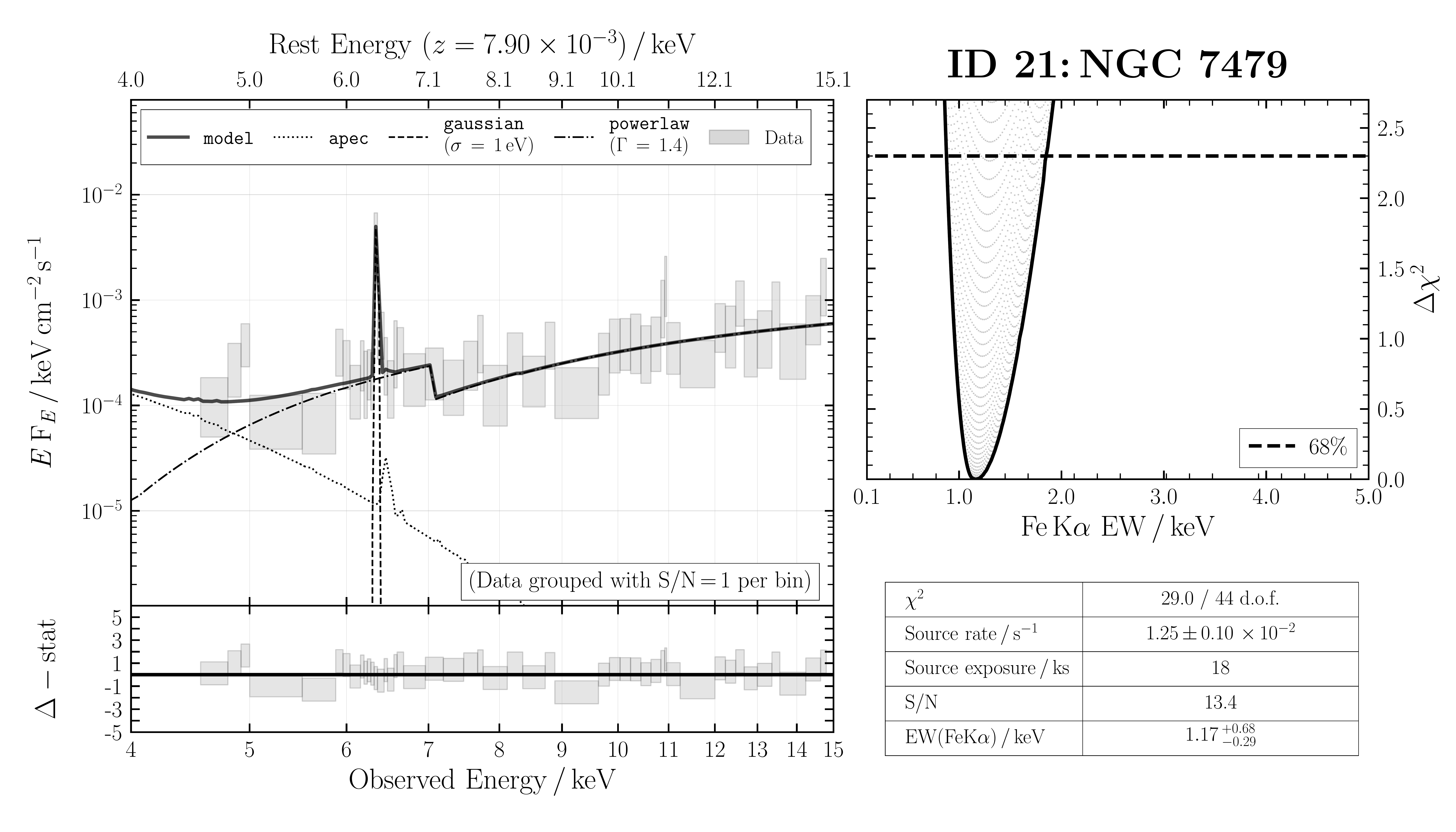}\\
\caption{\label{fig:NGC7479} ID 21: NGC 7479}
\end{center}
\end{figure}

\begin{figure}
\begin{center}
\includegraphics[angle=0,width=\columnwidth]{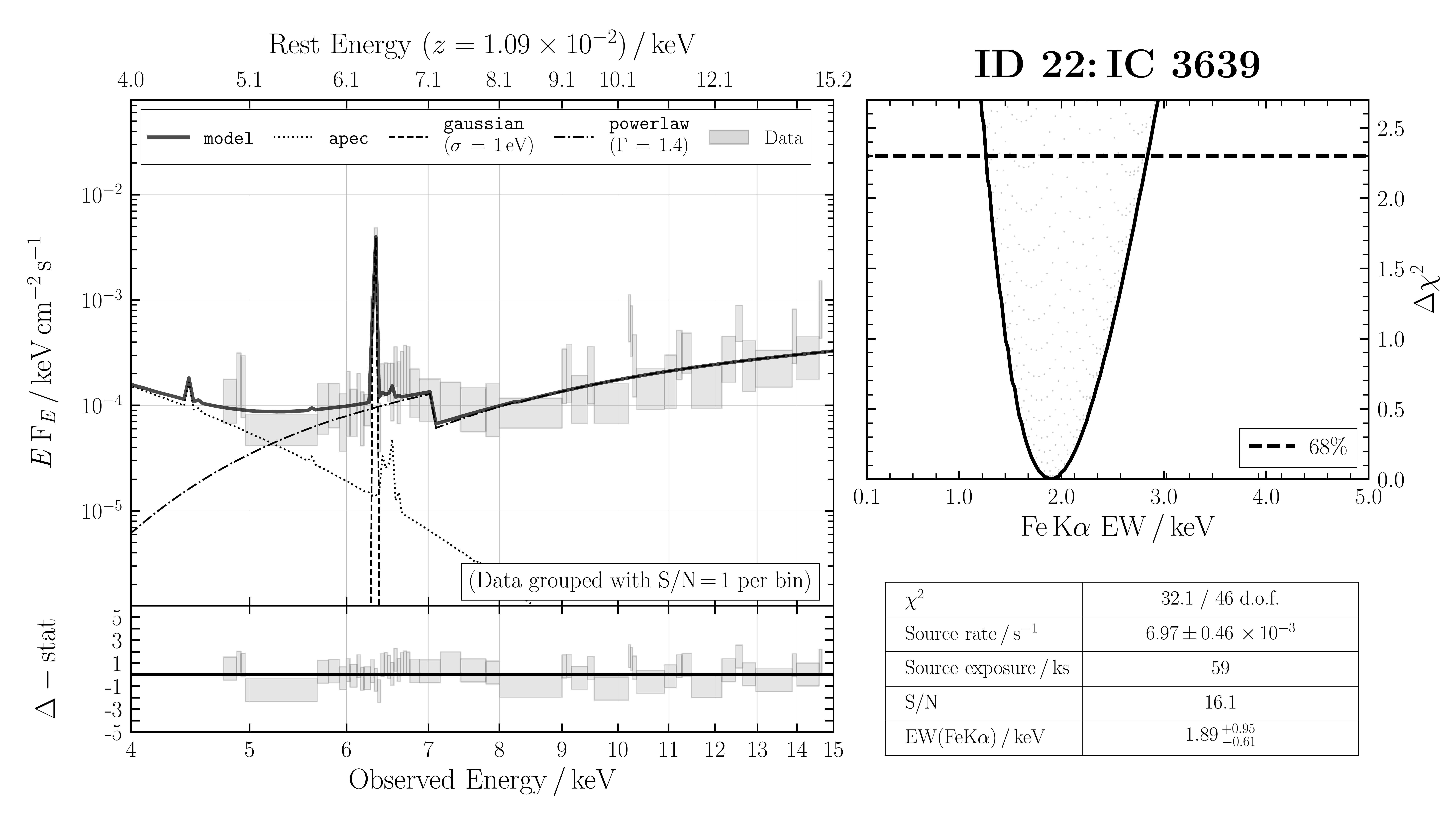}\\
\caption{\label{fig:IC3639} ID 22: IC 3639}
\end{center}
\end{figure}

\begin{figure}
\begin{center}
\includegraphics[angle=0,width=\columnwidth]{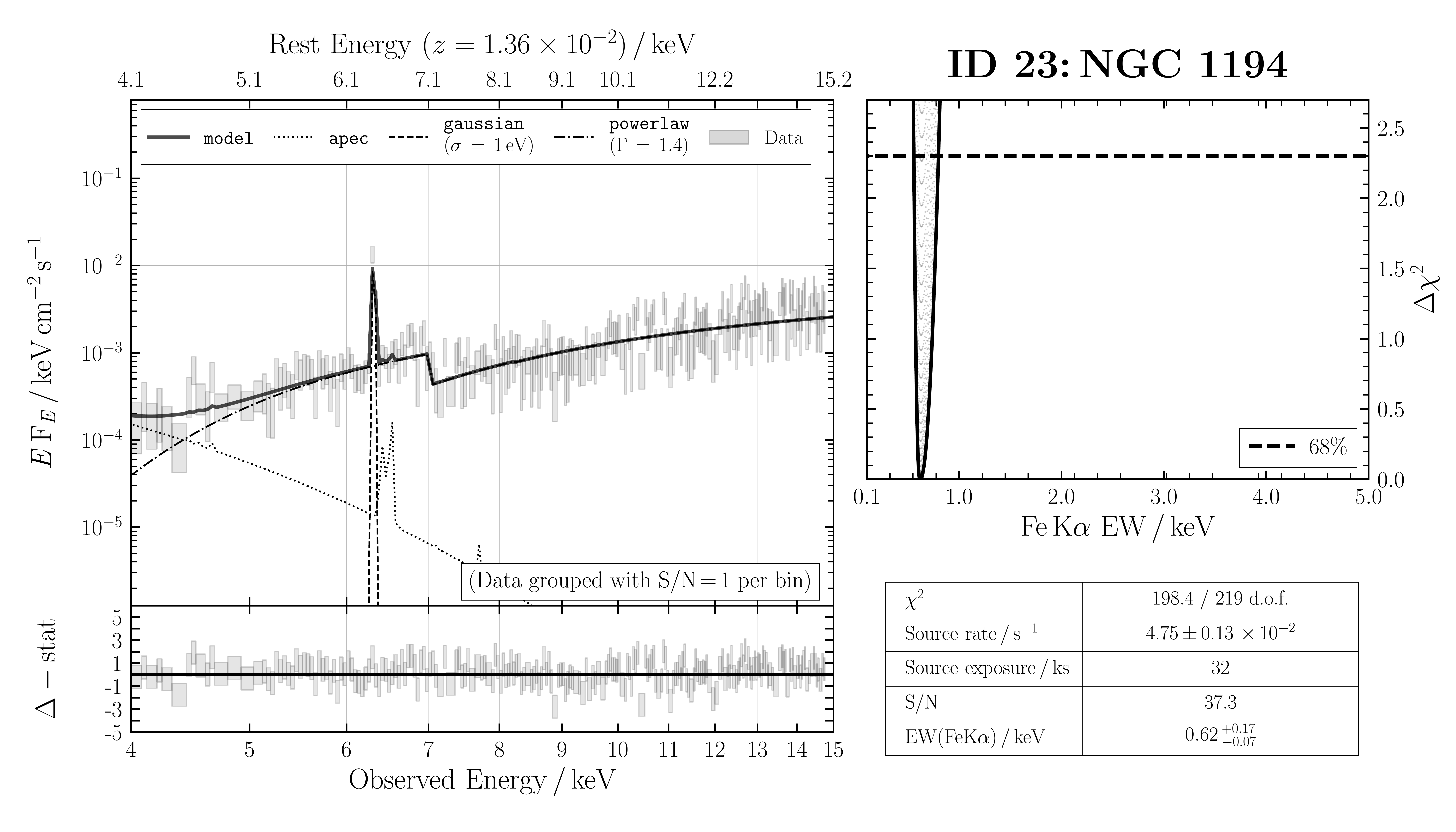}\\
\caption{\label{fig:NGC1194} ID 23: NGC 1194}
\end{center}
\end{figure}

\begin{figure}
\begin{center}
\includegraphics[angle=0,width=\columnwidth]{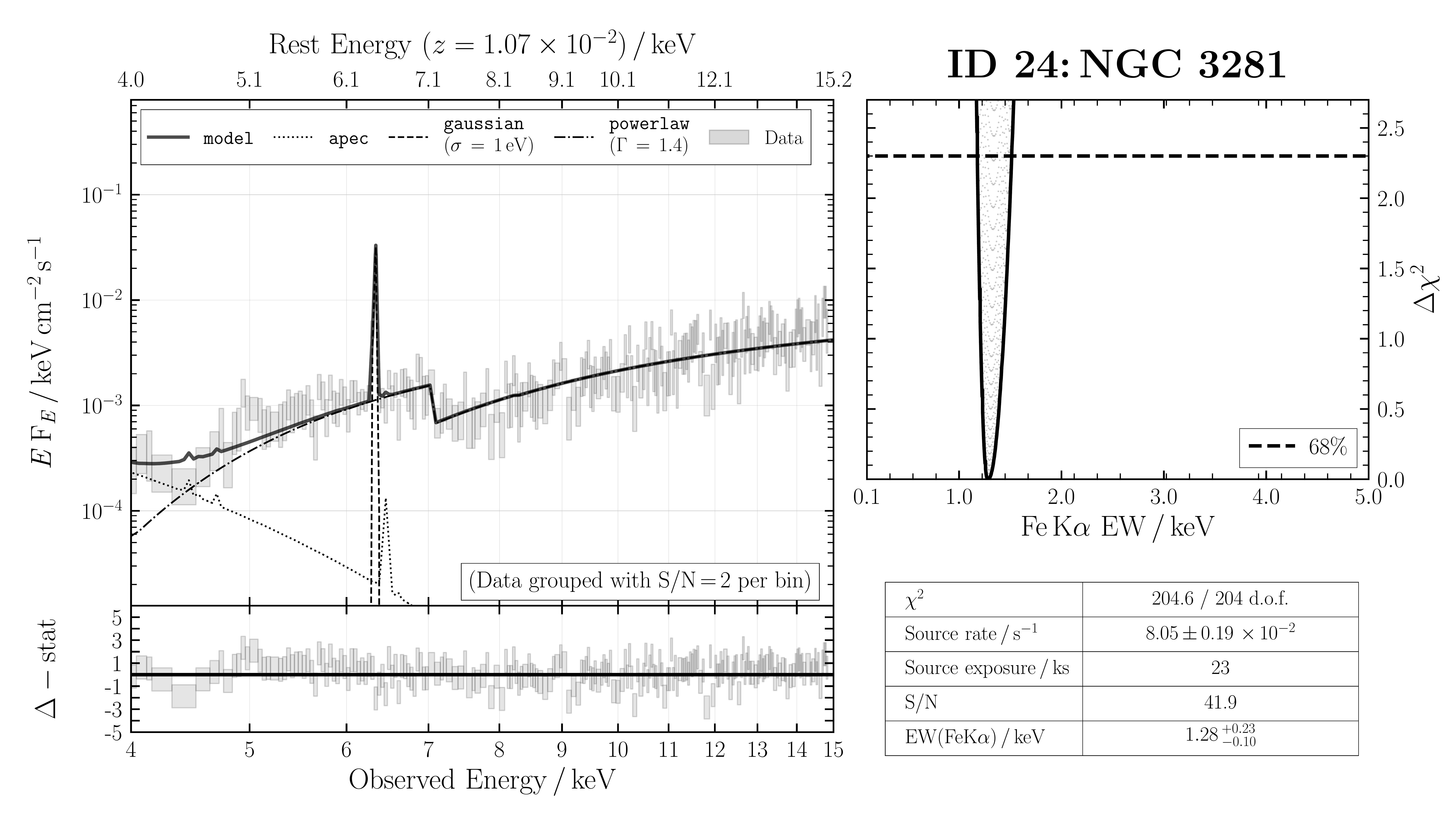}\\
\caption{\label{fig:NGC3281} ID 24: NGC 3281}
\end{center}
\end{figure}

\begin{figure}
\begin{center}
\includegraphics[angle=0,width=\columnwidth]{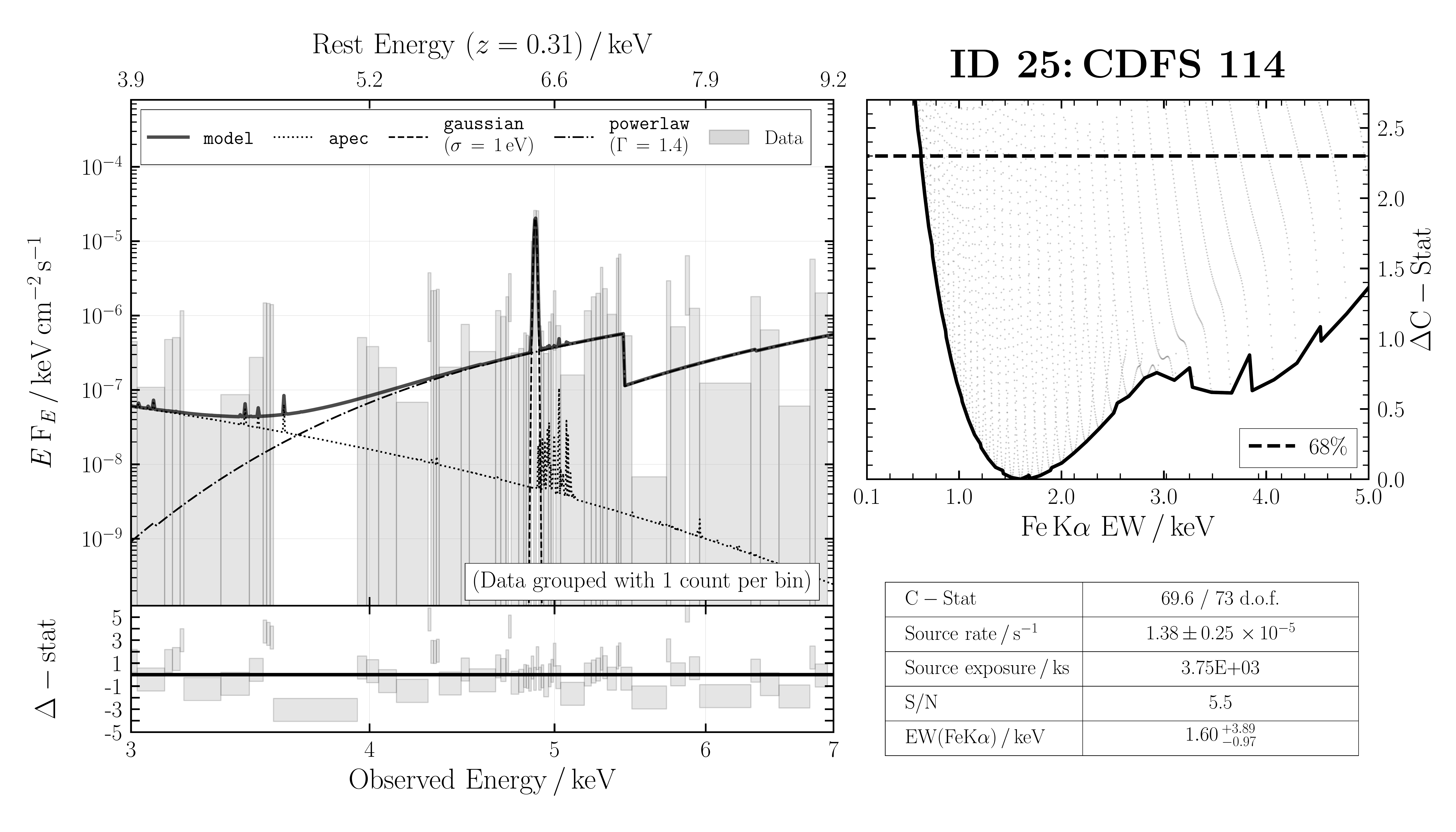}\\
\caption{\label{fig:CDFS114} ID 25: CDFS 114}
\end{center}
\end{figure}

\begin{figure}
\begin{center}
\includegraphics[angle=0,width=\columnwidth]{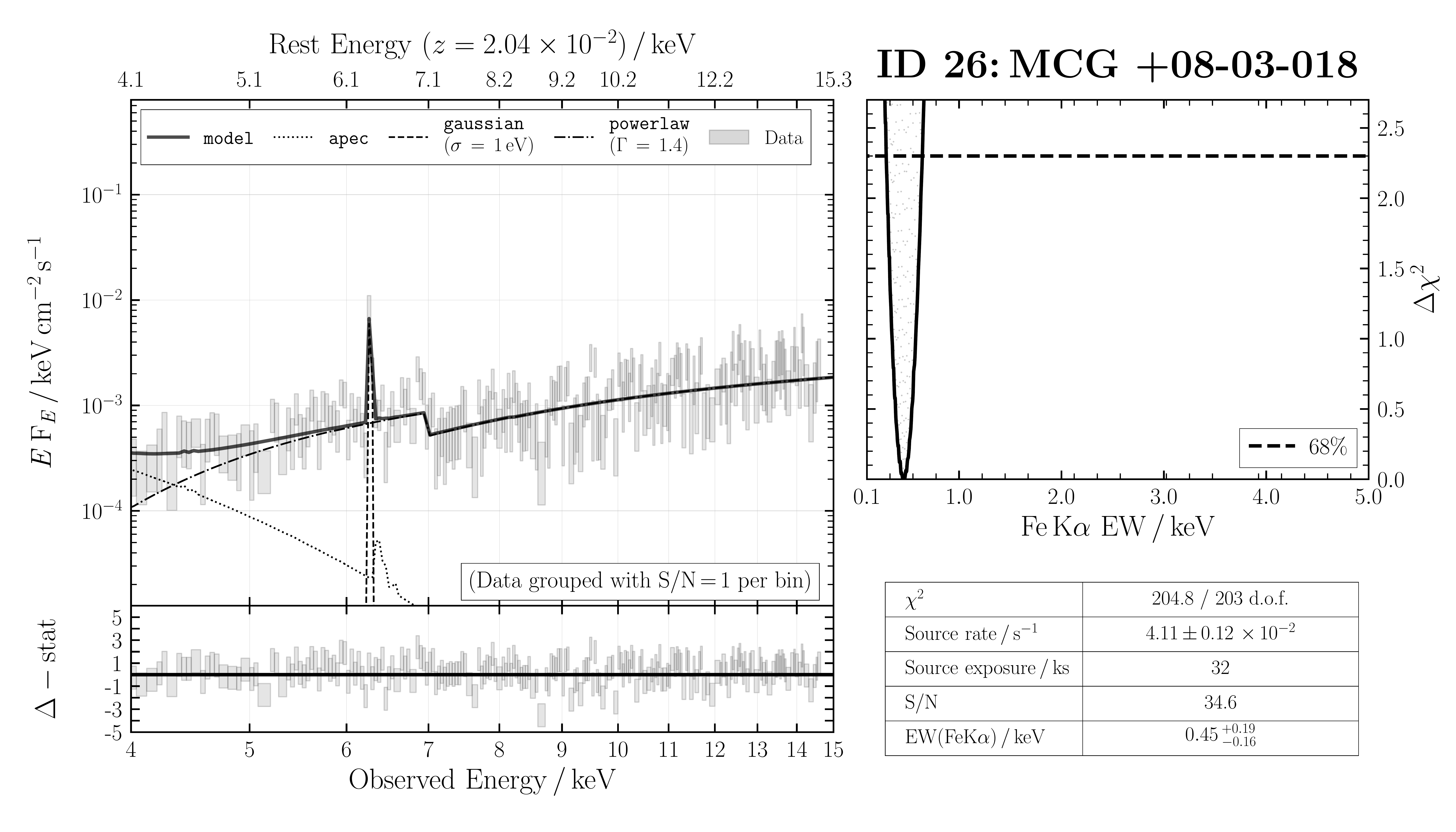}\\
\caption{\label{fig:MCG+08-03-018} ID 26: MCG +08-03-018}
\end{center}
\end{figure}

\begin{figure}
\begin{center}
\includegraphics[angle=0,width=\columnwidth]{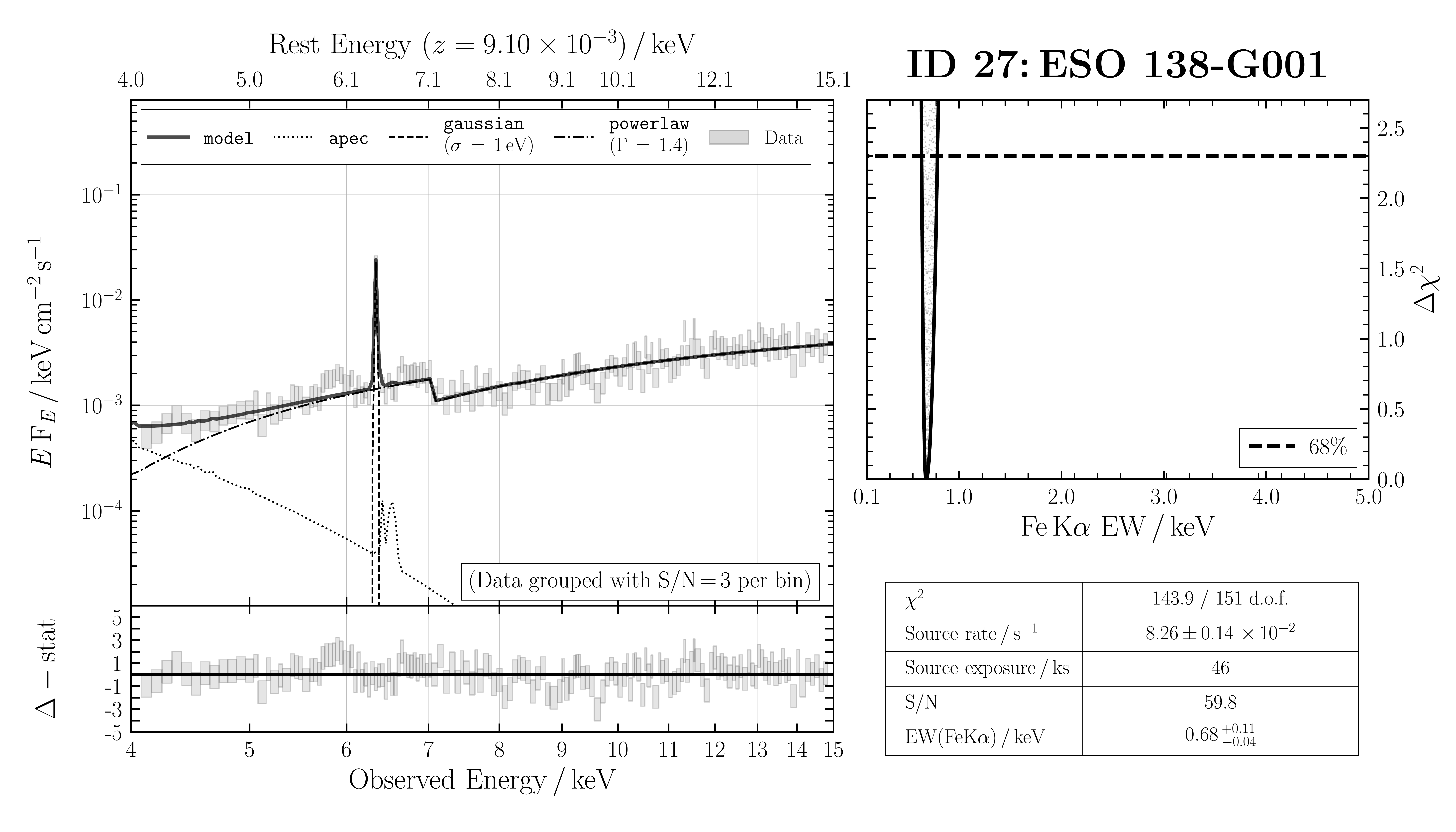}\\
\caption{\label{fig:ESO138-G001} ID 27: ESO 138-G001}
\end{center}
\end{figure}

\begin{figure}
\begin{center}
\includegraphics[angle=0,width=\columnwidth]{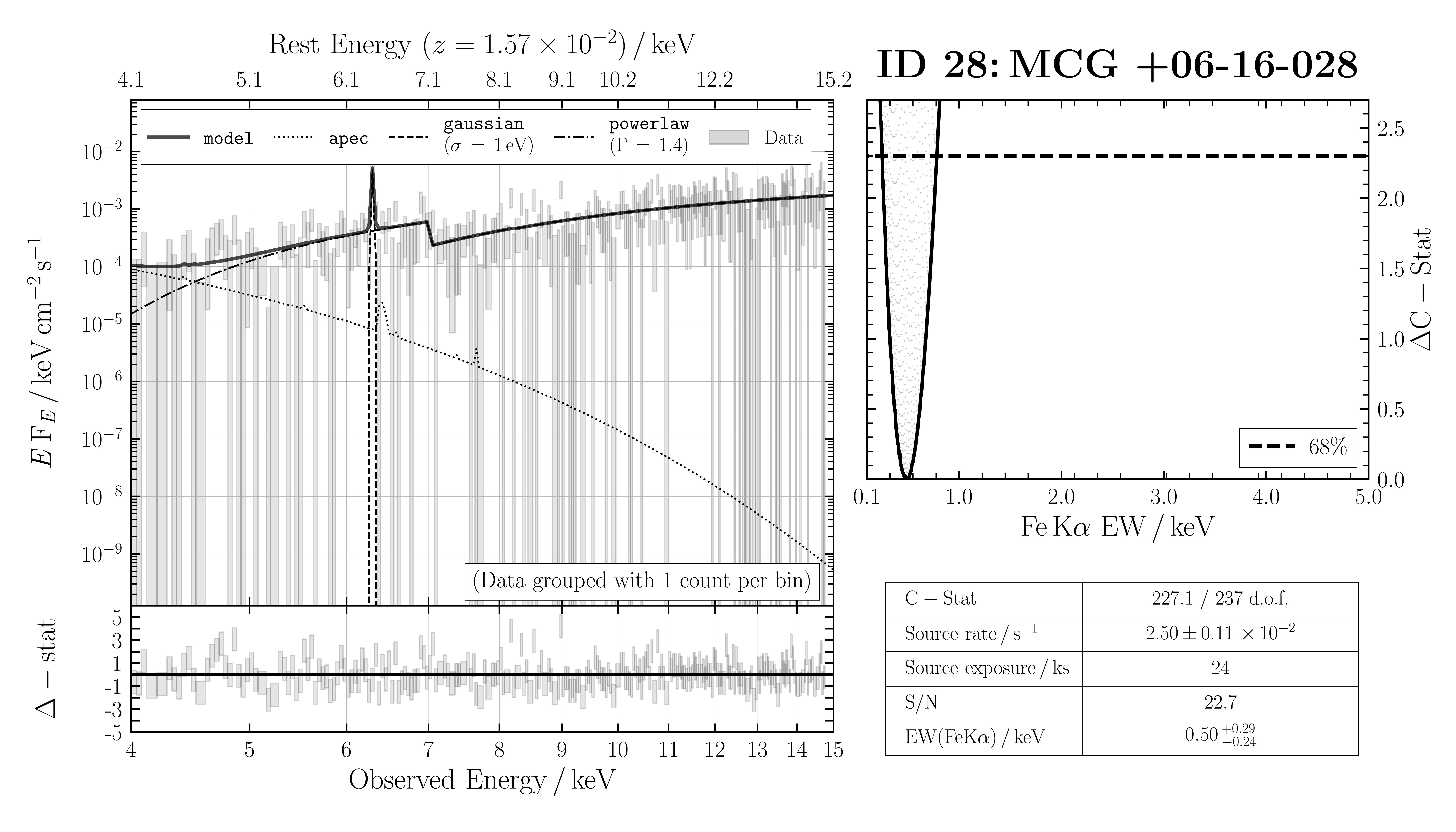}\\
\caption{\label{fig:MCG+06-16-028} ID 28: MCG +06-16-028}
\end{center}
\end{figure}

\begin{figure}
\begin{center}
\includegraphics[angle=0,width=\columnwidth]{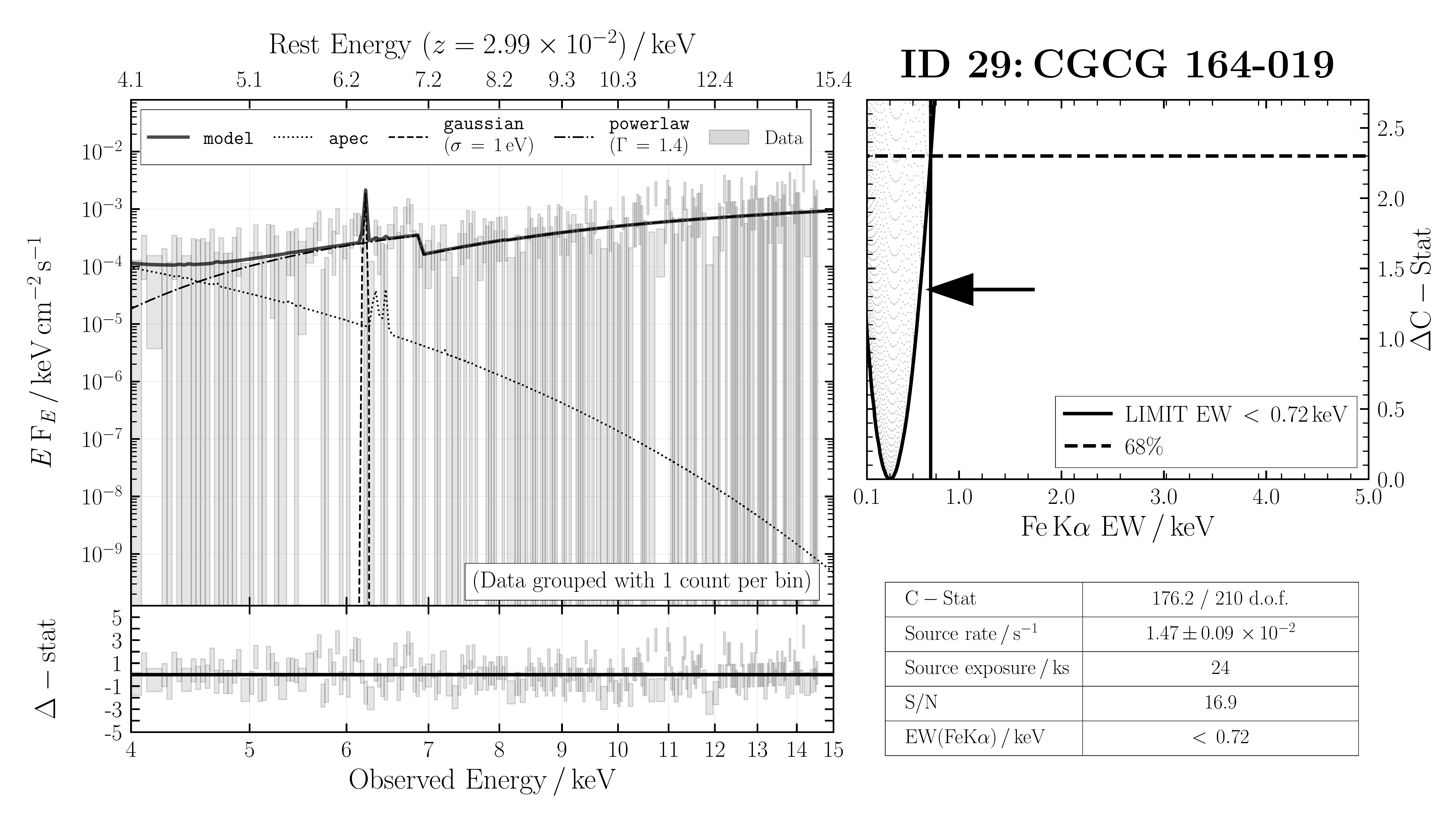}\\
\caption{\label{fig:CGCG164-019} ID 29: CGCG 164-019}
\end{center}
\end{figure}

\begin{figure}
\begin{center}
\includegraphics[angle=0,width=\columnwidth]{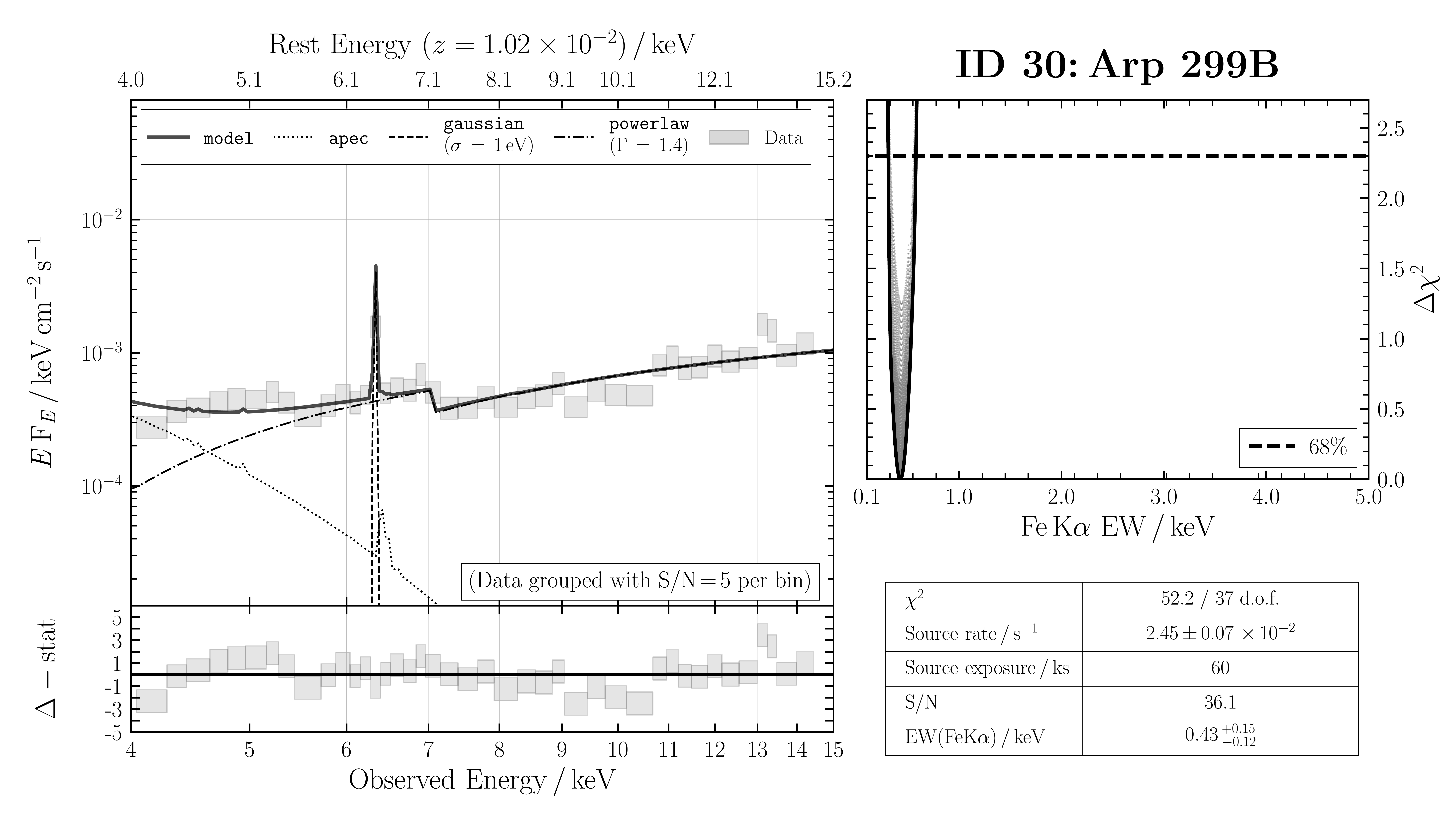}\\
\caption{\label{fig:Arp299B} ID 30: Arp 299B}
\end{center}
\end{figure}

\begin{figure}
\begin{center}
\includegraphics[angle=0,width=\columnwidth]{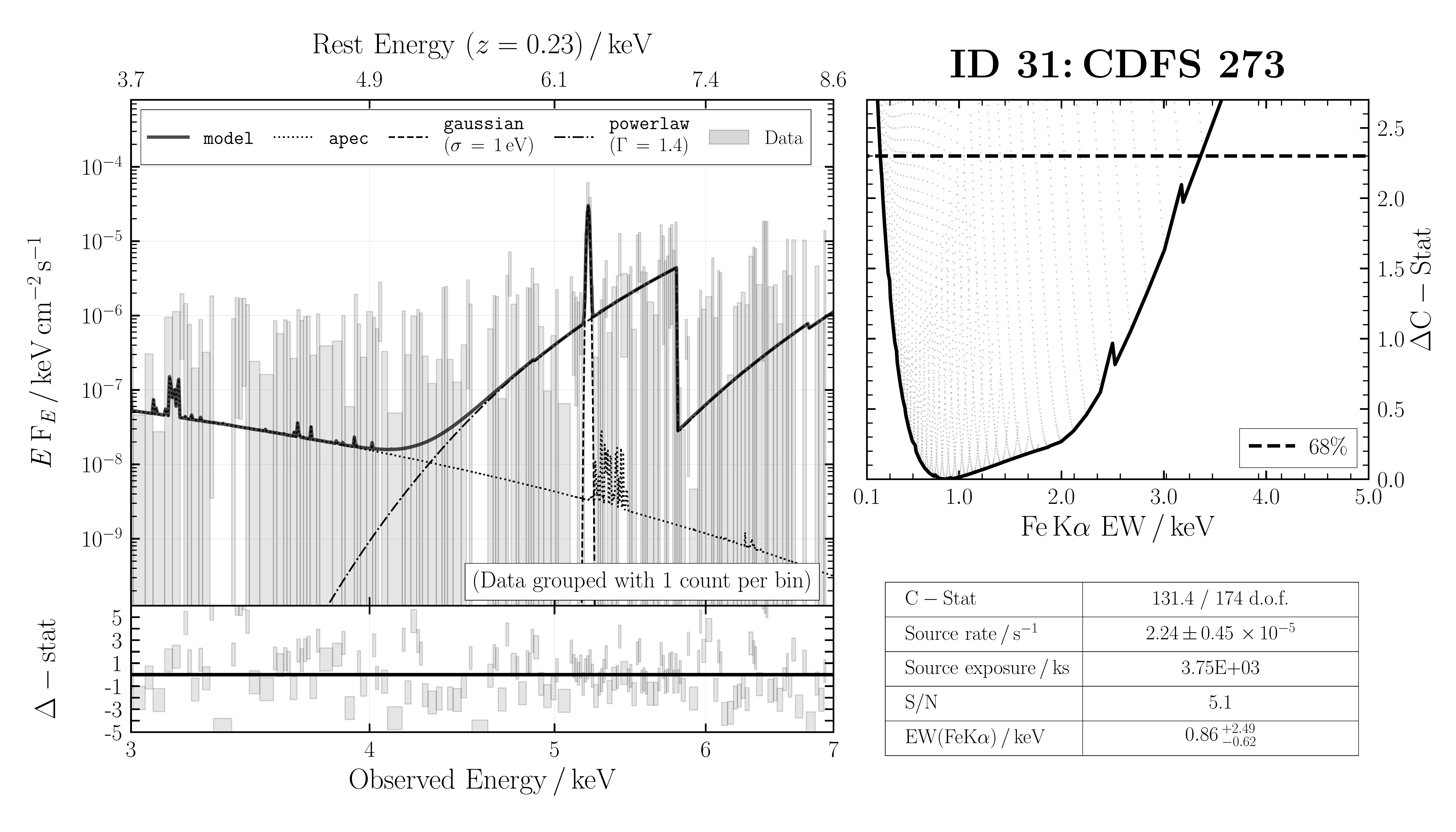}\\
\caption{\label{fig:CDFS273} ID 31: CDFS 273}
\end{center}
\end{figure}

\begin{figure}
\begin{center}
\includegraphics[angle=0,width=\columnwidth]{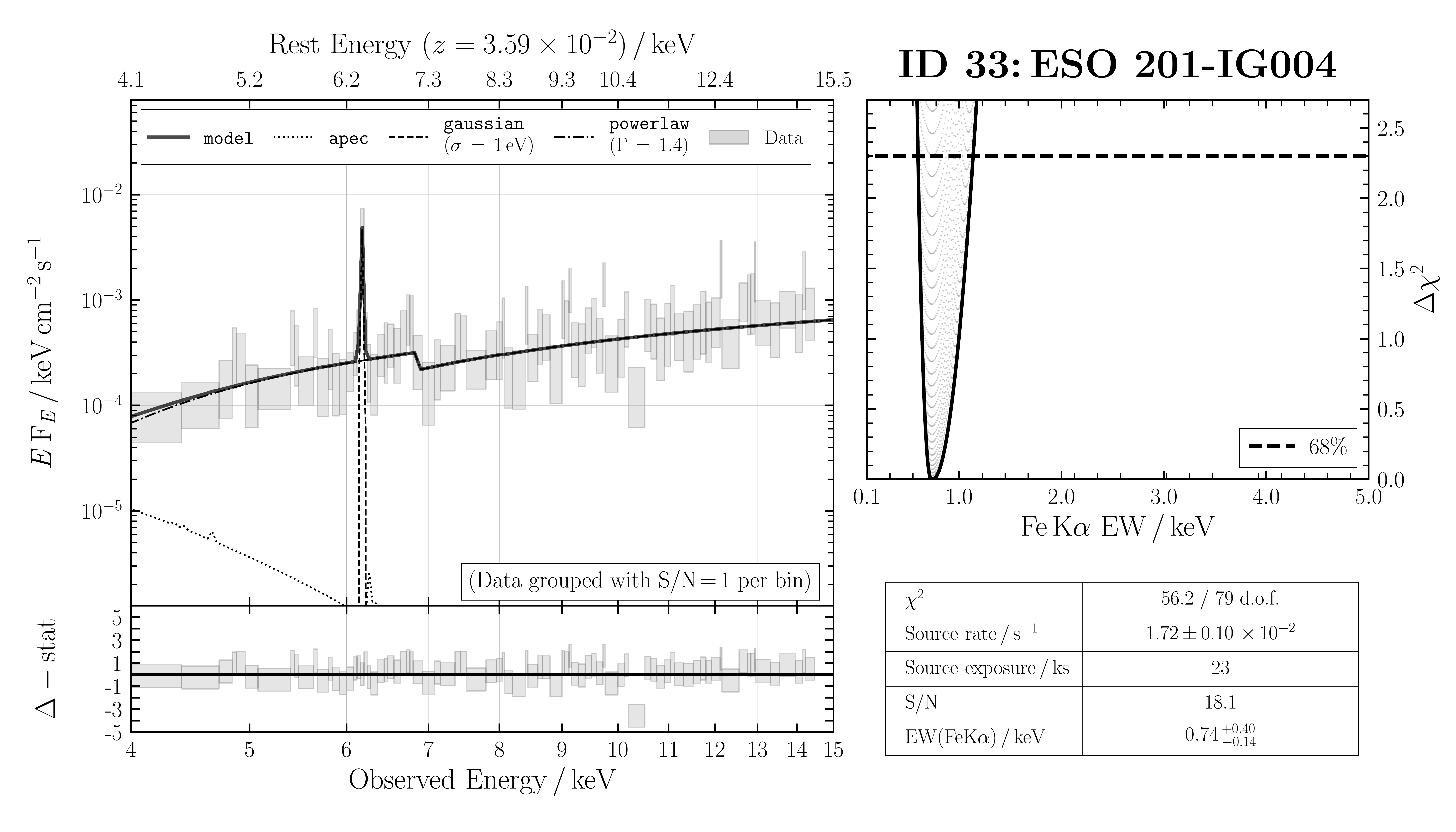}\\
\caption{\label{fig:ESO201-IG004} ID 33: ESO 201-IG004}
\end{center}
\end{figure}

\begin{figure}
\begin{center}
\includegraphics[angle=0,width=\columnwidth]{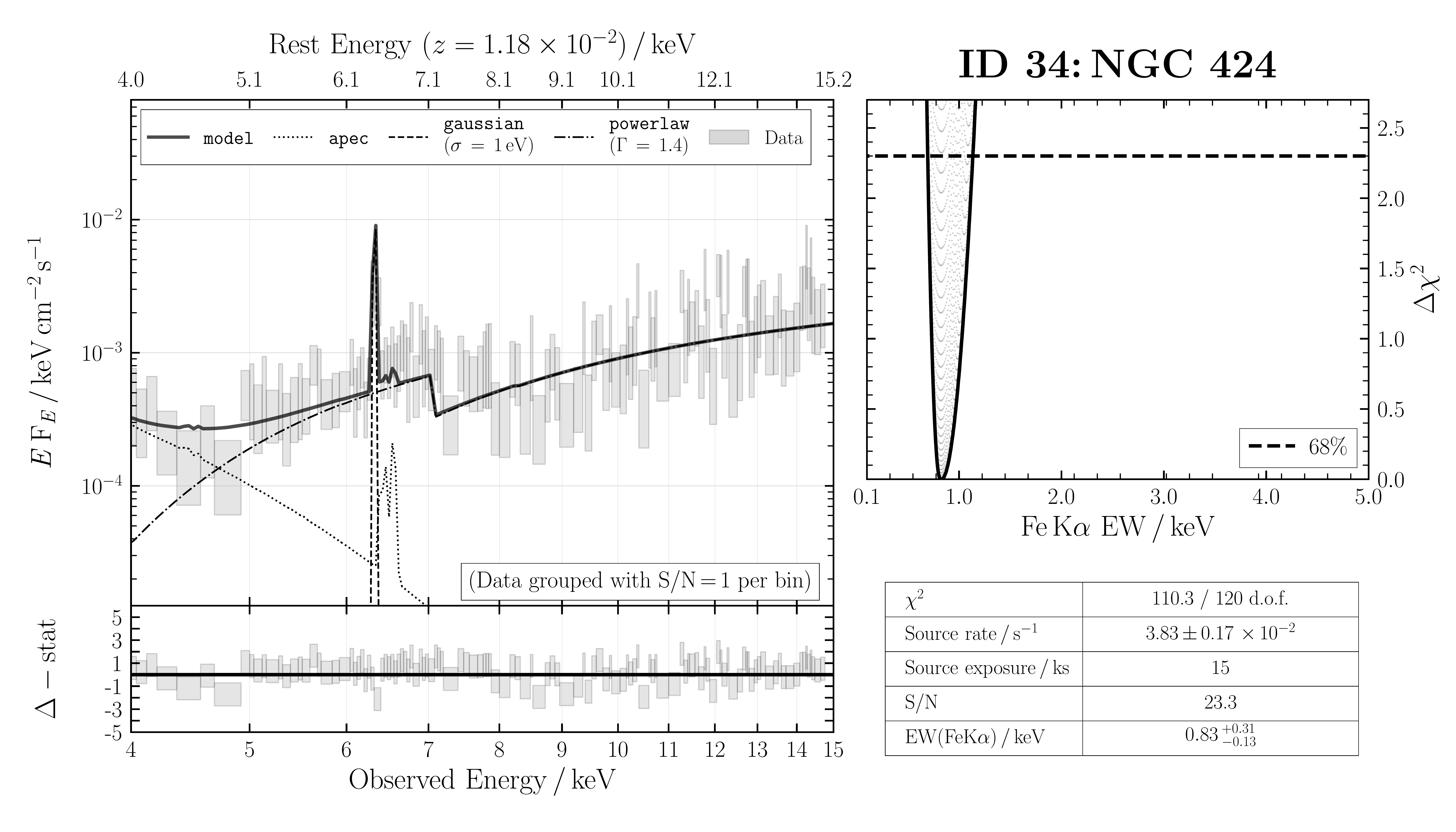}\\
\caption{\label{fig:NGC424} ID 34: NGC 424}
\end{center}
\end{figure}

\begin{figure}
\begin{center}
\includegraphics[angle=0,width=\columnwidth]{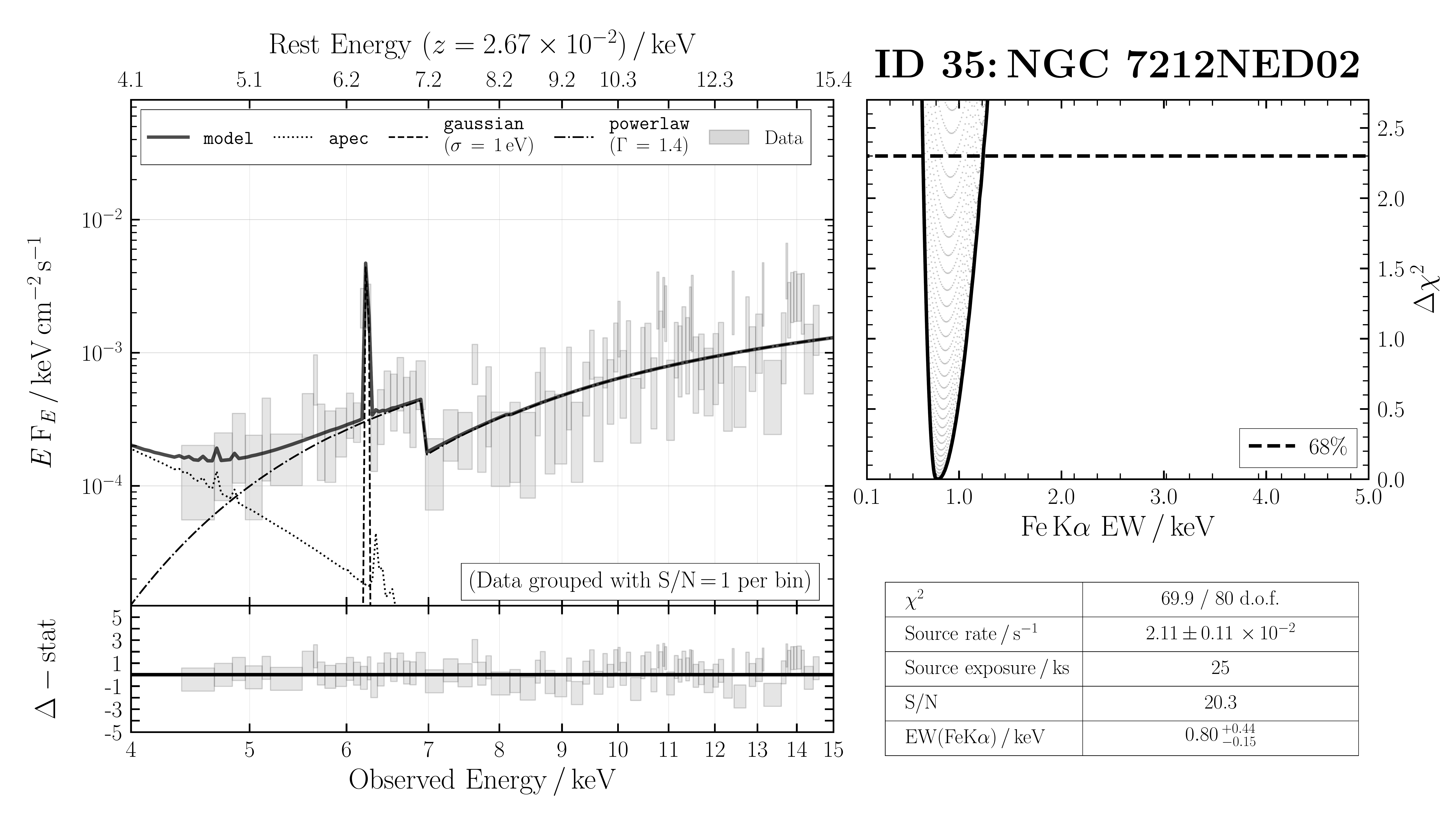}\\
\caption{\label{fig:NGC7212NED02} ID 35: NGC 7212NED02}
\end{center}
\end{figure}

\begin{figure}
\begin{center}
\includegraphics[angle=0,width=\columnwidth]{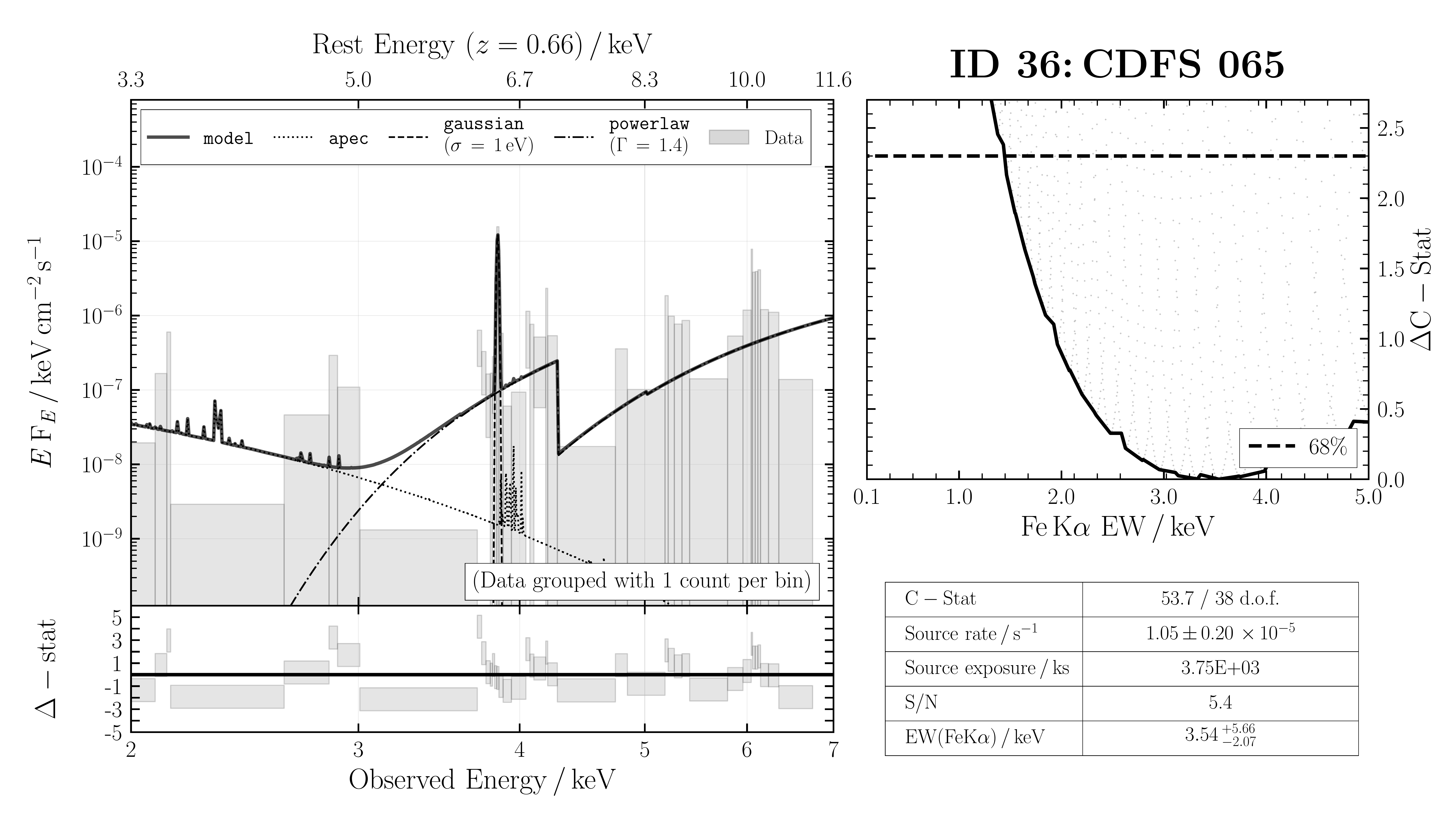}\\
\caption{\label{fig:CDFS065} ID 36: CDFS 065}
\end{center}
\end{figure}

\begin{figure}
\begin{center}
\includegraphics[angle=0,width=\columnwidth]{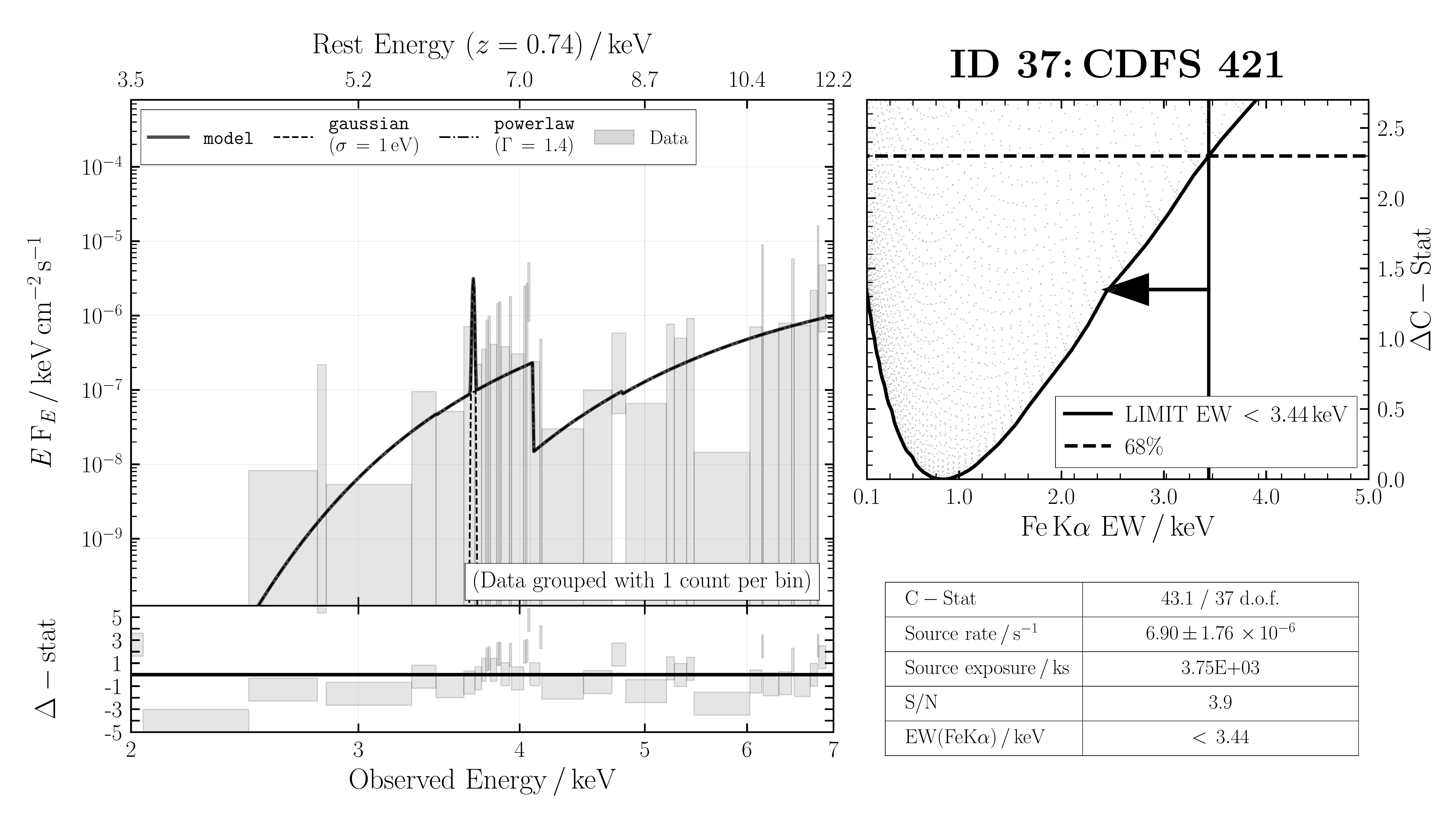}\\
\caption{\label{fig:CDFS421} ID 37: CDFS 421}
\end{center}
\end{figure}

\begin{figure}
\begin{center}
\includegraphics[angle=0,width=\columnwidth]{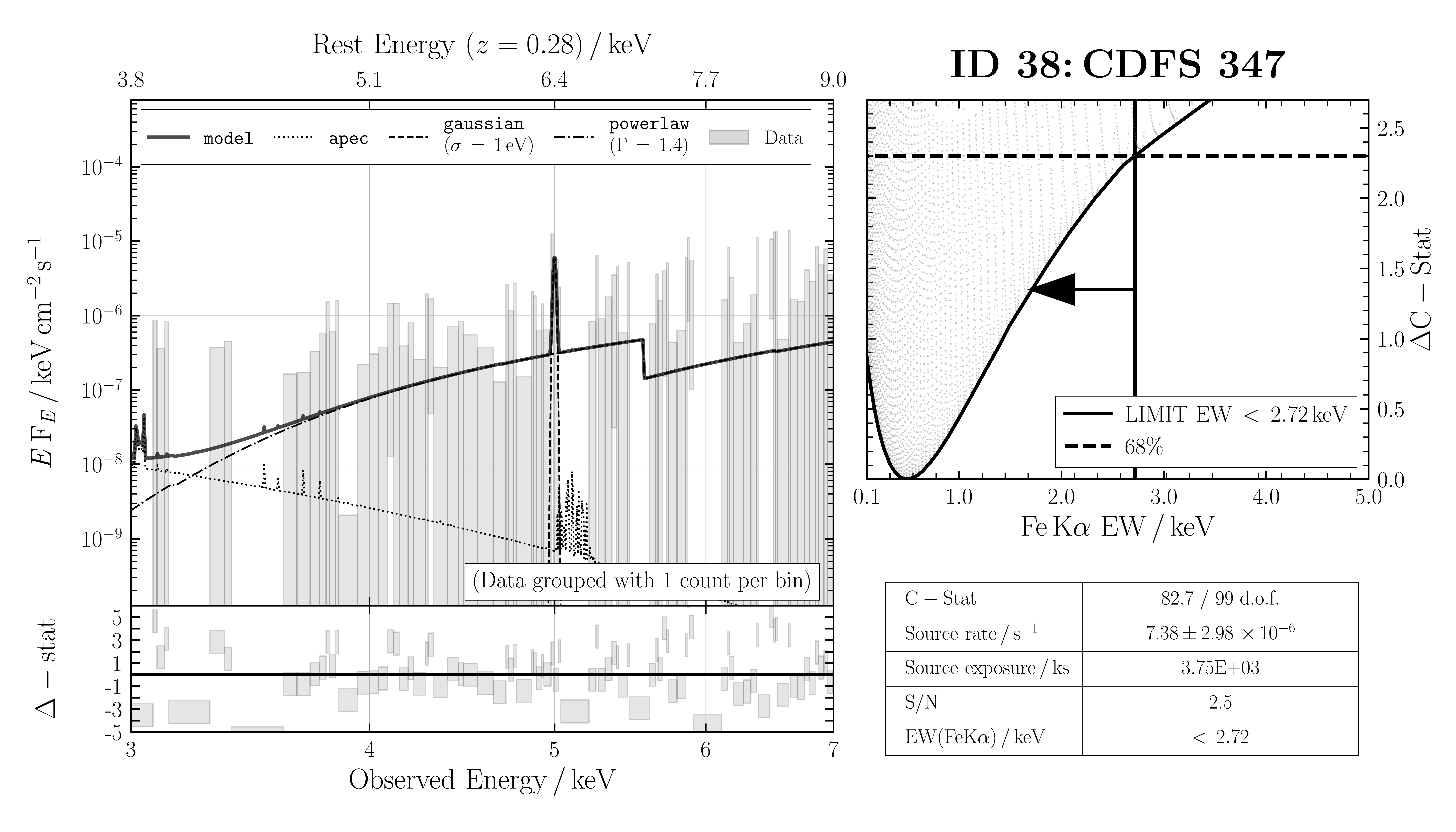}\\
\caption{\label{fig:CDFS347} ID 38: CDFS 347}
\end{center}
\end{figure}

\begin{figure}
\begin{center}
\includegraphics[angle=0,width=\columnwidth]{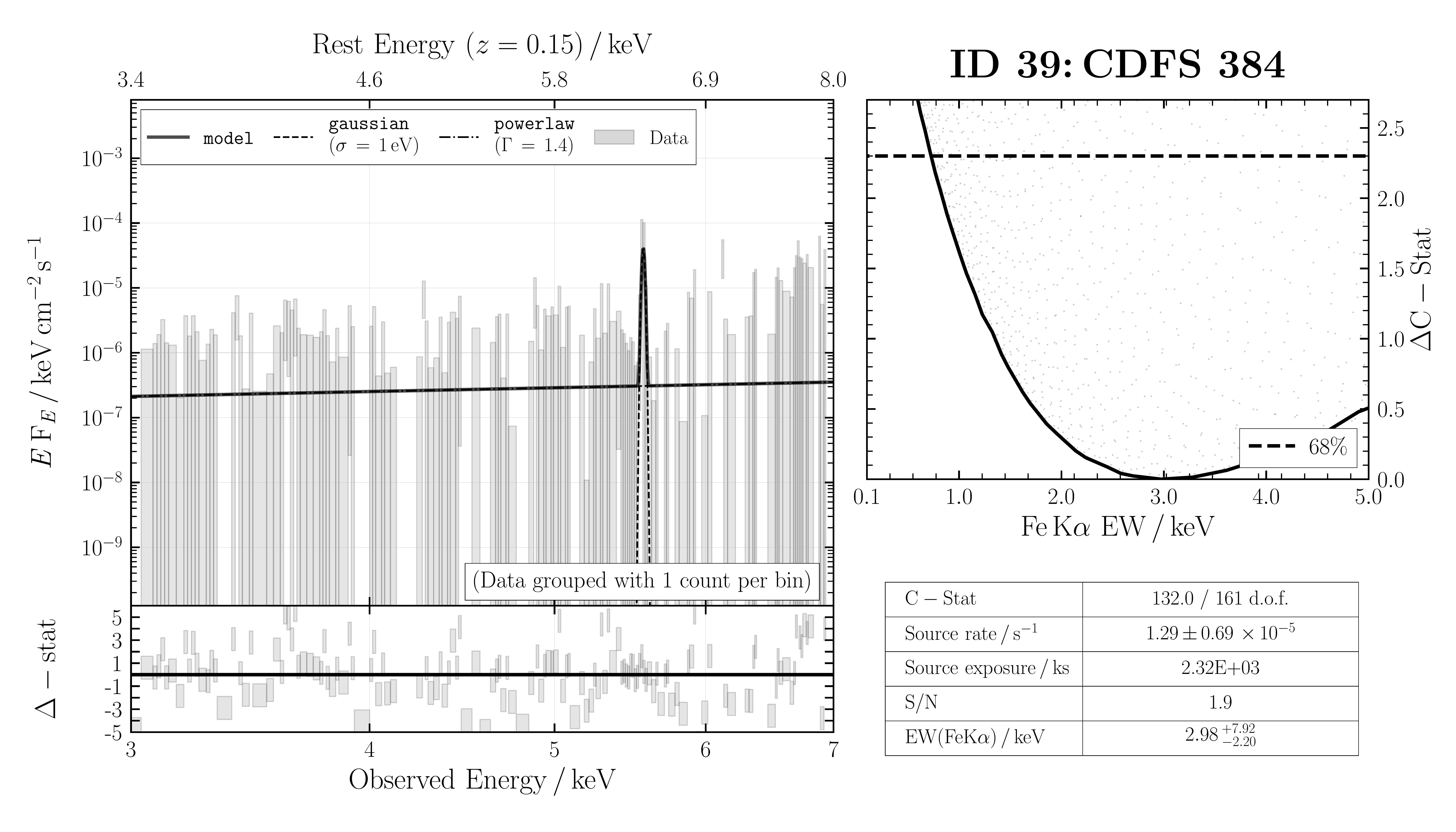}\\
\caption{\label{fig:CDFS384} ID 39: CDFS 384}
\end{center}
\end{figure}

\begin{figure}
\begin{center}
\includegraphics[angle=0,width=\columnwidth]{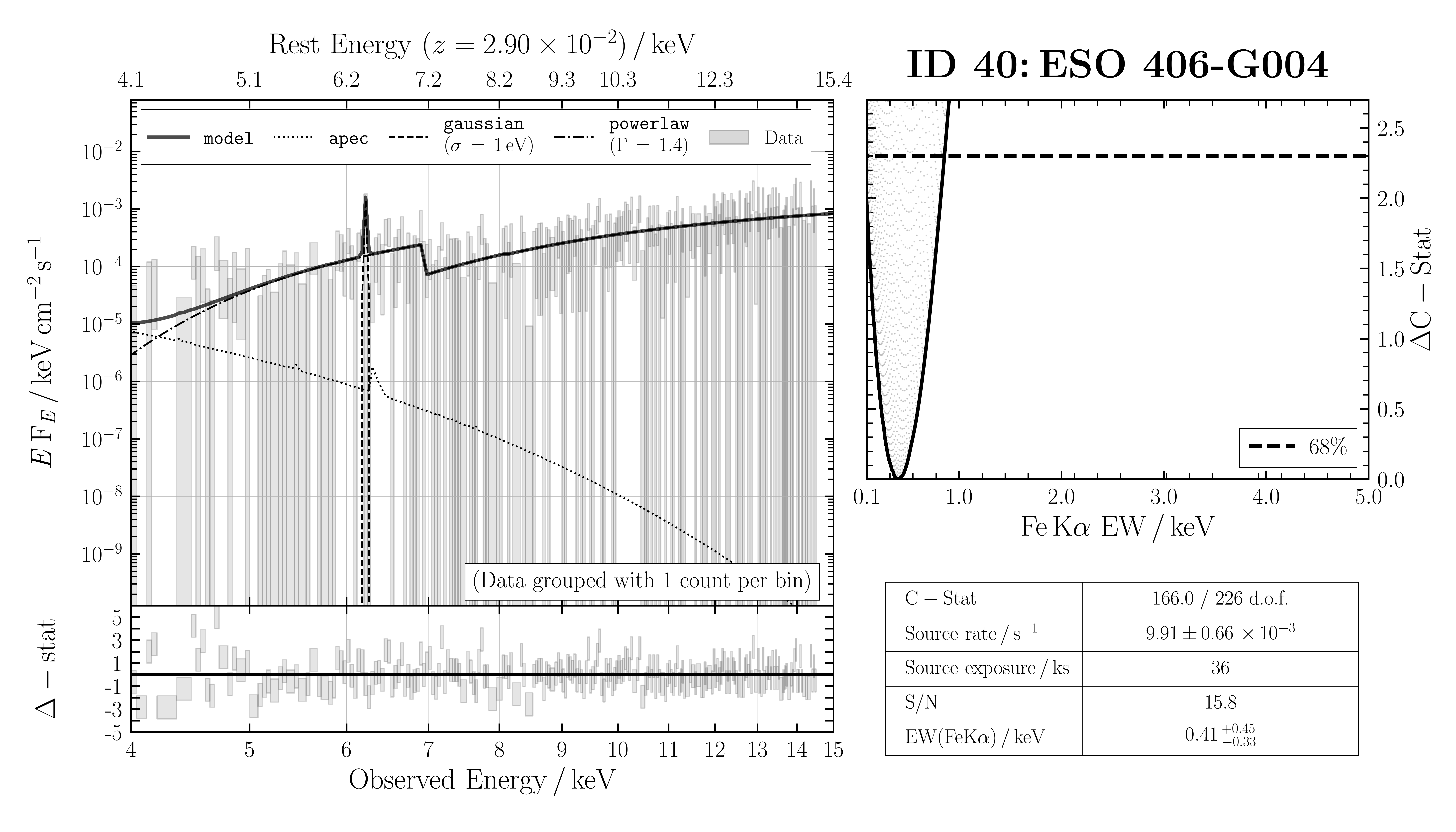}\\
\caption{\label{fig:ESO406-G004} ID 40: ESO 406-G004}
\end{center}
\end{figure}

\begin{figure}
\begin{center}
\includegraphics[angle=0,width=\columnwidth]{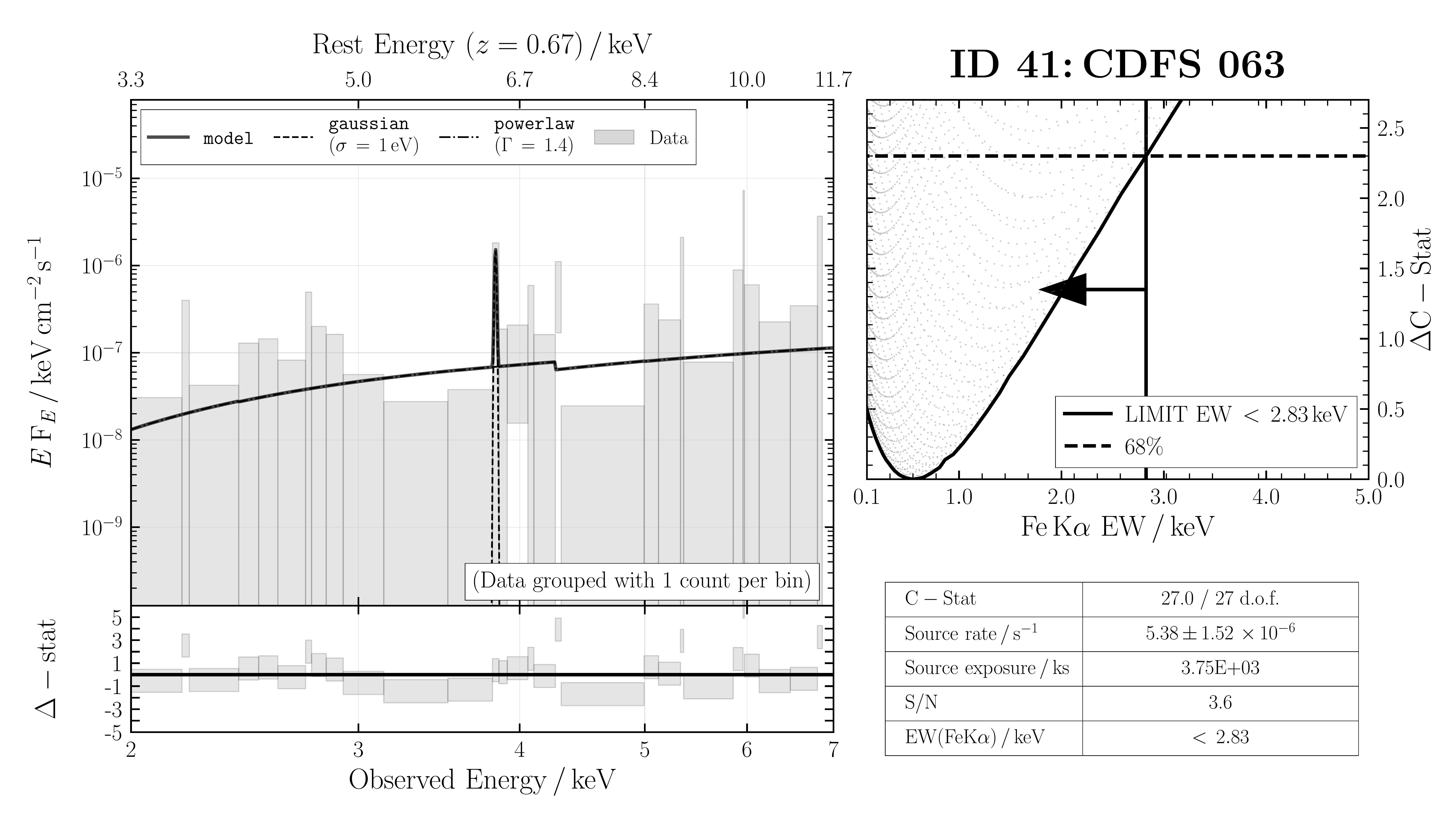}\\
\caption{\label{fig:CDFS063} ID 41: CDFS 063}
\end{center}
\end{figure}

\begin{figure}
\begin{center}
\includegraphics[angle=0,width=\columnwidth]{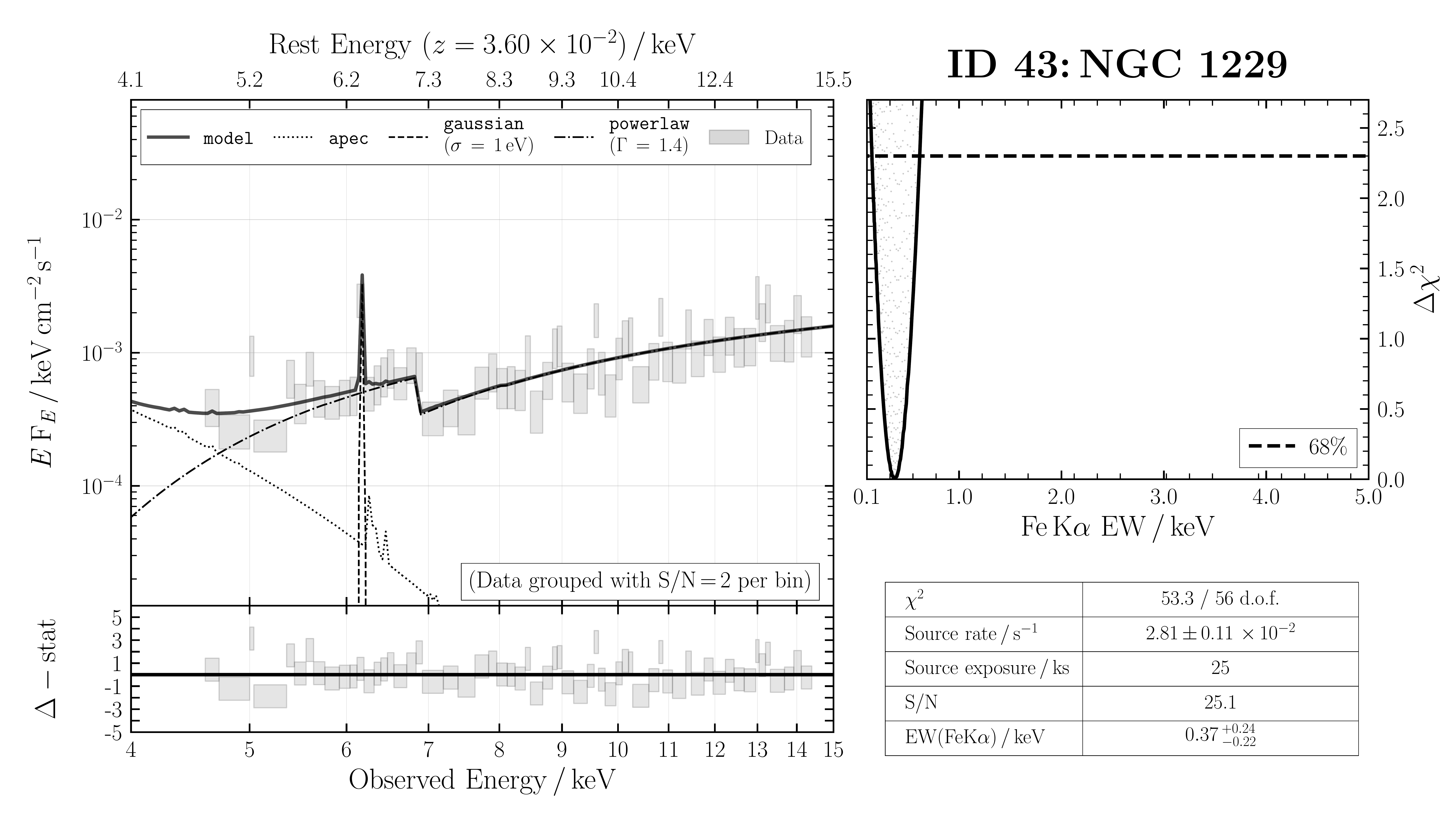}\\
\caption{\label{fig:NGC1229} ID 43: NGC 1229}
\end{center}
\end{figure}

\begin{figure}
\begin{center}
\includegraphics[angle=0,width=\columnwidth]{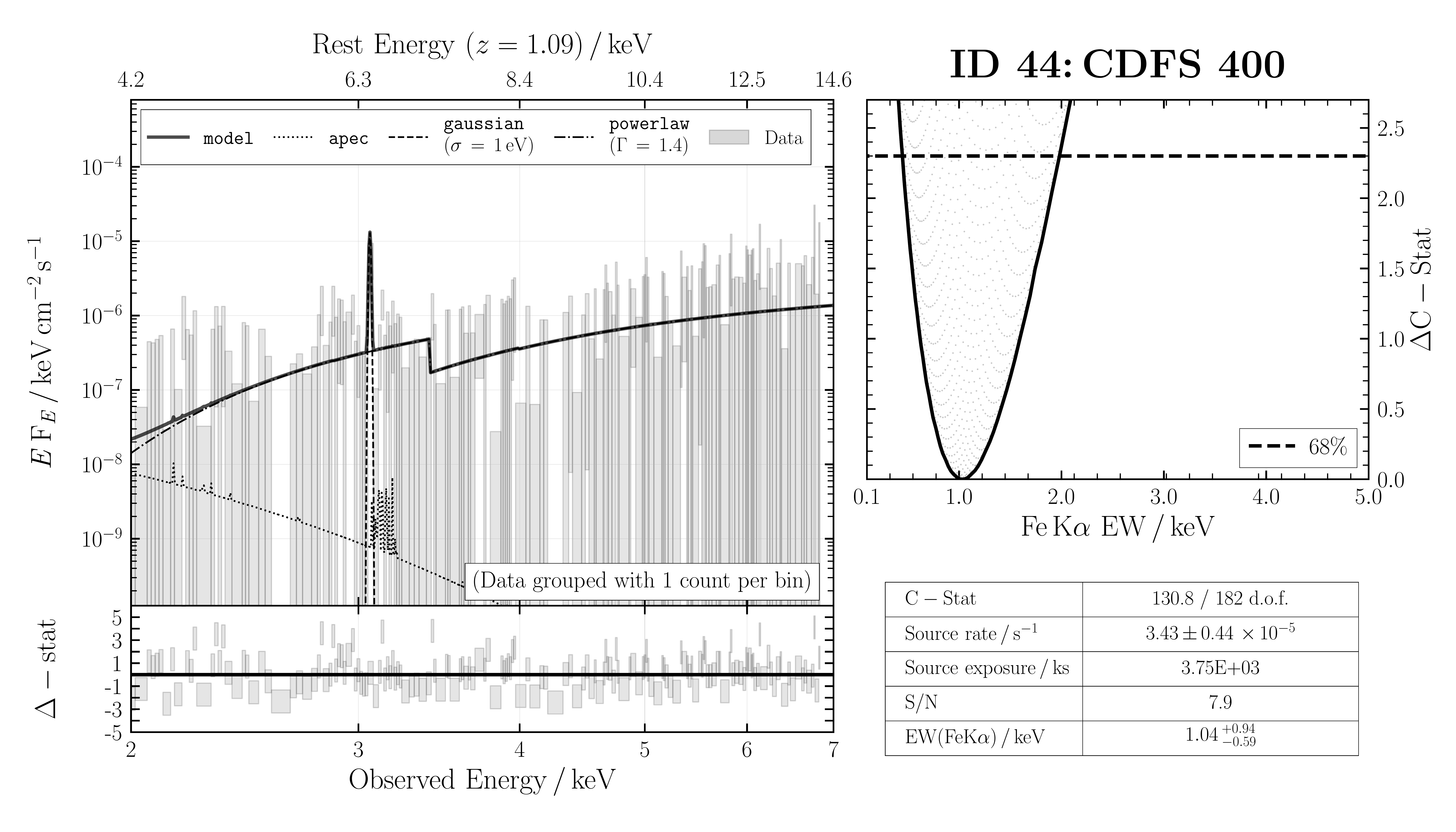}\\
\caption{\label{fig:CDFS400} ID 44: CDFS 400}
\end{center}
\end{figure}

\begin{figure}
\begin{center}
\includegraphics[angle=0,width=\columnwidth]{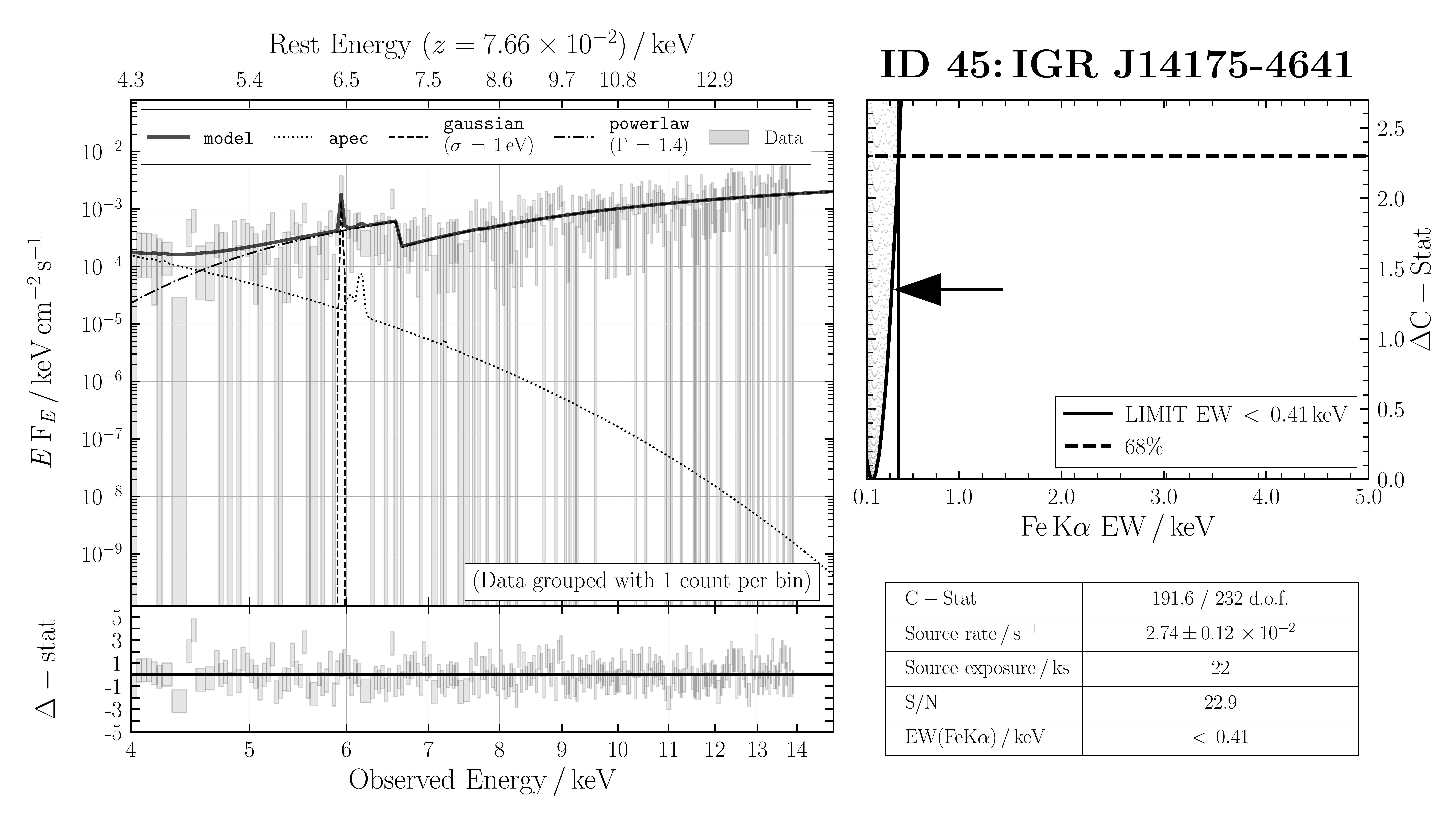}\\
\caption{\label{fig:IGRJ14175-4641} ID 45: IGR J14175-4641}
\end{center}
\end{figure}

\begin{figure}
\begin{center}
\includegraphics[angle=0,width=\columnwidth]{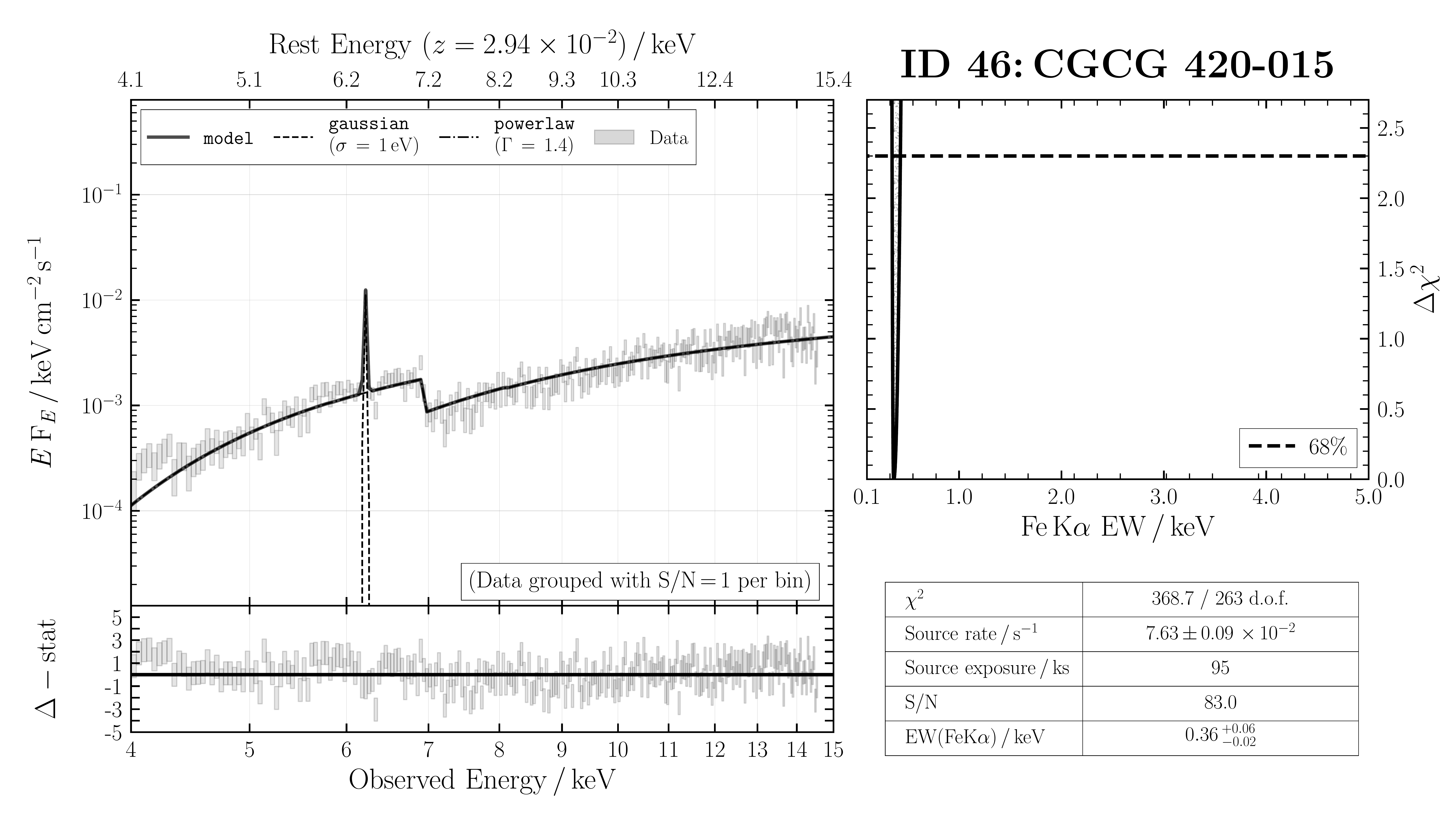}\\
\caption{\label{fig:CGCG420-015} ID 46: CGCG 420-015}
\end{center}
\end{figure}

\begin{figure}
\begin{center}
\includegraphics[angle=0,width=\columnwidth]{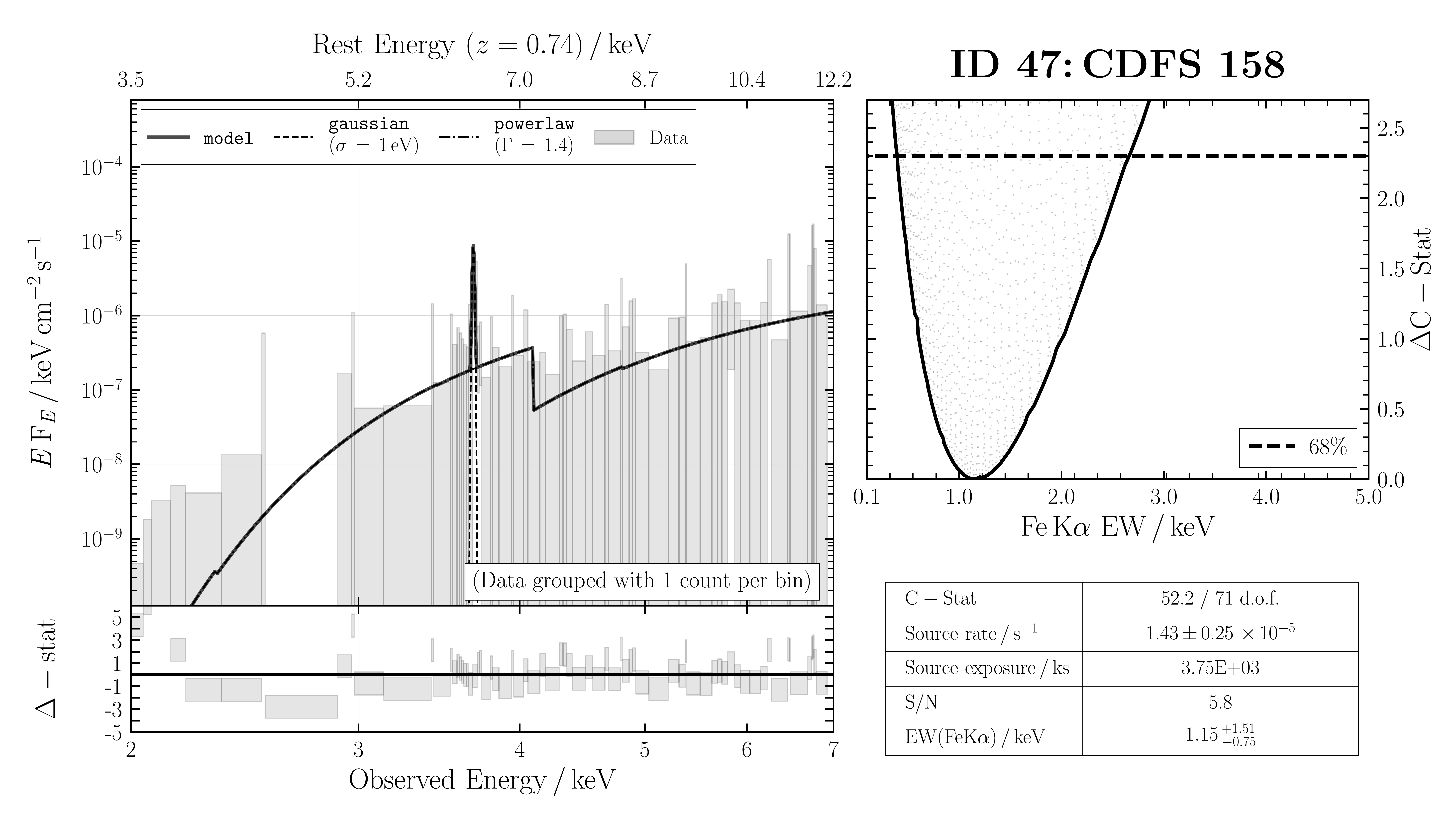}\\
\caption{\label{fig:CDFS158} ID 47: CDFS 158}
\end{center}
\end{figure}

\begin{figure}
\begin{center}
\includegraphics[angle=0,width=\columnwidth]{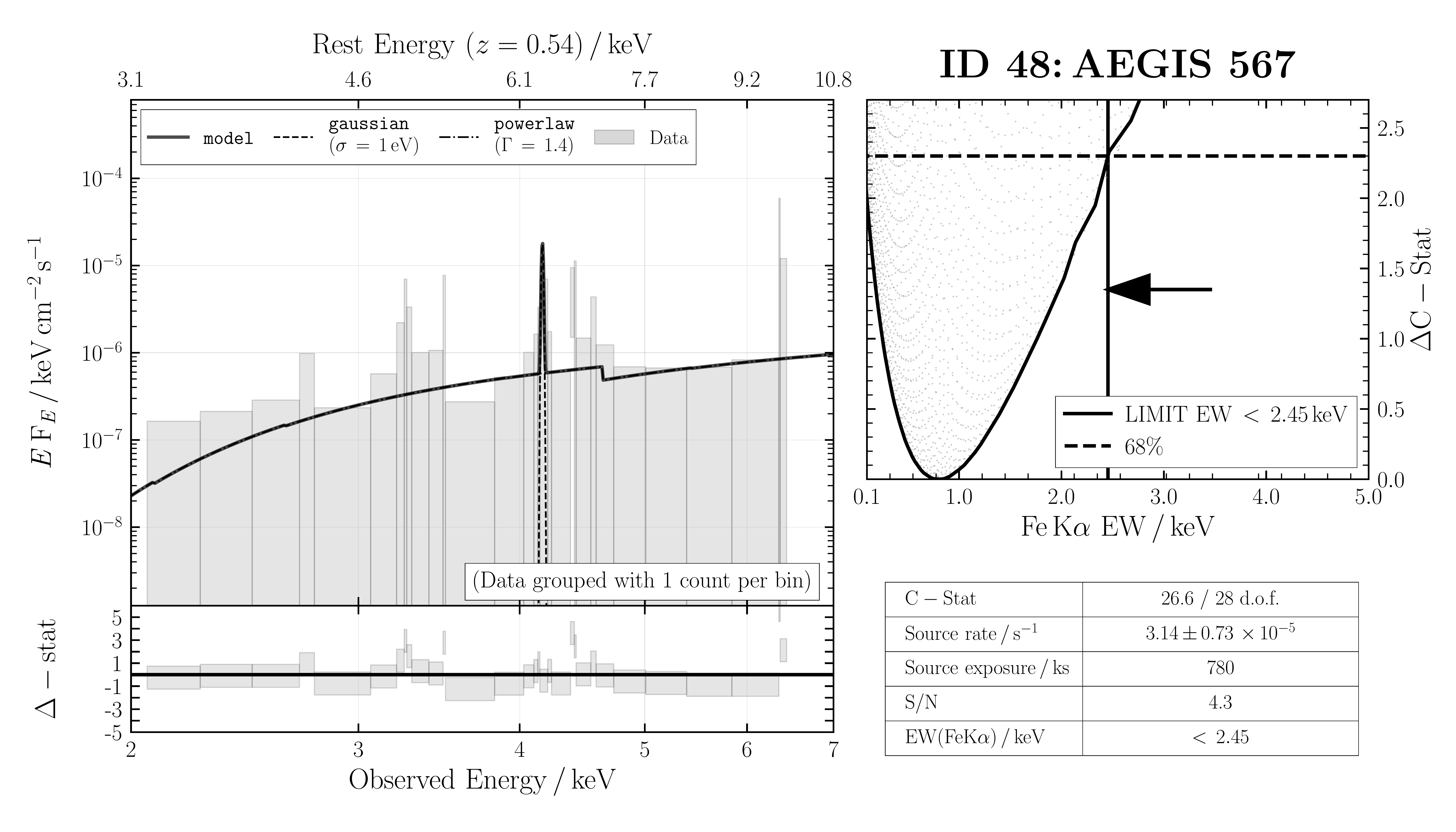}\\
\caption{\label{fig:AEGIS567} ID 48: AEGIS 567}
\end{center}
\end{figure}

\begin{figure}
\begin{center}
\includegraphics[angle=0,width=\columnwidth]{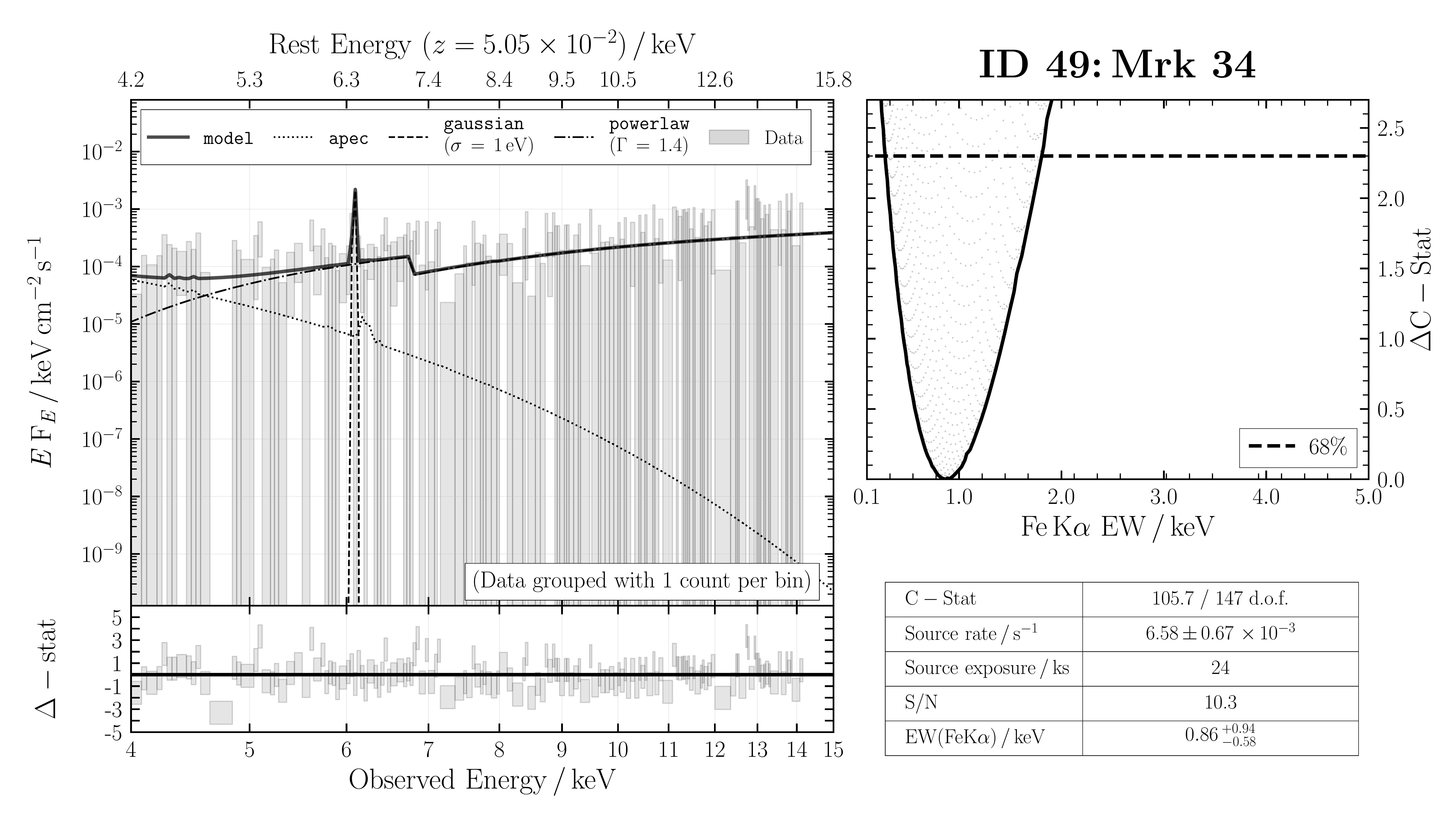}\\
\caption{\label{fig:Mrk34} ID 49: Mrk 34}
\end{center}
\end{figure}

\begin{figure}
\begin{center}
\includegraphics[angle=0,width=\columnwidth]{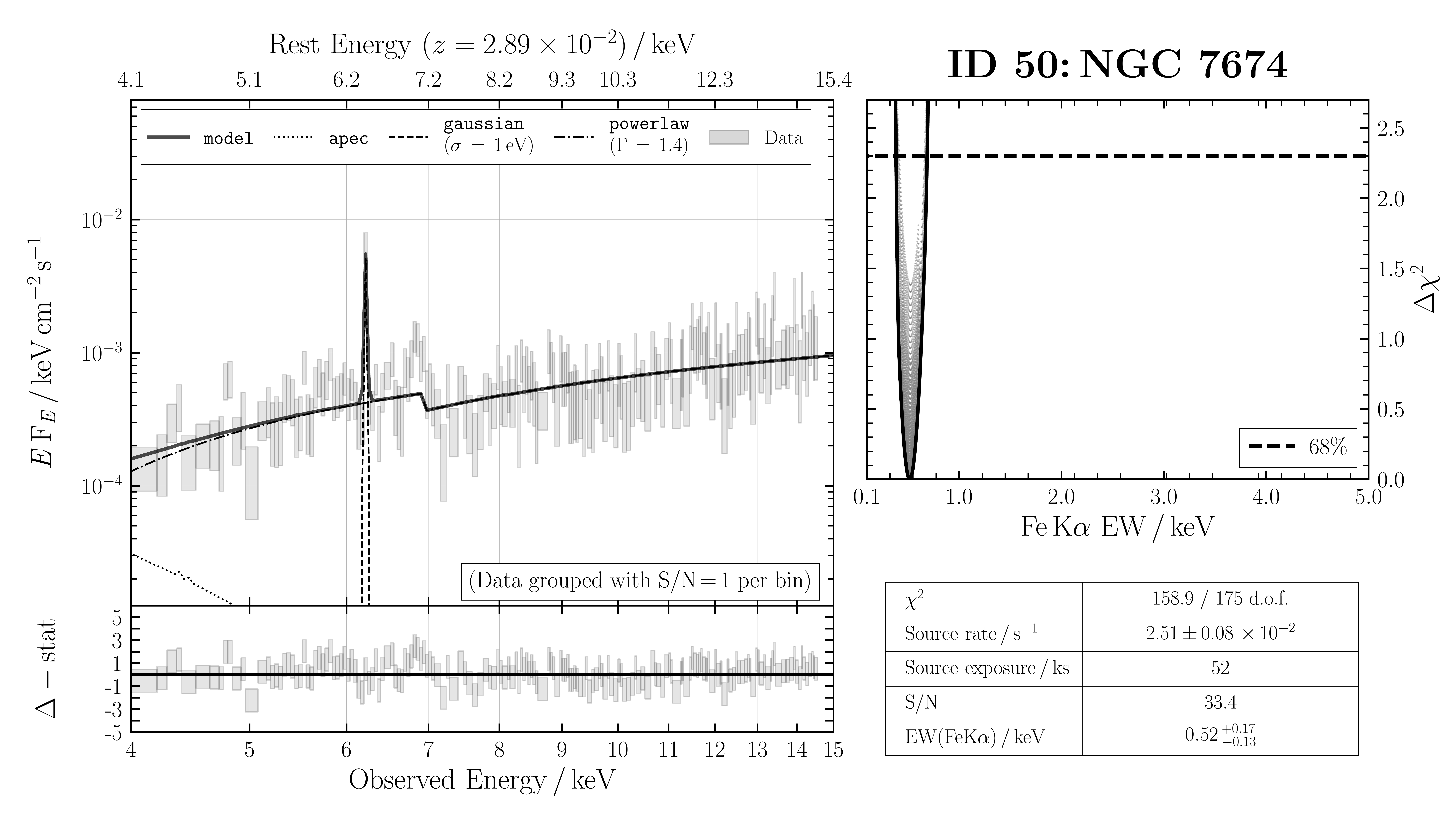}\\
\caption{\label{fig:NGC7674} ID 50: NGC 7674}
\end{center}
\end{figure}

\begin{figure}
\begin{center}
\includegraphics[angle=0,width=\columnwidth]{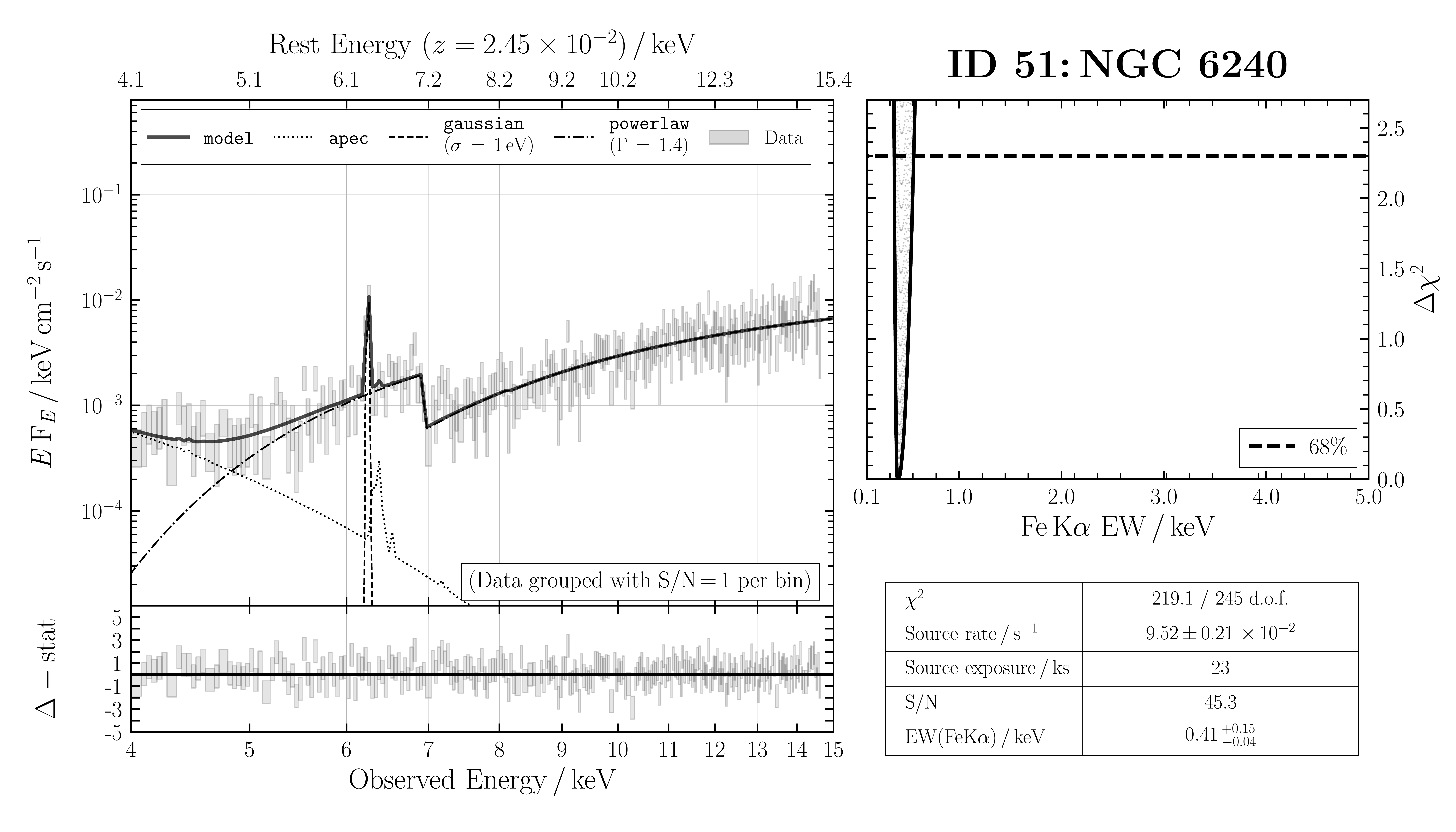}\\
\caption{\label{fig:NGC6240} ID 51: NGC 6240}
\end{center}
\end{figure}

\begin{figure}
\begin{center}
\includegraphics[angle=0,width=\columnwidth]{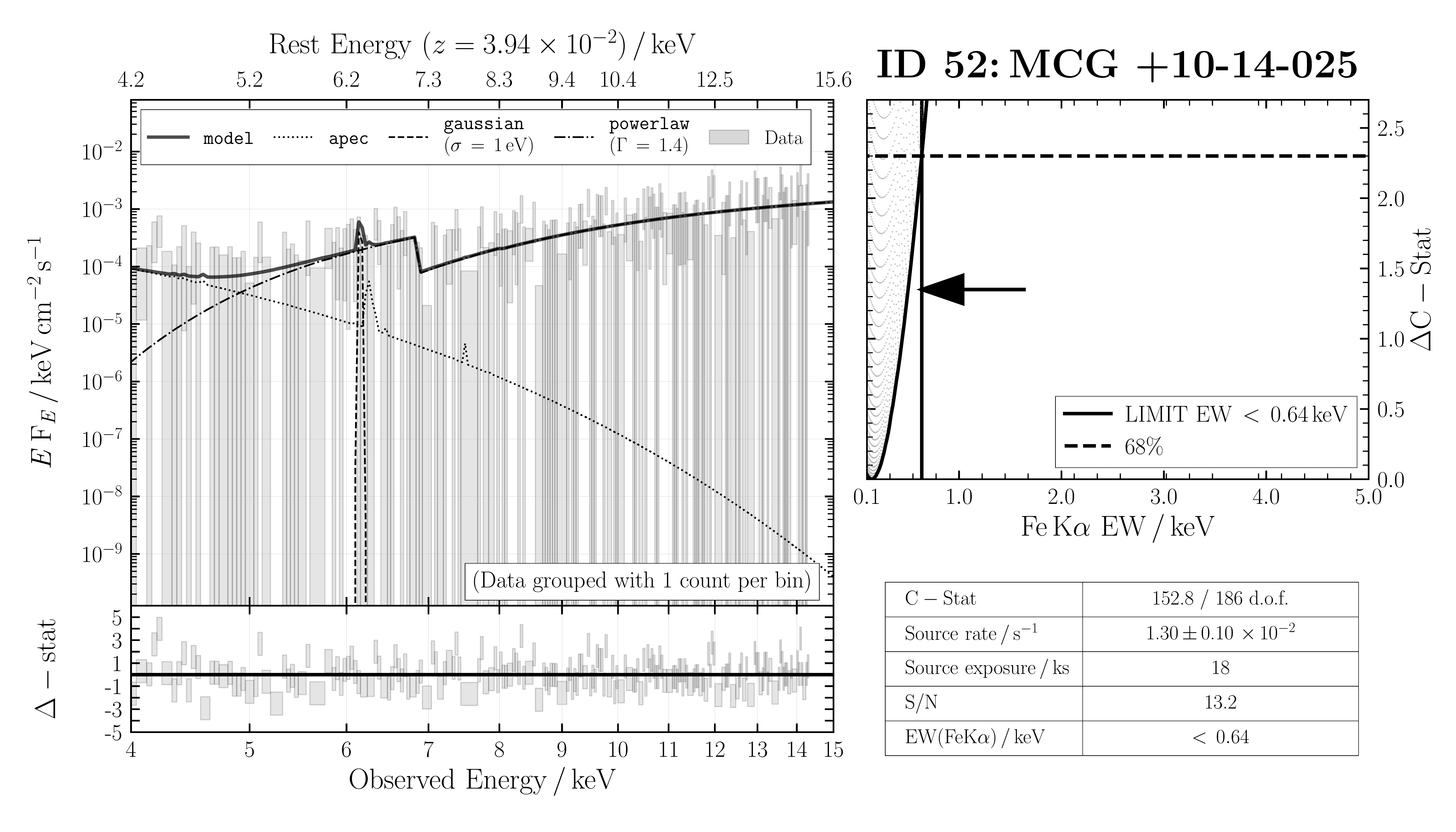}\\
\caption{\label{fig:MCG+10-14-025} ID 52: MCG +10-14-025}
\end{center}
\end{figure}

\begin{figure}
\begin{center}
\includegraphics[angle=0,width=\columnwidth]{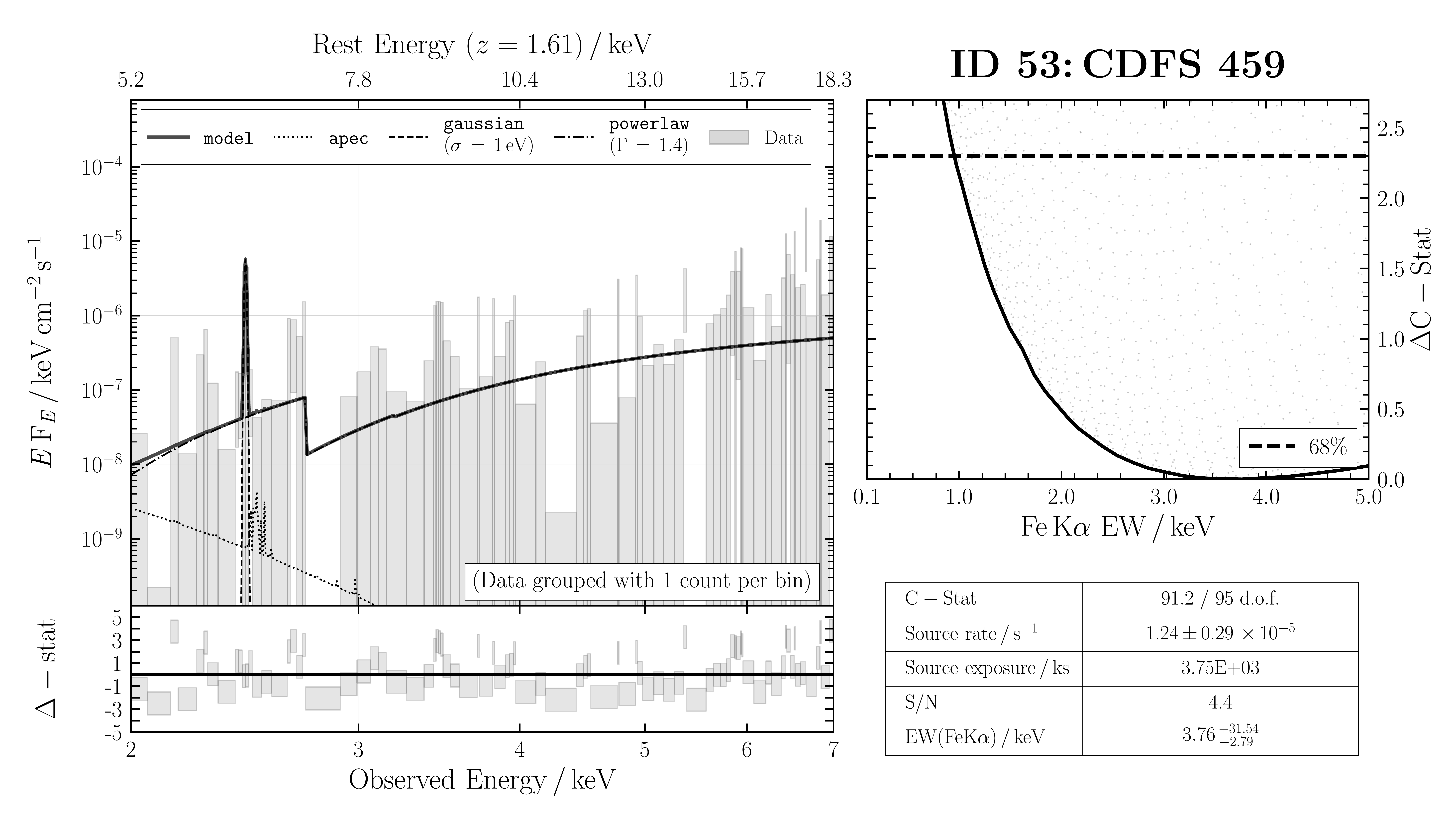}\\
\caption{\label{fig:CDFS459} ID 53: CDFS 459}
\end{center}
\end{figure}

\begin{figure}
\begin{center}
\includegraphics[angle=0,width=\columnwidth]{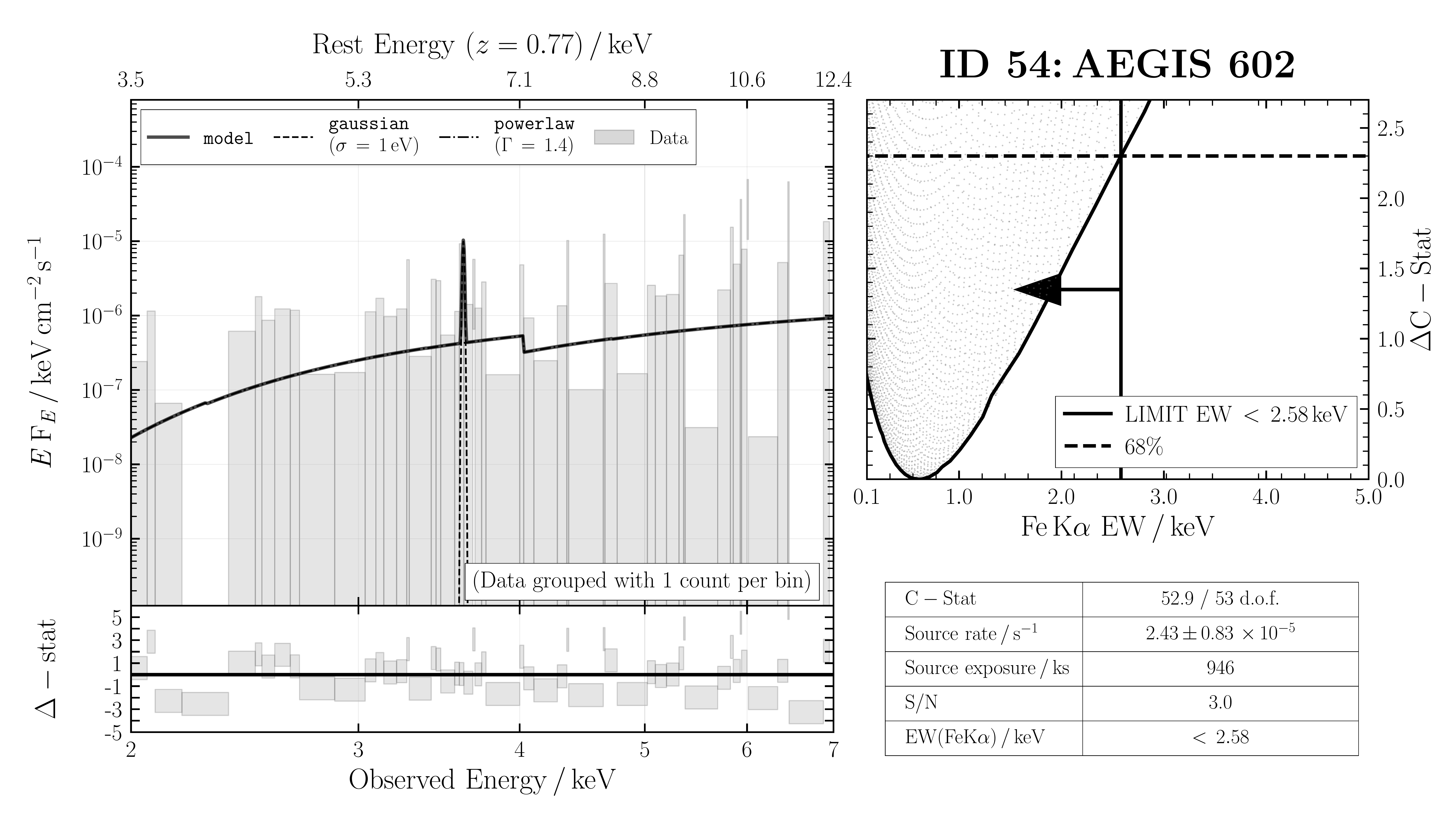}\\
\caption{\label{fig:AEGIS602} ID 54: AEGIS 602}
\end{center}
\end{figure}

\begin{figure}
\begin{center}
\includegraphics[angle=0,width=\columnwidth]{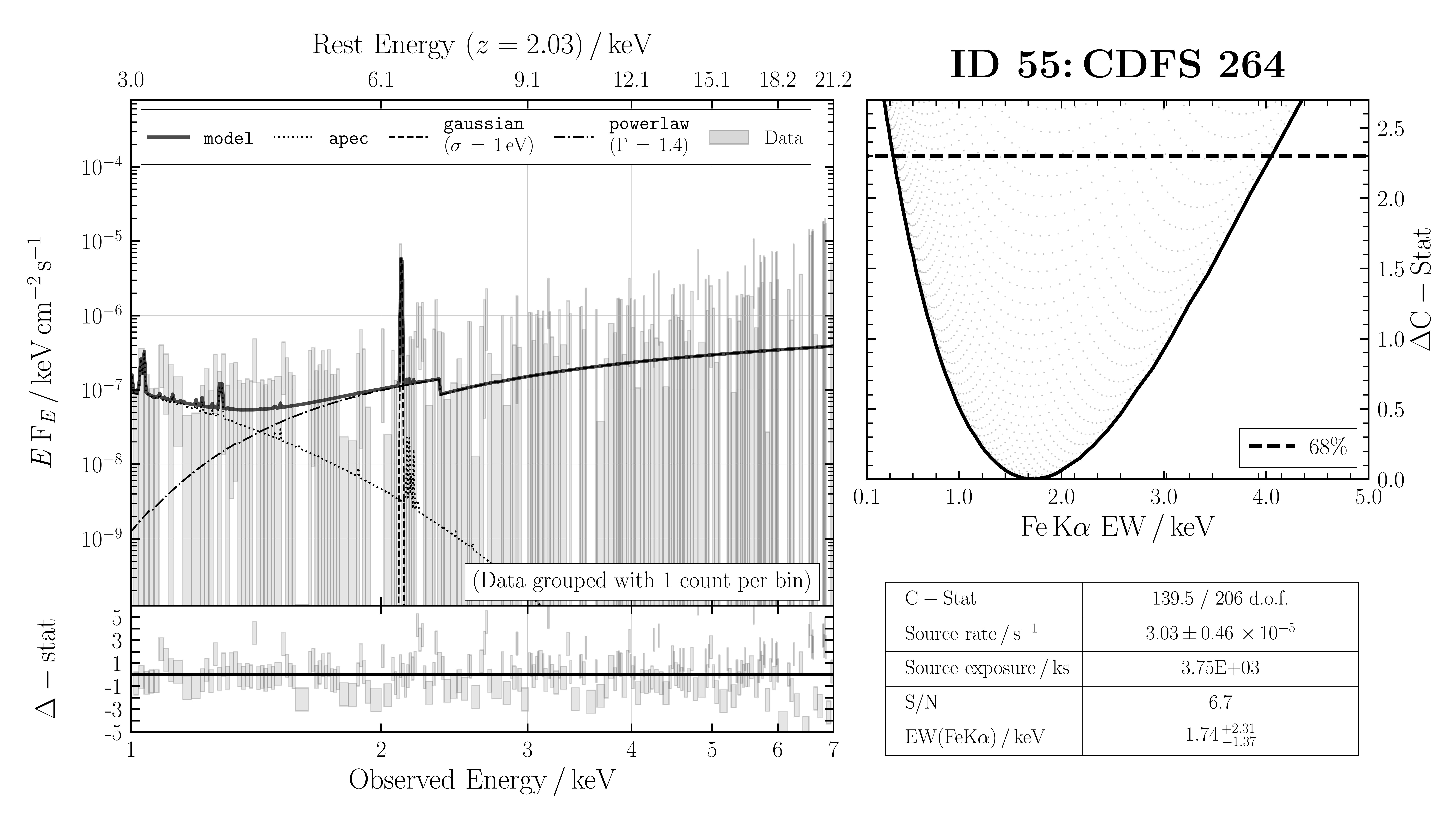}\\
\caption{\label{fig:CDFS264} ID 55: CDFS 264}
\end{center}
\end{figure}

\begin{figure}
\begin{center}
\includegraphics[angle=0,width=\columnwidth]{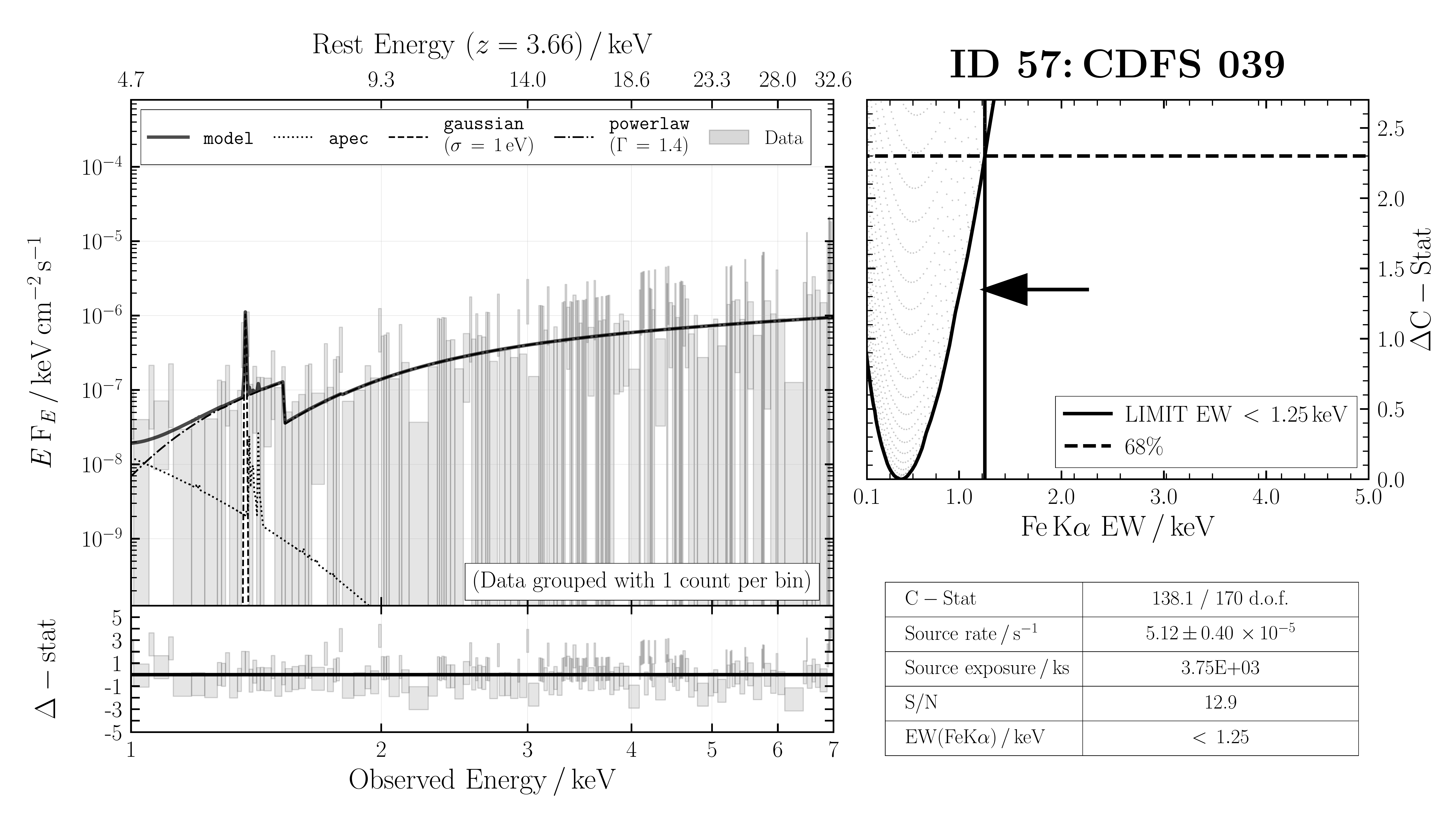}\\
\caption{\label{fig:CDFS039} ID 57: CDFS 039}
\end{center}
\end{figure}

\begin{figure}
\begin{center}
\includegraphics[angle=0,width=\columnwidth]{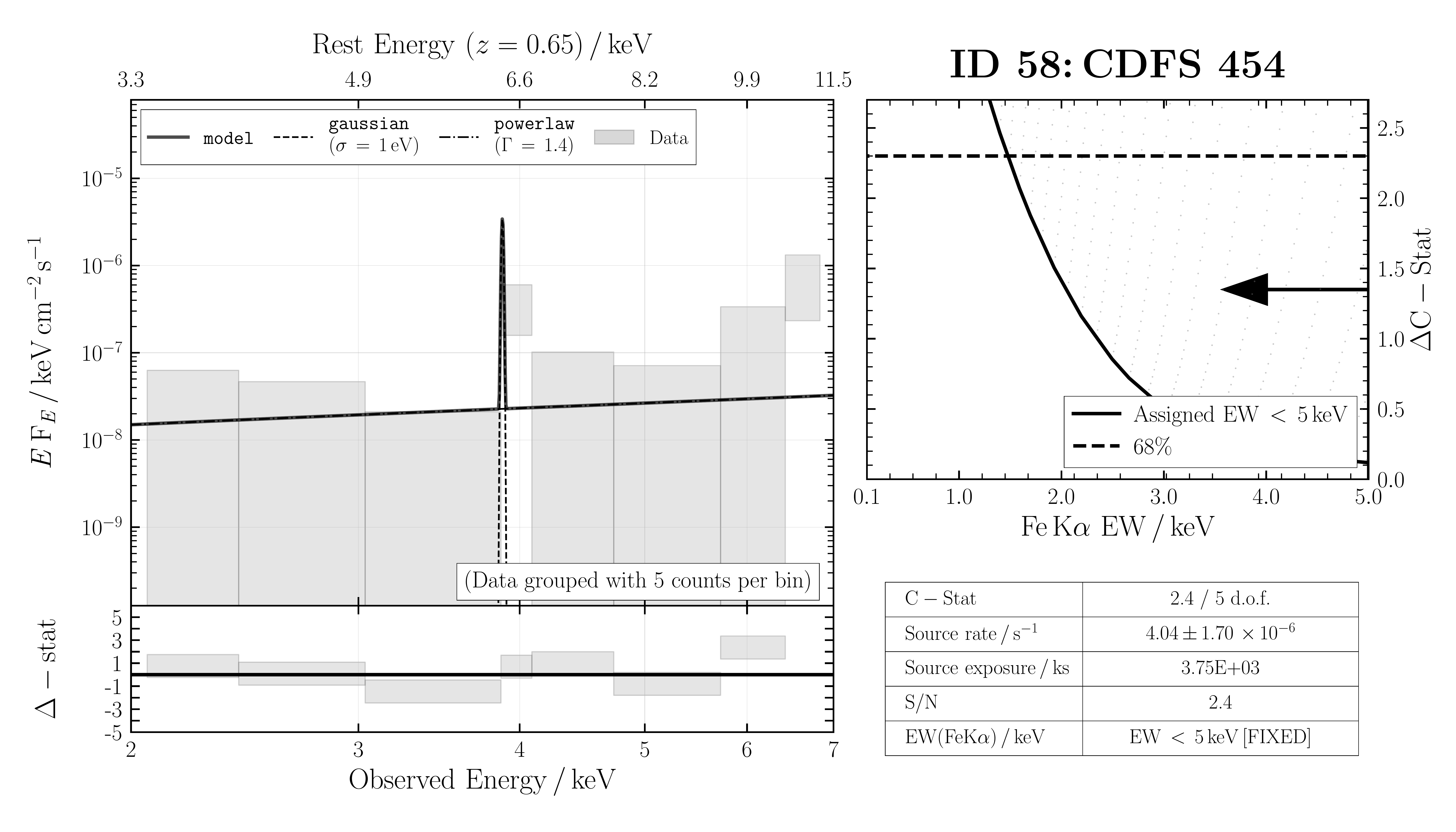}\\
\caption{\label{fig:CDFS454} ID 58: CDFS 454}
\end{center}
\end{figure}

\begin{figure}
\begin{center}
\includegraphics[angle=0,width=\columnwidth]{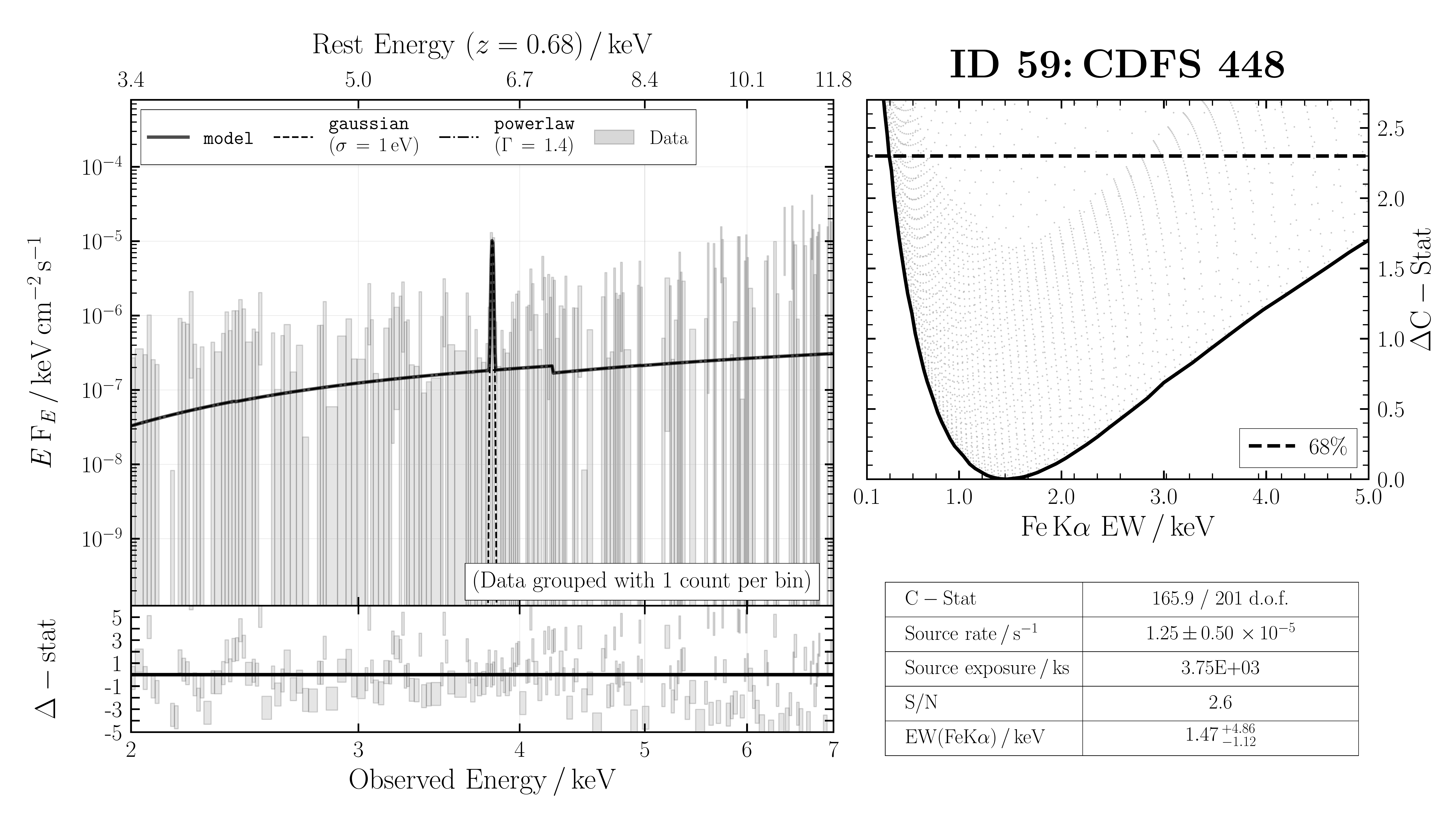}\\
\caption{\label{fig:CDFS448} ID 59: CDFS 448}
\end{center}
\end{figure}

\begin{figure}
\begin{center}
\includegraphics[angle=0,width=\columnwidth]{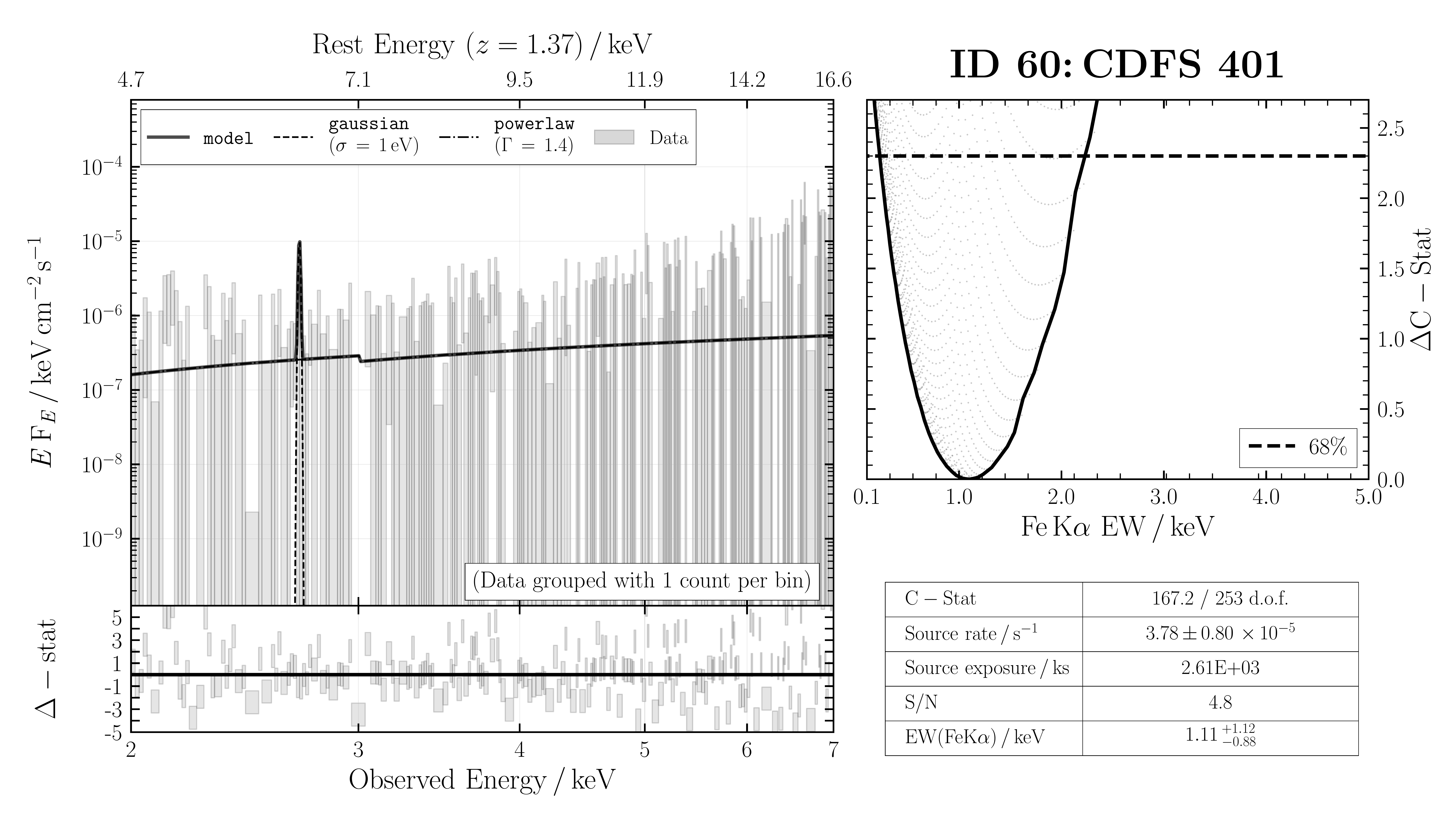}\\
\caption{\label{fig:CDFS401} ID 60: CDFS 401}
\end{center}
\end{figure}

\begin{figure}
\begin{center}
\includegraphics[angle=0,width=\columnwidth]{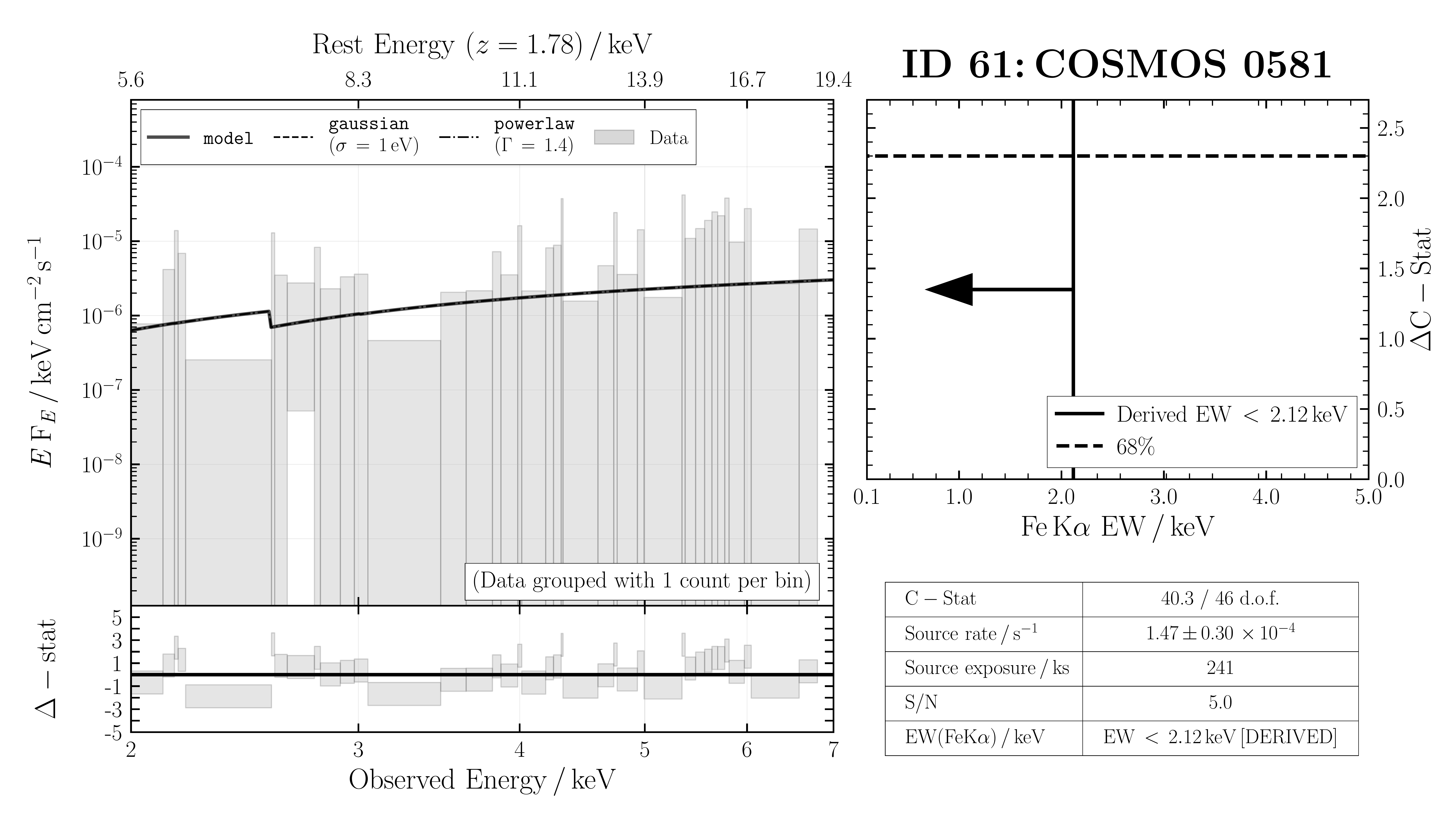}\\
\caption{\label{fig:COSMOS0581} ID 61: COSMOS 0581}
\end{center}
\end{figure}

\begin{figure}
\begin{center}
\includegraphics[angle=0,width=\columnwidth]{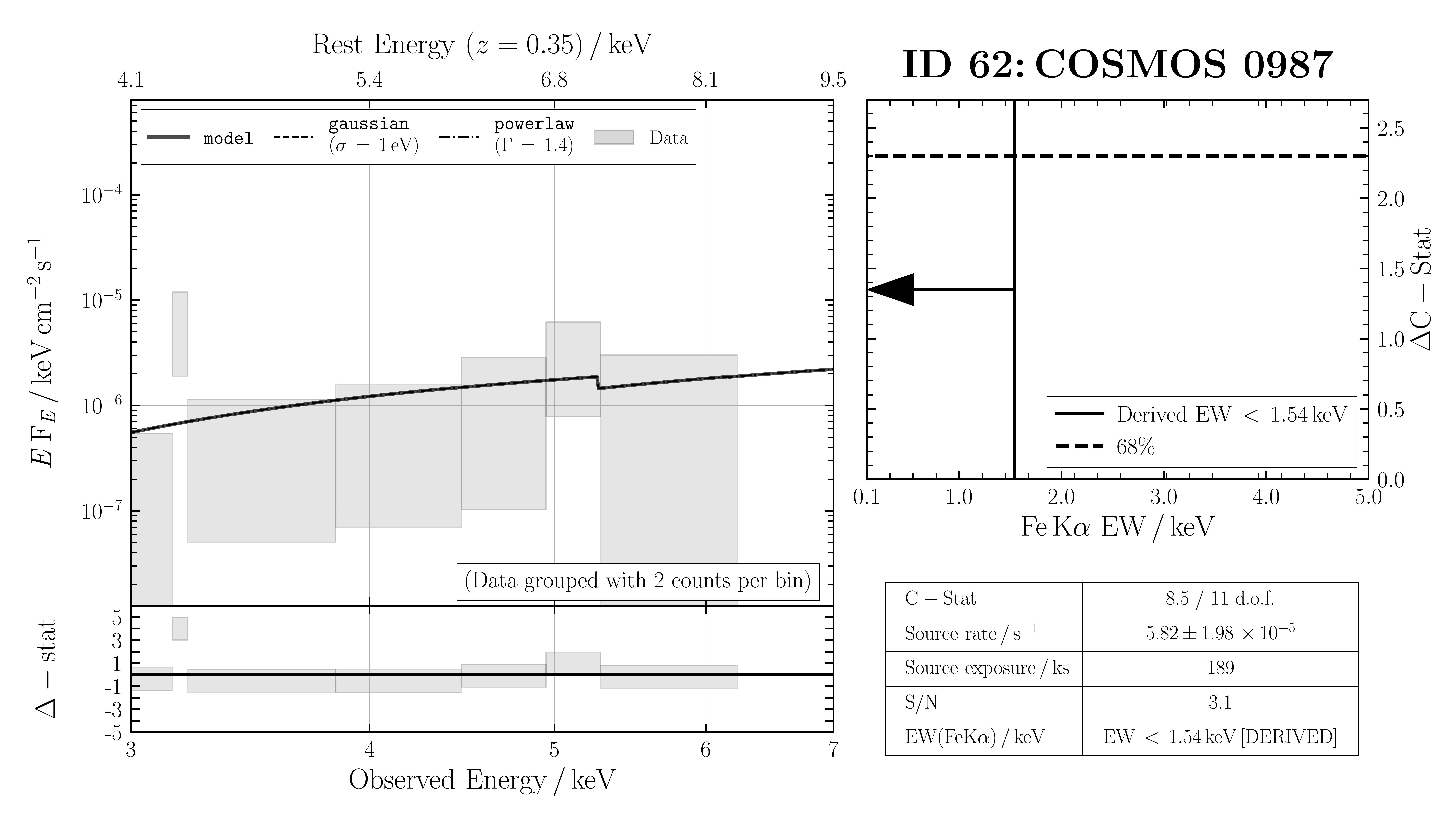}\\
\caption{\label{fig:COSMOS0987} ID 62: COSMOS 0987}
\end{center}
\end{figure}

\begin{figure}
\begin{center}
\includegraphics[angle=0,width=\columnwidth]{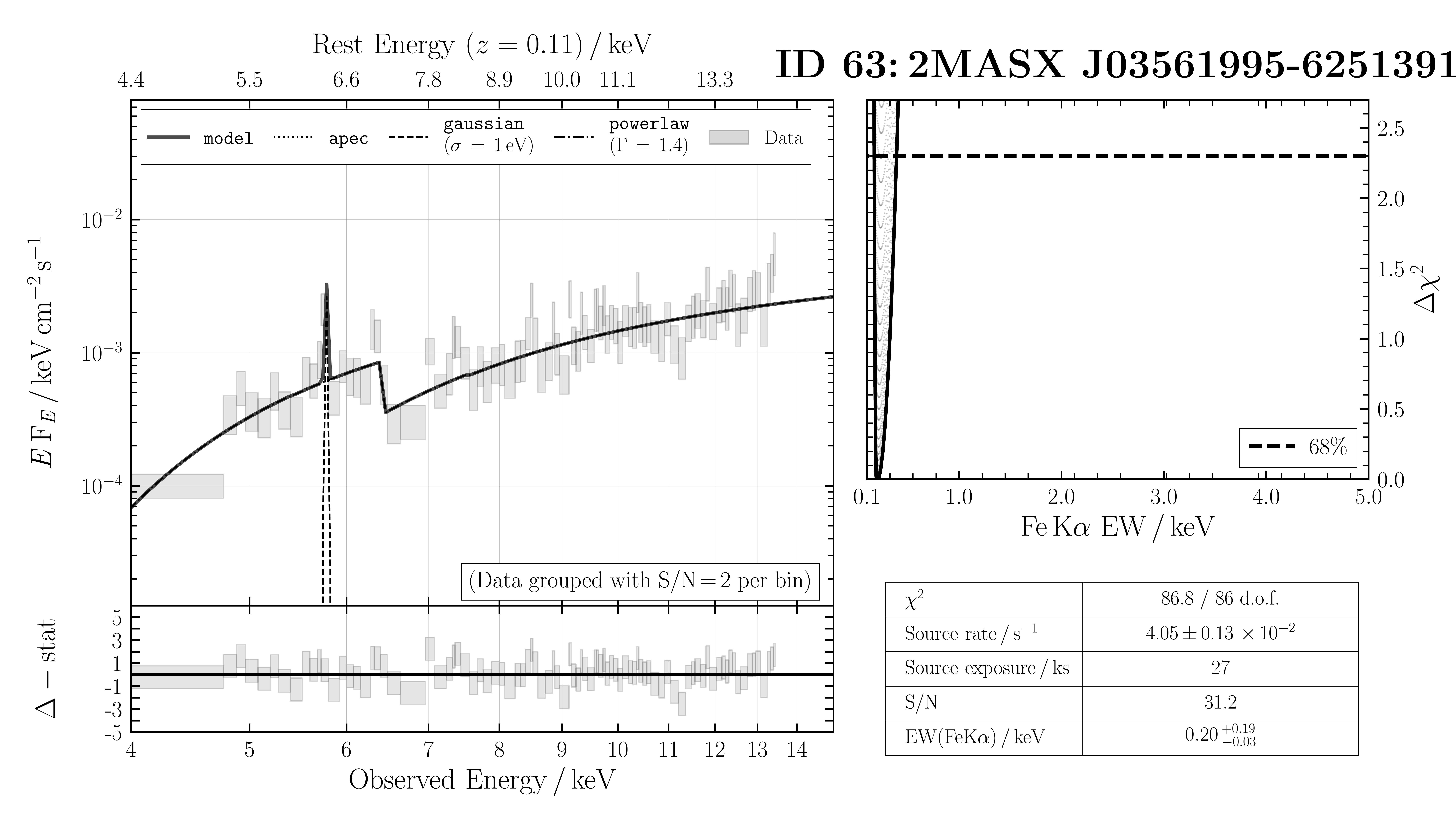}\\
\caption{\label{fig:2MASXJ03561995-6251391} ID 63: 2MASX J03561995-6251391}
\end{center}
\end{figure}

\begin{figure}
\begin{center}
\includegraphics[angle=0,width=\columnwidth]{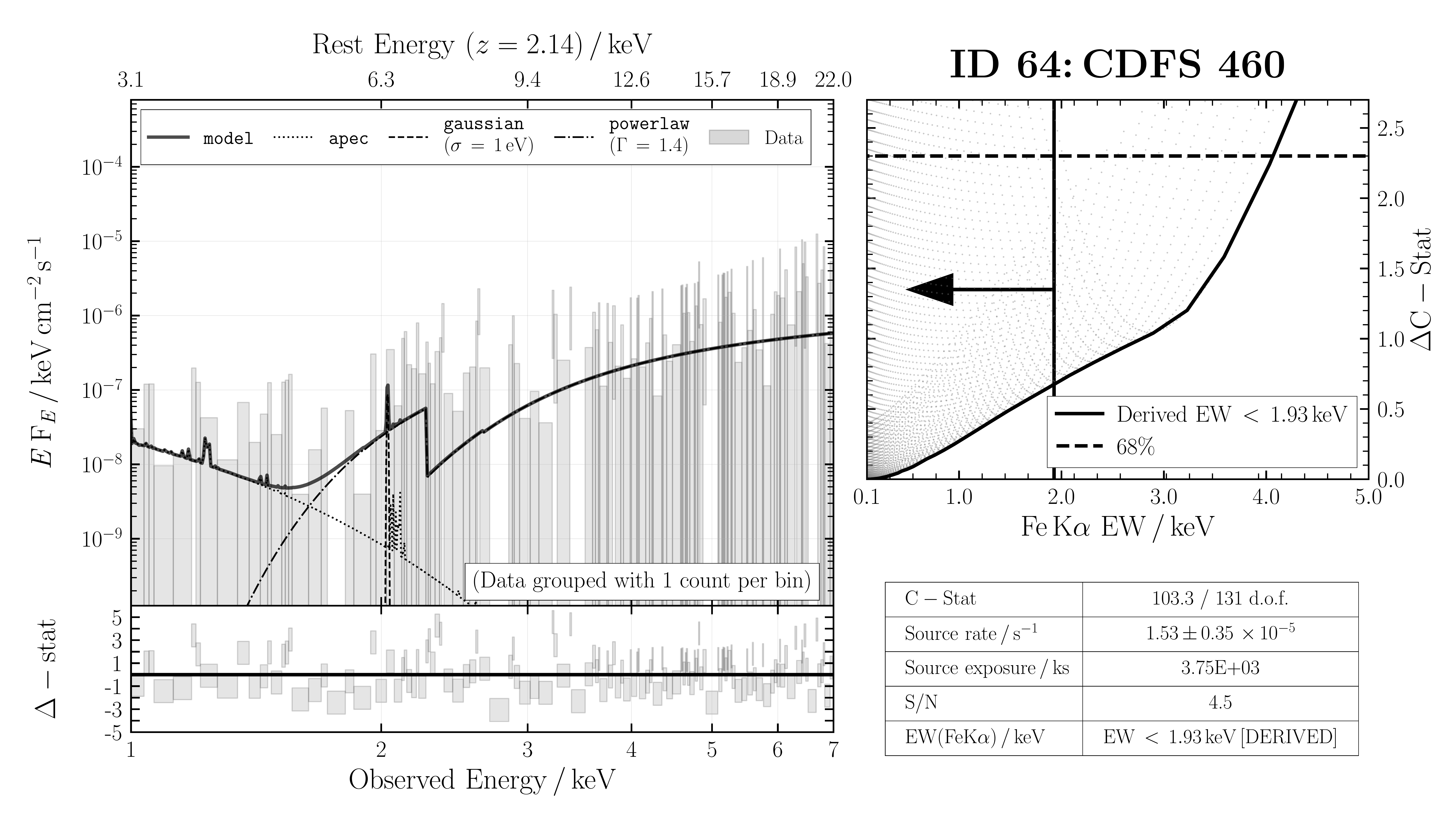}\\
\caption{\label{fig:CDFS460} ID 64: CDFS 460}
\end{center}
\end{figure}

\begin{figure}
\begin{center}
\includegraphics[angle=0,width=\columnwidth]{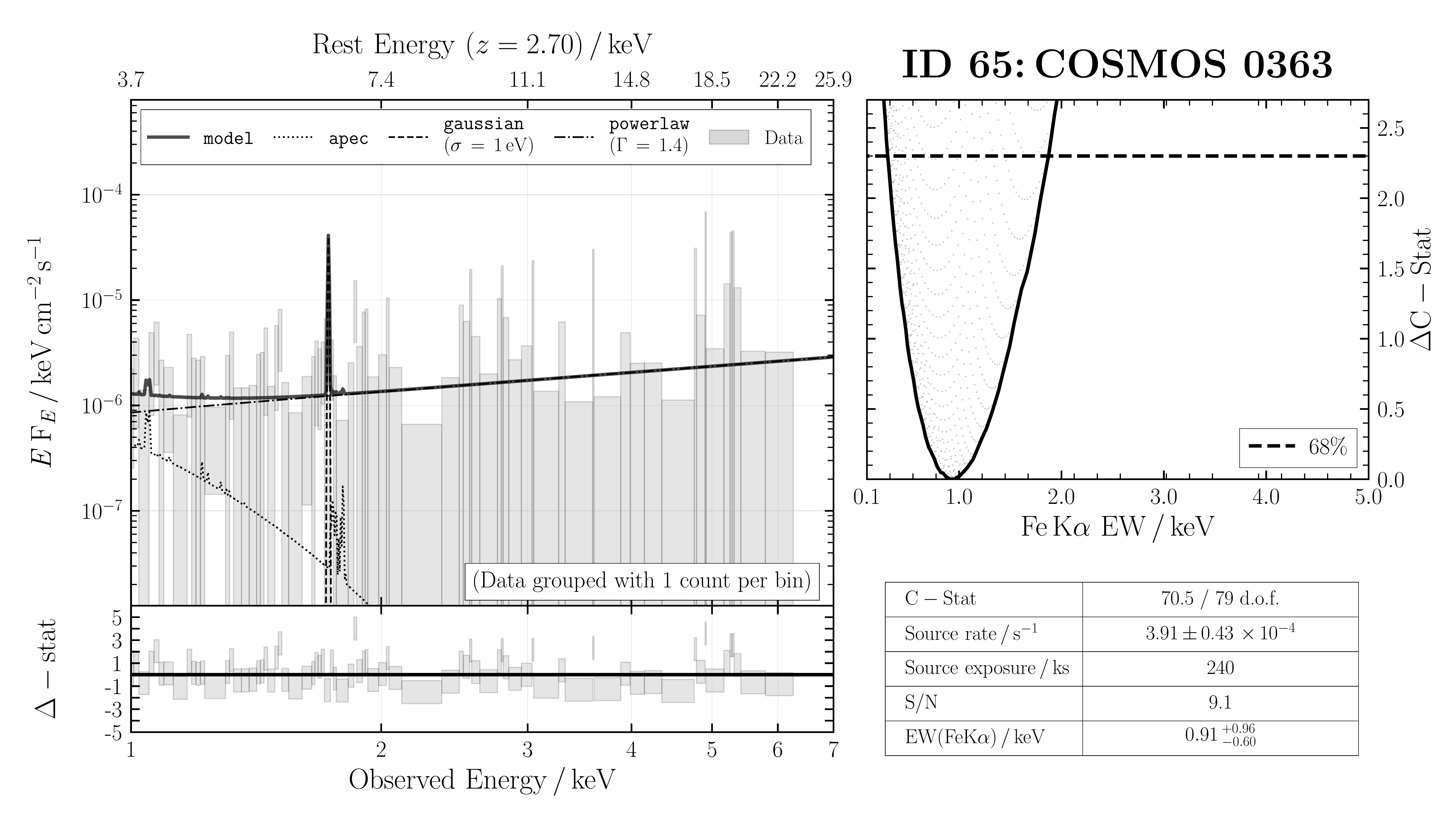}\\
\caption{\label{fig:COSMOS0363} ID 65: COSMOS 0363}
\end{center}
\end{figure}

\begin{figure}
\begin{center}
\includegraphics[angle=0,width=\columnwidth]{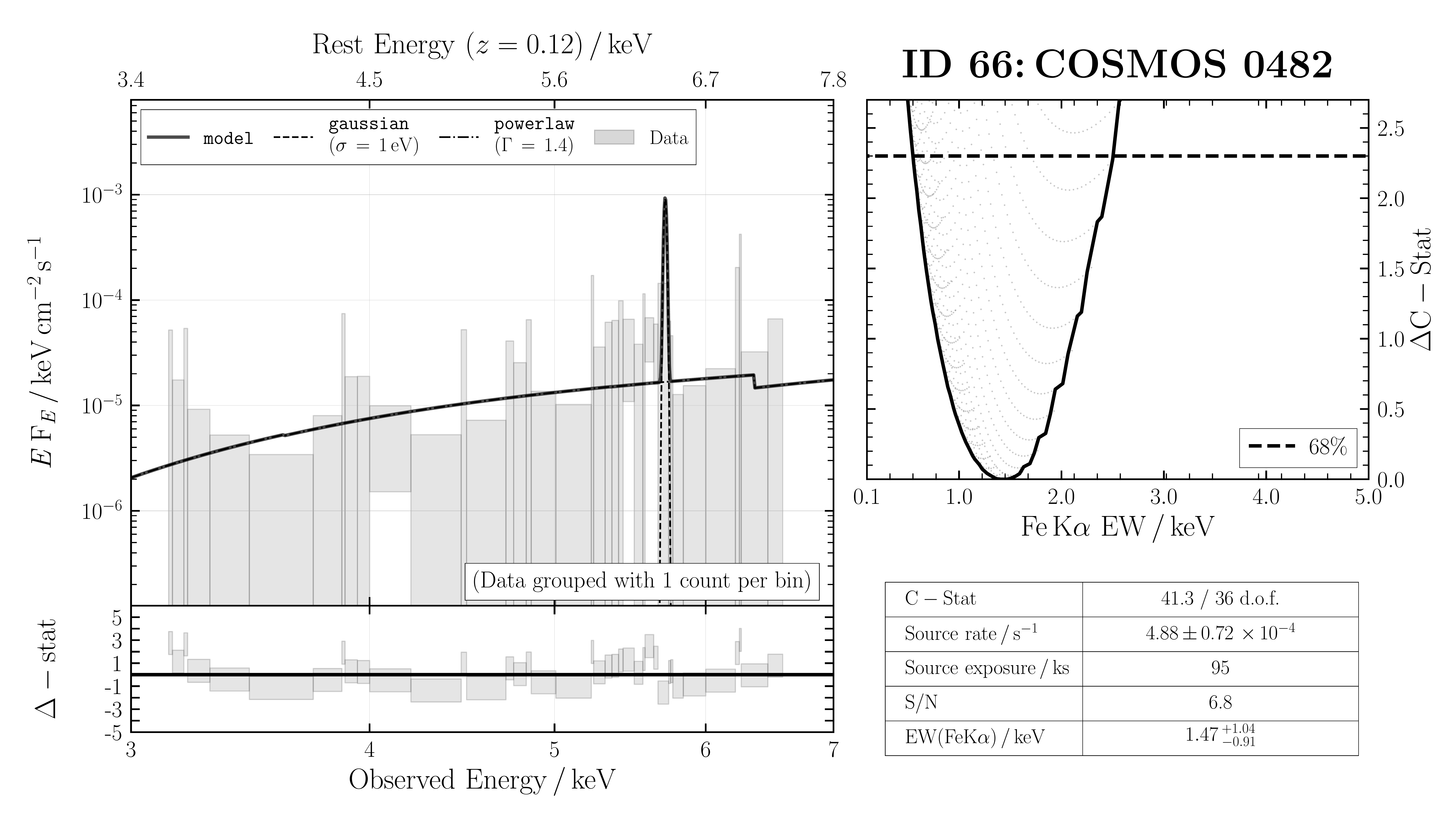}\\
\caption{\label{fig:COSMOS0482} ID 66: COSMOS 0482}
\end{center}
\end{figure}

\begin{figure}
\begin{center}
\includegraphics[angle=0,width=\columnwidth]{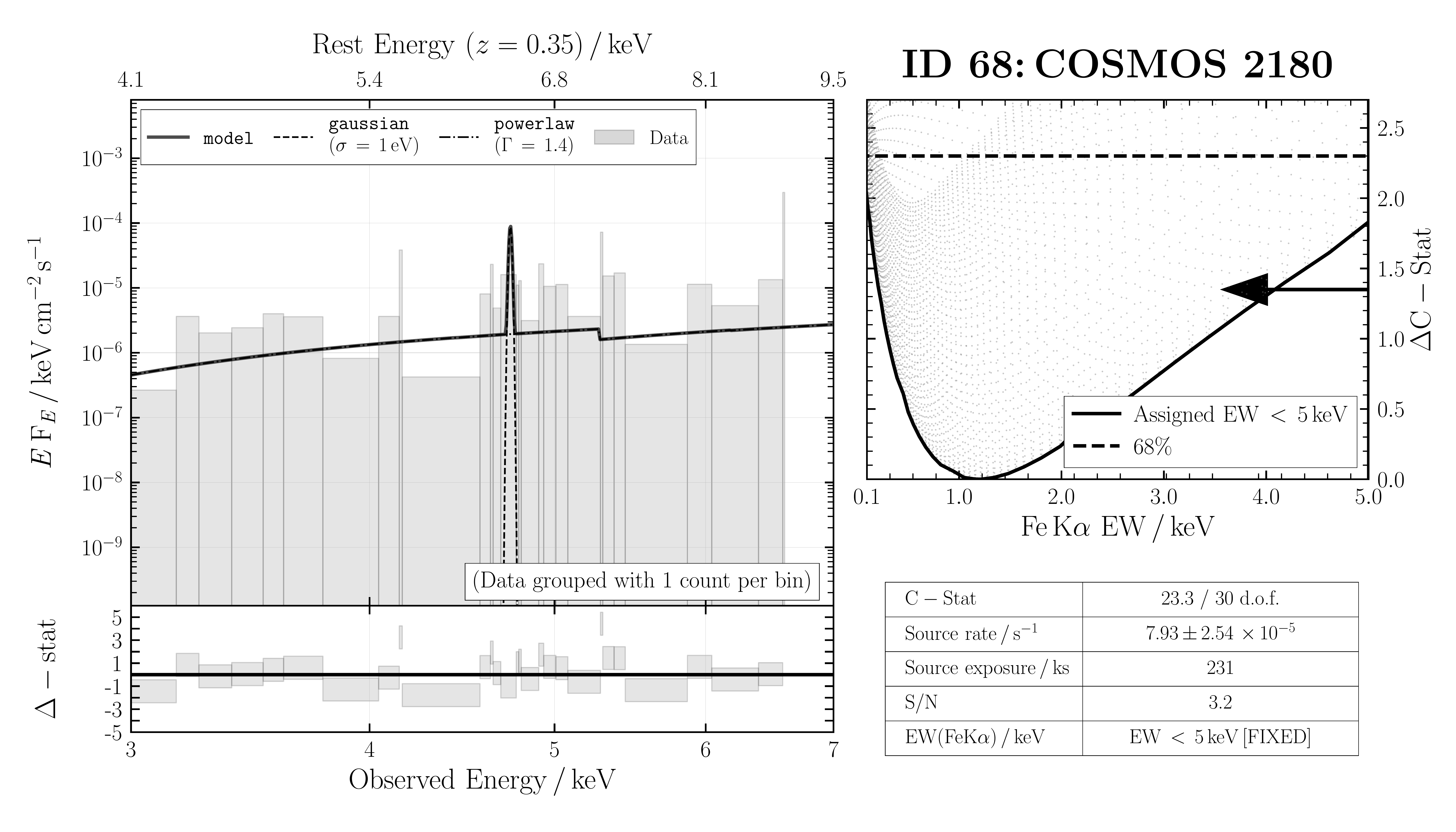}\\
\caption{\label{fig:COSMOS2180} ID 68: COSMOS 2180}
\end{center}
\end{figure}

\begin{figure}
\begin{center}
\includegraphics[angle=0,width=\columnwidth]{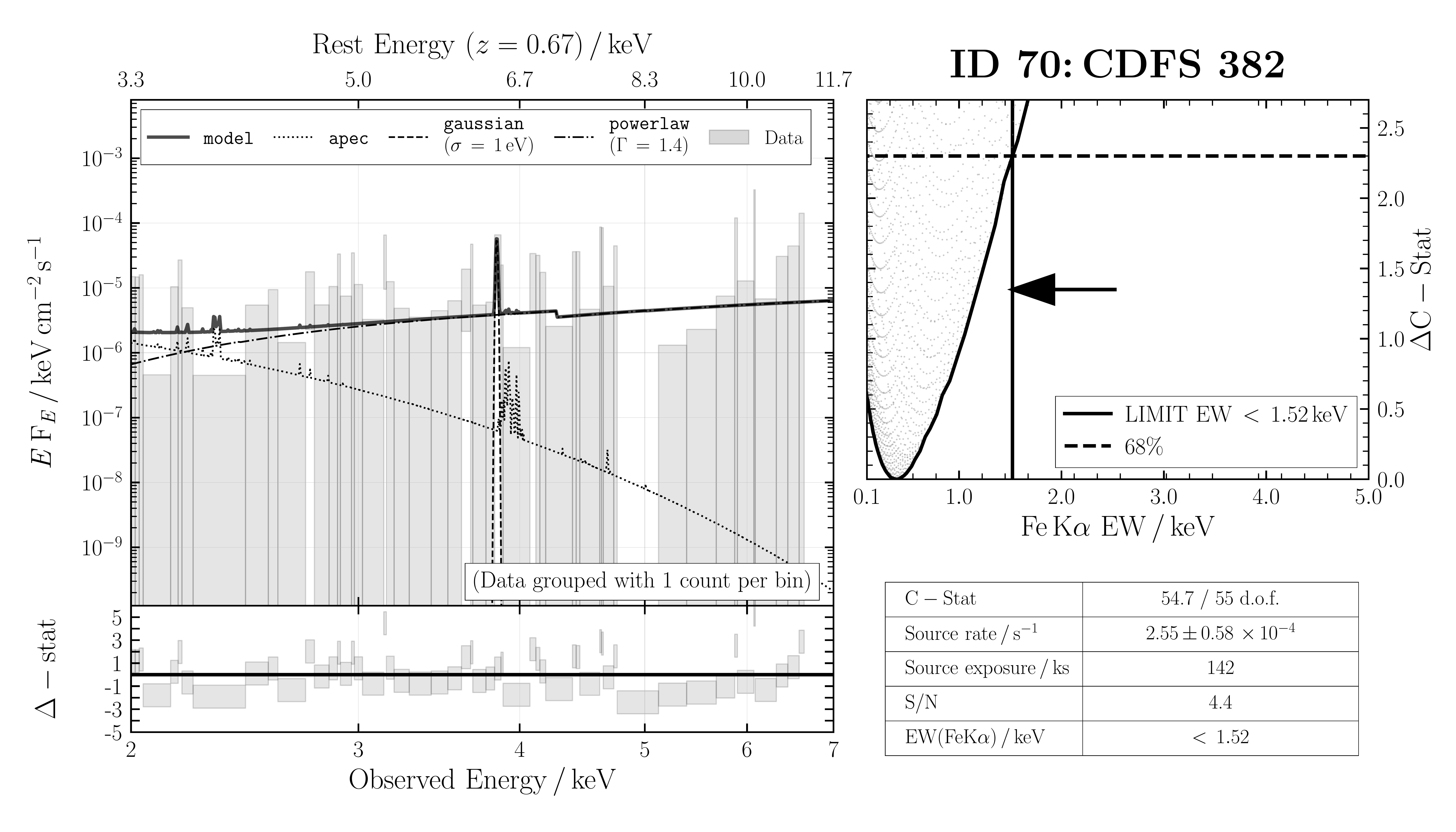}\\
\caption{\label{fig:CDFS382} ID 70: CDFS 382}
\end{center}
\end{figure}

\begin{figure}
\begin{center}
\includegraphics[angle=0,width=\columnwidth]{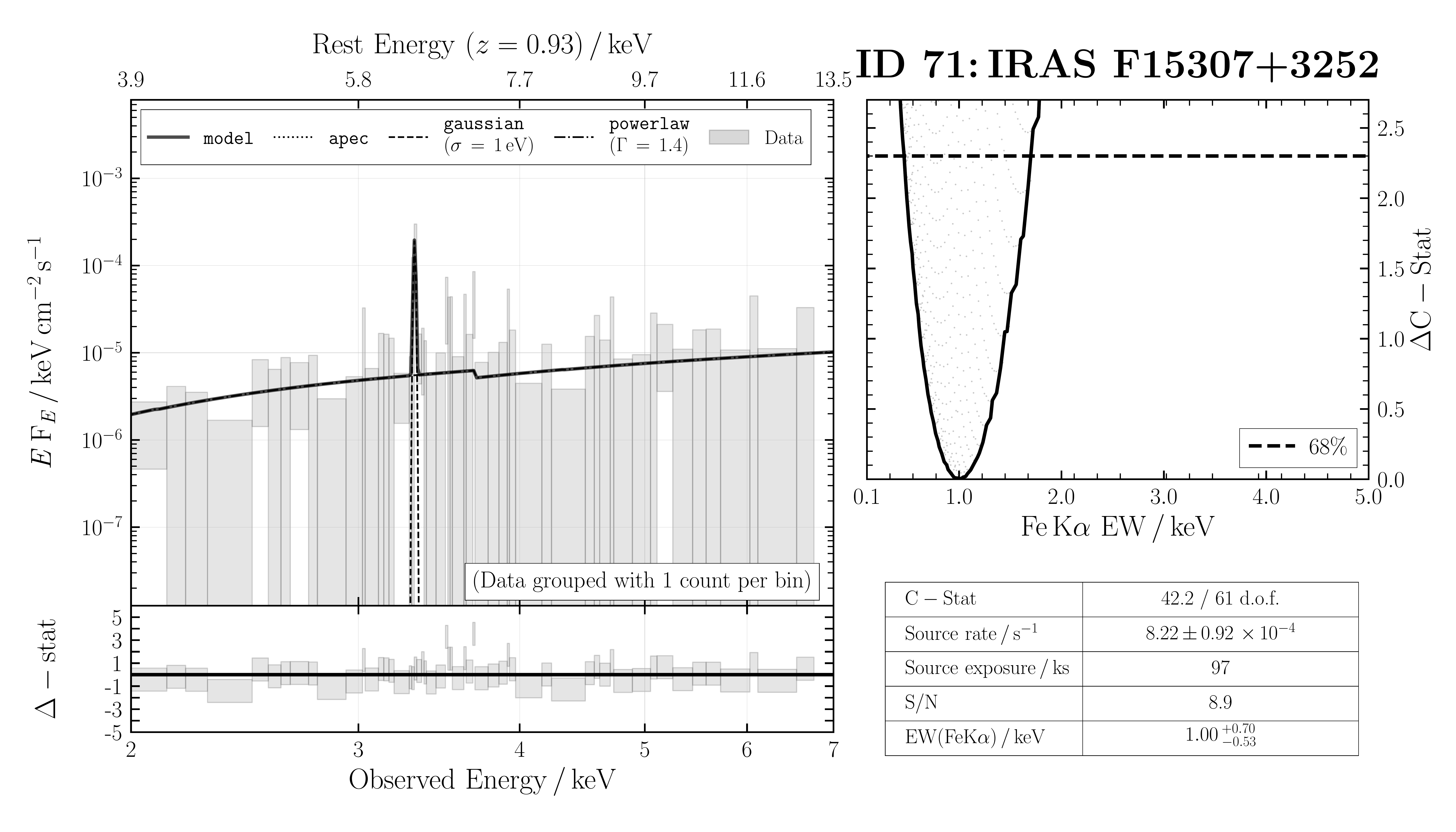}\\
\caption{\label{fig:IRASF15307p3252} ID 71: IRAS F15307+3252}
\end{center}
\end{figure}


\bsp	
\label{lastpage}
\end{document}